\documentclass[twoside]{book}
\usepackage{amsmath}
\usepackage{graphicx}
\usepackage{amsfonts}
\usepackage{amssymb}
\usepackage{setspace}
\usepackage[dvips]{epsfig}
\usepackage{helvet,euler}
\usepackage{fancyhdr}
\begin{document}
\frontmatter
\thispagestyle{empty}
\vspace*{1cm}

\begin{center}
{\bf \begin{large}I\end{large}NSTABILITIES, \begin{large}N\end{large}UCLEATION, AND \begin{large}C\end{large}RITICAL \begin{large}B\end{large}EHAVIOR\\
IN \begin{large}N\end{large}ONEQUILIBRIUM \begin{large}D\end{large}RIVEN \begin{large}F\end{large}LUIDS\\}
\vspace*{0.5cm}
{\bf \begin{large}T\end{large}HEORY AND \begin{large}S\end{large}IMULATION}
\end{center}

\newpage
\thispagestyle{empty}
\vspace*{10cm}
\newpage
\thispagestyle{empty}
\begin{figure}
\centerline{\includegraphics[width=4cm]{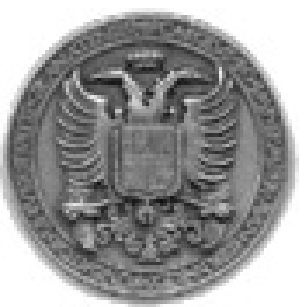}}
\end{figure}
\begin{center}
Institute `Carlos I' for Theoretical and Computational Physics \\
and Depto. de Electromagnetismo y F\'isica de la Materia \\
Universidad de Granada, Spain \\ 
\vspace*{0.75cm}

\begin{center}
{\bf \begin{large}I\end{large}NSTABILITIES, \begin{large}N\end{large}UCLEATION, AND \begin{large}C\end{large}RITICAL \begin{large}B\end{large}EHAVIOR\\
IN \begin{large}N\end{large}ONEQUILIBRIUM \begin{large}D\end{large}RIVEN \begin{large}F\end{large}LUIDS\\}
\vspace*{0.25cm}
{\bf \begin{large}T\end{large}HEORY AND \begin{large}S\end{large}IMULATION}
\end{center}


\end{center}

\vspace*{0.75cm}
\begin{center}
\large
{\bf Ph.D. Thesis}
\end{center}
\begin{center}
{MANUEL D\'IEZ MINGUITO}
\end{center}
\begin{center}
email: mdiezm@ugr.es\\
web: http://www.ugr.es/$\sim$mdiezm
\end{center}

\vspace*{0.65cm}
\begin{center}
\large
{\bf Supervisor}
\end{center}
\begin{center}
Prof. Dr. Joaqu\'in Marro
\end{center}

\vspace*{0.65cm}
\begin{center}
\large
{\bf Thesis Committee}\\
\end{center}
\begin{center}
Prof. Dr. J. Javier Brey (president)\\
Dr. Elena S\'anchez-Badorrey (secretary)\\
Prof. Dr. Margarida Telo da Gama (vocal)\\
Prof. Dr. Juan Colmenero de Le\'on (vocal)\\
Dr. Pep Espa\~nol (vocal)
\end{center}

\vspace*{0.85cm}
\begin{center}
Granada, February 21st, 2007
\end{center}

\newpage
\thispagestyle{empty}
\vspace*{10cm}

\newpage
\thispagestyle{empty}
\vspace*{6cm}
\begin{flushright}
\large
To Linda
\end{flushright}

\newpage
\thispagestyle{empty}
\vspace*{10cm}

\newpage
\thispagestyle{empty}
\vspace*{15cm}
\begin{flushright}
Then be it ours with steady mind to clasp\\ 
The purport of the skies- the law behind \\
The wandering courses of the sun and moon; \\
To scan the powers that speed all life below.
\end{flushright}
\begin{flushright}
Lucretius. \textit{De rerum natura.}
\end{flushright}





\newpage
\thispagestyle{empty}
\vspace*{10cm}

\chapter{Acknowledgments}

This stage of my life ends with two births: in March, my daughter Cora and now this thesis. While the sensations remain within the family in the former case, I wish to express my gratitude to those who contributed to the latter.

First of all, I wish to thank Joaqu\'in Marro, my professor, director, and mentor for all the unconditional support he offered me from the outset, knowing what was best for me at all times. Also, I wish to express my appreciation to Pedro Garrido for all the time he dedicated, for his special way of motivating me, for the many doubts he resolved and those he posed as well, and for his frankness. Part of this work is owed to the intuition of Paco de los Santos, my office mate, with whom I shared many good times and sincere chats, and whose intelligent humour I have enjoyed. Also, my gratitude goes to Miguel \'Angel Mu\~noz, for me a model to follow, his interest, support, and advice at right moments. And to all the rest of the members of the group: Omar, my travel companion during these last four years, always cordial, altruistic, and always willing to help; David Navidad, Jes\'us Cort\'es, Pablo Hurtado, Abdelfattah Achahbar, Antonio L. Lacomba, Joaqu\'in Torres, Don Jes\'us Biel, and Paco Ramos. Without this atmosphere of work and camaraderie in the group, this thesis simply would not have reached term.

Also, I feel a special gratitude, both at the professional and personal level, to Baruch Meerson. His commitment and dedication still overwhelm me; his sincere counselling, as well as his warmth and closeness helped me believe in this thesis and remain beyond any accolade. Nor could I ever forget the fine reception that he and his wife gave me in their marvellous city of Jerusalem.

During these years, I have had two good collaborators in Thorsten P\"oschel and Thomas Schwager, whom I thank for their interest and doubly for the warm welcome they offered me in Berlin.

How important friends are ---always present and about who there is little that can be put into words and much to appreciate. In the course of this undertaking, I have had the good fortune of knowing Juan Antonio, also a member of the group. With him, I have shared innumerable conversations, concerns, confidence and friendship. Thank you. To my now old friends: Fernando, who also knows what it is to write a thesis, struggling day to day, I am indebted for his experience and support at difficult times; to Miguel, for me more than a friend, for the enthusiasm and time that we invested in so many projects; to Jes\'us, for his exemplary honesty and faithfulness; and to the rest of the members of the team.

To my parents, who have done everything for me. To Gabi, Holger, and Christoph, for their unflagging interest and affection.

To Cora, because, despite having delayed the writing of this work, you have brought happiness into my life. And finally, to Linda, my wife and my guide, to whom I dedicate this thesis.

Thank you for everything, the expressible and the ineffable.

\vspace{2cm}
\begin{flushright}
Granada, a 17 de Septiembre de 2006.
\end{flushright}

\tableofcontents
\listoffigures
\listoftables
\mainmatter
\chapter{General Introduction\label{cap0}}
Nature has a hierarchical structure, with time, length, and energy scales ranging from the \textit{microscopic} to the \textit{macroscopic}. Surprisingly, it is often possible to discuss these levels independently, e.g., in the case of molecular fluids. The macroscopic level, i.e. the world directly perceived at our human senses, is described hydrodynamically. That is, by \textit{continuous} (or piecewise continuous) functions of the spatial coordinates \textbf{r} and time $t$ (hydrodynamic \textit{fields}). As a result, the great disciplines of macroscopic physics, such as fluid mechanics, elasticity, acoustics, electromagnetism, are field theories. These fields are determined by integro-differential equations involving unknown functions of \textbf{r} and $t$. The microscopic level is described as a collection of a very large number of constituents, such as atoms, molecules, or other more complex entities interacting with each other through well-defined electro-mechanical forces\footnote{The difference between \textit{microscopic} and \textit{macroscopic} levels is essentially relative. The key concept here is the large number of constituents, not their small size. For instance, a galaxy (macroscopics) is a \textit{object} composed by a large number of stars (microscopics), a colony of organisms is a large set of individuals, a gas comprised by a large number of molecules, etc.}. Their evolution with time is provided by the laws of quantum mechanics, but in many cases classical mechanics results a good approximation.

If there is a clear scale separation between microscopic and macroscopic, then an intermediate scale can be introduced. Such a level of description is referred to as the \textit{mesoscopic} level. It is a coarse--grained representation to capture the physics of microscopic models on the large length and time scales, which is described by stochastic partial differential equations \cite{vkampen,gardiner}. By concreteness, the \textit{mesoscopic} description of the system in question, whose dynamics is on the average governed by the laws of macroscopic time evolution, is a result of building up of microscopic fluctuations. This approach aims at understanding the slowly-varying long wavelength, low frequency properties of many-body systems and form the basis of the field theoretical analyses of critical phenomena. In the literature it is common include in this intermediate scale the kinetic description of gases, e.g., Boltzmann equation. Thus, this level is also referred to as the \textit{kinetic} level\footnote{Formally, both stochastic differential equations and kinetic equations may belong to the same level, although they involve different (mesoscopic) length and time scales.}. This is in fact a reduced microscopic level, described by one-particle coarse-grained distribution functions and the asymptotic, irreversible kinetic equations. 

The three different levels of description\footnote{There may be larger scale structures (e.g., the Karman vortex train) produced by a large scale fluid dynamics, but we do not pay attention to such large scales.} are, in principle, equivalent approaches of the same reality. Nevertheless, the form of the laws derived at these three levels are so different from each other that there soon appeared a stringent need for an explanation of hydrodynamics in terms of the microscopic evolution of the underlying collection of constituents. The bridges between them are provided by \textit{Statistical Mechanics}. Statistical mechanics aims to derive macroscopic behavior of matter originated in the cooperative behavior of interacting (microscopic) individual entities. Some of the phenomena are simple synergetic effects of the actions of individuals, e.g., the pressure exerted by a molecular gas on the walls of its container and thousands of fireflies flashing in synchrony; while others are paradigms of emergent behavior, having no direct counterpart in the properties or dynamics of individual constituents, e.g., the transition from laminar to turbulent flows in fluids and the one million of atoms which give rise to the program of life: the deoxyribonucleic acid or DNA for short. In the latter cases, the behavior of the constituents become singular, but very different from what they would exhibit in the absence of the others.

The most successful achievement of statistical mechanics is \textit{Ensemble Theory} \cite{libros_eq}, which yields the connection between the macroscopic properties of \textit{equilibrium} systems from the laws governing the microscopic interactions of the individual particles. It is said that a certain system stay at equilibrium when it is isolated, shows no hysteresis, and reaches a steady state (all its macroscopic properties are fixed) \cite{biel}. In that case, the macroscopic properties are expressed in terms of \textit{thermodynamic} variables. Indeed, the \textit{thermodynamics} is in this sense a kind of hydrodynamic description, which consists only of laws and relations between thermodynamic quantities. From the mathematical point of view, ensemble theory can be completely axiomatized. This provides us, at least in principle, analytic expressions for these thermodynamic quantities allowing us to derive all the relevant macroscopic information of the thermal system in question\footnote{Once the microscopic Hamiltonian (say $\mathcal{H}$) of the system is specified the ``canonical" stationary probability distribution $\mathcal{P}$ is known in terms of the Boltzmann factor $\mathcal{P}=e^{-\mathcal{H}/k_{B}T}/Z$, where $Z$ is the partition function, $k_{B}$ is the Boltzmann's constant, and $T$ the temperature. Thus, averages over such a distribution of time-independent observables can be computed. The remaining difficulties are merely technical.}, e.g., the free energy, equation of state, etc.

\begin{figure}
\centerline{\includegraphics[width=5.0cm]{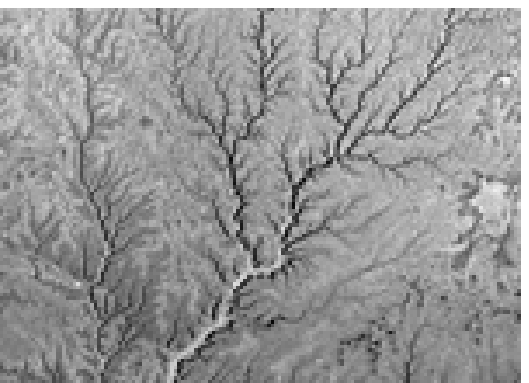}
\includegraphics[width=5.0cm]{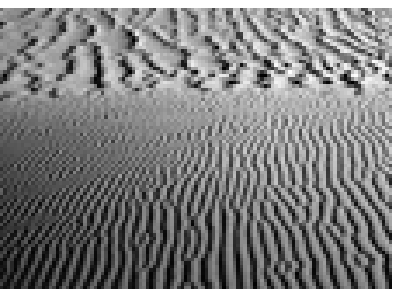}}
\centerline{\includegraphics[width=5.0cm]{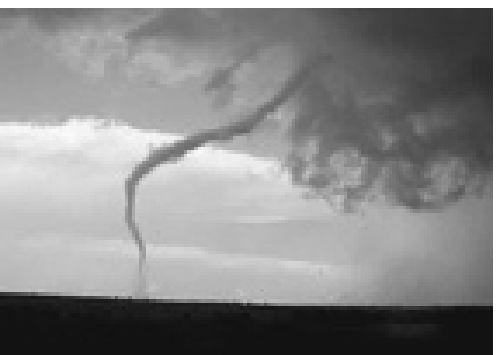}
\includegraphics[width=5.0cm]{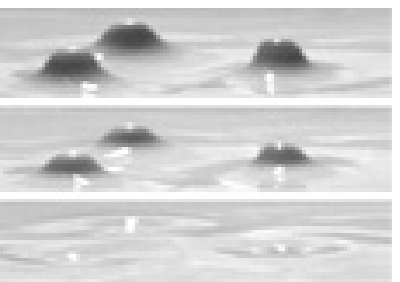}}
\caption[Complexity of the macroscopic behavior]{Complexity arises at all levels. Some amazing examples are the following. Upper left: river networks in Yemen (source NASA). Upper right: aeolian patterns in Colorado, USA (photo by Bob Bauer). Lower left: tornado as a rotating geophysical flow. Lower right: \textit{oscillons} in a colloidal suspension\cite{clay}.}
\label{examples}
\end{figure}

In nature, by contrast, equilibrium systems are an exception (even an idealization) rather than the rule: \textit{non-equilibrium} phenomena are overwhelmingly more abundant. Galaxies, human beings, chemical reactions, geophysical flows, carriers in semiconductor devices, financial stock markets, ratchet-effect transport, traffic flow on a highway, to cite just a few, are (many-body) systems under nonequilibrium conditions\footnote{Nonequilibrium phenomena are also encountered whenever systems are relaxing towards an equilibrium steady state. Nevertheless, in most of cases these can be understood in terms of equilibrium concepts.}. Out-of-equilibrium systems are characterized by not being closed, i.e., by having an exchange of energy, particles, and/or information with their environment. In general, the state of a nonequilibrium system is not determined solely by external constraints, but also depends upon its history (hysteretic systems). Consequently, this gives rise to the enormous complexity of our world which occurs to all levels of description \cite{kadanoff_complex} and manifests in pattern formation, absorbing states, self-organization, chaos, ageing, avalanches, morphogenesis, oscillations, fractals, etc \cite{cross,gollub,barabasi,bak}. All these amazing and complicated phenomena are associated to \textit{instabilities}\footnote{Nonequilibrium instabilities are attended by ordering phenomena analogous to those of equilibrium statistical mechanics; one may therefore refer to \textit{nonequilibrium phase transitions} \cite{marro}. Although nonequilibrium phase transitions represent a more varied picture than their equilibrium counterparts.}, which are variously described as nonequilibrium phase transitions \cite{cross,marro,haken,gar}, bifurcations or synergetics, with the aim of connecting microscopics with the coherent structures observed.

However, most of studies of nonequilibrium systems adopt a macroscopic (phenomenological) point of view. In fact, little is known in general on why are there such an interesting structures or how this complexity arrises from interaction at a microscopic level. Consequently, no theory exists and the development of a solid theoretical background is nowadays in an early stage compared with equilibrium, where the ensemble theory holds successfully. As a matter of fact, the key point of difference between equilibrium and nonequilibrium statistical mechanics is that whereas in the former case the stationary probability distribution is known, out of equilibrium one must find the time-dependent distribution, which obey a general evolution equation, i.e., a \textit{master equation}. In most of soluble cases ---which in fact are just a few---, this can be done only approximately. One can only provide some unified view when the systems are \textit{not too far}\footnote{Nevertheless, a precise definition of \textit{not too far} is rather difficult to give; the borderline is rather unclear.} from equilibrium \cite{mazur}, where the system can be treated perturbatively around the equilibrium state and studied using linear response theory. Nevertheless, our attention here lies at systems far from equilibrium where such schemes break down.\\

\textbf{\textit{Within the framework given above, the main subject of this thesis rests on the study ---at different levels of description--- of instabilities in systems which are} driven\textit{, i.e., maintained far from equilibrium by an external forcing. We focus here on two main classes, namely, }driven--diffusive fluids \textit{and} driven granular gases.}\\

The first family (\textit{driven--diffusive fluids}\footnote{We should mention here that in the literature these systems are commonly referred as driven--diffusive \textit{systems} instead of driven--diffusive \textit{fluids}, which is adopted in this thesis. As we shall describe in some detail in Chapter~\ref{cap1}, we reserve the former denomination for a mean--field mesoscopic description.}) corresponds to systems which are\footnote{Following Schmittmann \& Zia in Ref.~\cite{zia}.} coupled to two reservoirs of energy in such a way that there is a steady energy flow through the system. Clearly, this definition comprises a bunch of very varied systems. We restrict ourselves to systems in which a (non-zero) current of particles (whose number is a conserved quantity) is set up through the system, there is spatial anisotropies associated with an external driving field, and eventually a nonequilibrium steady state is reached. The simplest example seems to be a resistor gaining energy from a battery and losing it to the atmosphere. But even for this reduced class of systems the stationary probability distribution is unknown.

The basic motivation behind the introduction of \textit{driven--diffusive fluids} corresponds to a need of unravel the key ingredients of nonequilibrium behavior. A reasonable approach consist in investigating systems as simple as possible, which aim for capture the microscopic essentials yielding to the complicate, macroscopic nonequilibrium ordering. These microscopic models are usually oversimplified representations of real systems, and consider entities as particles interacting via simple rules. Between them, \textit{lattice} models ---which are based in discretization of space into lattice sites--- have played a important role due to the fact that they sometimes allow for exact results, and allow one to isolate the specific features of a system. In addition, lattice models allow us to obtain the intuition one needs in order to develop the appropriate theoretical tools and, from a experimental point of view, are easier to be implemented in a computer. Such an approach has given rise to a lively research activity in the last decade \cite{marro,zia,liggett,privman,droz,hinrichsen,odor} and a bunch of emerging techniques may now be applied to lattice systems, including nonequilibrium statistical \textit{field theory}. A general amazing result from these studies is that lattice models often capture some essentials of social organisms \cite{ferreira,treves,migrating}, formation of river networks \cite{river_networks}, epidemics \cite{epidemics}, glasses, electrical circuits, traffic \cite{traffic,helbing}, hydrodynamics, colloids and foams, enzyme biology \cite{enzyme}, living organisms \cite{lanes}, and markets \cite{econophys}, for example. In particular, lattice models have succeed in understanding (equilibrium) phase transitions and critical phenomena \cite{liggett,onsager,thompson} due, in many aspects, to renormalization group methods \cite{amit,renormalization}.
 
The central result in (equilibrium) critical phenomena is \textit{universality}, i.e., the behavior of disparate systems in the vicinity of critical points are determined solely by basic features ---dimensionality, range of forces, symmetries, etc.--- and does not seem to depend to any great extent on the particular system \cite{amit,ma,binney}. One can therefore hope to assign all systems to classes each of them being identified by a set of critical indices (exponents and some amplitude ratios). How much of these concepts apply to nonequilibrium phase transitions are only just emerging, and one expects lattice models to be equally important here. These issues are considered in Chapter~\ref{cap1}, in which a particular driven-diffusive lattice model, prototype for nonequilibrium phase transitions, is investigated. However, a well-known disadvantage of lattice models is that when they are compared directly with experiment result too crude. This has to be understood in the sense that they often do not account for important features of the corresponding nonequilibrium phase diagram, such as structural, morphological, and even critical properties. Furthermore, theoreticians often tend to consider them as prototypical models for certain behavior, a fact which is in many cases not justified. This will be discussed in Chapter~\ref{cap2}, where we introduce a novel, \textit{realistic} model for computer simulation of anisotropic fluids.\\

The second class of systems we consider in this thesis concerns \textit{driven granular gases}. A granular material is a large conglomerate of discrete macroscopic\footnote{In the sense that they are directly perceived at our human eyes.} (classical) particles. Examples, which occur at very different length scales, may range from powders to intergalactic dust clouds; they include sand, concrete, rice, volcanic flows, Saturn ring's, and many others (for a review, see \cite{jaeger}). They have important applications in industrial sectors as pharmaceutical, construction and civil engineering, chemical, food and agricultural, etc. On a larger scale, they are also relevant in understanding many geological and astrophysical processes. In addition, they are closely related to a broader class of systems, such as foams, colloids and glasses. The collective behavior of an assembly of granular particles exhibits an impressive variety of phenomena\footnote{The size range of phenomena goes from $\approx 10^{-6} m$ in powders, to $\approx 10^{3}m$ in deserts, up to $\approx 10^{20}m$ in planetary rings.} which include a plethora of pattern forming instabilities \cite{melo,umbanhowar,ristow,aransonRMP}, clustering \cite{macnamara,goldhirsch2,olafsen}, flows and jamming in hoppers \cite{hoppers}, avalanches and slides \cite{avalanches}, mixing and segregation \cite{segrega}, convection and heaping \cite{convect1,convect2}, eruptions \cite{eruptions,thorsten3}, to cite just a few. Clearly, their relevancy is beyond any doubt, and understanding their properties is not only an urgent industrial need, but also an important challenge for physicists. 

Since grains have a macroscopic size, friction and restitutional losses from collisions give rise to dissipative interactions ---kinetic energy is continually transferred into heat---. Although the dominant interactions depend on particle size. The relevant energy scale for a typical, say, rice grain of length $l\approx 5mm$ and mass $m$ is its potential energy $mgl$, where $g$ is the gravitational acceleration. For much smaller grain sizes other kind of interactions may become important\footnote{For instance, it may occur charging and surface coating for $d\approx 0.1$, or magnetization and surface adhesion for $d\approx 0.01$.}. Hence, in granular materials the thermal energy $k_BT$ is irrelevant, i.e. $k_BT\ll mgl$. This implies that dynamical effects outweigh entropy considerations \cite{jaeger} and, therefore, neither equilibrium statistical--mechanics nor thermodynamics arguments are useful. Furthermore, due to the fact that granulates are comprised by a finite
number of particles\footnote{In contrast with molecular systems $N<<N_A$, where $N_A$ is the Avogadro number. Nevertheless, in certain cases a few thousands or even hundreds of particles are enough to allow for statistical approaches.}
usually there is a strong effect of fluctuations and, therefore, statistics may be dominated by \textit{rare events}. Other important effects involve the interstitial fluid, e.g. air or water, although in many situations one can ignore it with confidence and consider the grain--grain interactions alone. In this thesis (specifically in Part~\ref{part:Granos}) we will restrict ourselves to \textit{dry} granular systems, where the interaction between particles take place mainly via contact forces. 

However, in spite of the simple (classical) features which characterize them, granular materials behave in a rather unconventional way: their phases ---solid, liquid, and gas--- have complex--collective properties that distinguish them from molecular fluids and solids. As a result, their statistics and dynamics are poorly understood. In fact, from a theoretical point of view, these systems are still only understood in terms of predictions of a general nature and many open questions remain. Linear elasticity does not apply for granular \textit{solids}, which are highly histeretic, show static stress indeterminacy, and possess static yield shear stress. 
The interlocking of grains leads to force chains, jamming, dilation on shearing, etc. \textit{Liquids} are viscous, show avalanches (possess a critical slope), size segregation in mixtures, shear bands (narrow regions separating blocks moving with different velocities), pattern formation. \textit{Gases}, which can only persist with continuous energy supply, are highly compressible, show clustering, long range correlations, inelastic collapse in computational models, non-maxwellian distribution functions, pattern formation, and lack of scale separation between macroscopic scales and microscopic ones. All of this phenomenology makes difficult to cast them in the framework of the classical three different phases of matter. Some of the interesting open issues and controversies that we address in Chapters~\ref{cap3} and \ref{cap4} are: What are the statistical properties of granular systems, how do granular phase changes occur, and what are the optimum continuum models and when do they apply. In particular, we focus on the continuum description of clustering, symmetry breaking, and phase separation instabilities in granular gases. As we shall see, these instabilities provide sensitive tests to models of granular flows and contribute to the understanding of pattern formation far from equilibrium.

\section{Overview\label{cap0:over}}
In this thesis, we investigate the two aforementioned types of systems, which have become of paramount importance in recent years. The study is divided into four parts. The Part~\ref{part:DDF}, which comprises Chapters~\ref{intro:DDF}, \ref{cap1} and \ref{cap2}, is devoted to \textit{driven-diffusive fluids}. Part~\ref{part:Granos} deals with \textit{driven granular gases}, and consists of Chapters~\ref{intro:Granos}, \ref{cap3} and \ref{cap4}. Part~\ref{part:apend} includes the Appendix~\ref{apendA}, \ref{apendC}, \ref{apendB}, and \ref{apendD}. Finally, we close (Part~\ref{part:Concl}) with a detailed summary and the main conclusions.\\

The first part begin in \textbf{Chapter~\ref{intro:DDF}} by introducing the \textit{driven lattice gas} ---prototype for nonequilibrium anisotropic phase transitions---. This model may be viewed as an oversimplification of certain traffic and flow problems. The particles interact via a local anisotropic rule, which induces a preferential hopping along one direction, so that a net current sets in if allowed by boundary conditions. We discuss some of its already known properties and controversies. We shall describe in greater detail the computational and field--theoretical methods employed in Chapters~\ref{cap1} and \ref{cap2}.\\

One of the most important questions we can ask about any model is whether the behavior that it displays is universal. In \textbf{Chapter~\ref{cap1}} we address this question and study the phase segregation process in the prototypical lattice (microscopic) model introduced in Chapter~\ref{intro:DDF}. The emphasis in our study is on the influence of dynamic details on the resulting (non-equilibrium) steady state, although we also pay some attention to kinetic aspects. In particular, we shall discuss on the similarities and differences between the \textit{driven lattice gas} and related lattice and off--lattice models, from a simulational and theoretical point of view. The comparison between them allows us to discuss some exceptional, hardly realistic features of the original \textit{driven lattice gas}. In addition, we test the validity of the two competing mesoscopic (Langevin--type) approaches on describing the critical behavior of these models.\\

We introduce in \textbf{Chapter~\ref{cap2}} a novel, \textit{realistic} off--lattice driven fluid for anisotropic behavior. We describe short--time kinetic and steady--state properties of the non--equilibrium phases, namely, solid, liquid and gas anisotropic phases. This is a computationally--convenient model which exhibits a net current and striped structures at low temperature, thus resembling many situations in nature. We here focus on both critical behavior and some details of the nucleation process. Concerning the critical behavior, we discuss on the role of symmetries on computer and field--theoretical modeling of non-equilibrium fluids. In addition, some important comparisons with experiments are drawn.\\

\textbf{Chapter~\ref{intro:Granos}} serves as an introduction to Part~\ref{part:Granos}, which is devoted to driven granular gases. We briefly outline the current situation with regard to research on granular gases. In particular, we discuss on the applicability of a hydrodynamic theory to granular flows. We also present the basic ingredients of the studied models and the computer simulation methods.\\

In \textbf{Chapter~\ref{cap3}} we address phase separation instability of a monodisperse gas of inelastically colliding hard disks confined in a two-dimensional annulus, the inner circle of which represents a ``thermal wall". Our main objectives are to characterize the instability and compute the phase diagram by using granular hydrodynamics ---we employed for hydrodynamics Enskog--type constitutive relations, which are derived from first principles--- and event driven molecular dynamics simulations. We also discuss on clustering and symmetry-breaking instabilities. By focusing on the annular geometry, we hope to motivate experimental studies of the granular phase separation which may be advantageous in this geometry.\\

This \textbf{Chapter~\ref{cap4}} addresses granular hydrodynamics and fluctuations in a simple two--dimensional granular system under conditions when existing hydrodynamic descriptions break down because of large density, \textit{not} large inelasticity. We study a model system of inelastically colliding hard disks inside a circular box, driven by a thermal wall in which a dense granular cluster behave like a \textit{macro-particle}. Some features of the macro--particle are well described by the stationary solution of granular hydrostatic equations by using phenomenological constitutive relations. Hydrostatic predictions are tested by comparing with molecular dynamics simulations. We are able to develop an effective Langevin description for the macro-particle, confined by a harmonic potential and driven by a delta-correlated noise. We are also concerned about whether the crystal structure depends on the geometry.\\

The Part~\ref{part:apend} is comprised by four appendices: \textbf{Appendix~\ref{apendA}} is devoted to the most important properties of the \textit{Lattice Gas}\footnote{The \textit{lattice gas} is the conserved version of the Ising model \cite{thompson,ising}. As a result, they are isomorphic.}, which is the equilibrium counterpart of the \textit{driven lattice gas} studied in Chapter~\ref{cap1}; in \textbf{Appendix~\ref{apendC}} we introduce a model which aim at characterizing the stability of the interface in driven lattice models; \textbf{Appendix~\ref{apendB}} details some field theoretical derivations performed in Chapter~\ref{cap1} concerning Langevin-type equations; and in \textbf{Appendix~\ref{apendD}} we propose a set of constitutive relations for the hydrostatic description of three-dimensional granular gases.\\

Finally, after the appendices, we present in Part~\ref{part:Concl} our main conclusions, summing up our original contributions and pointing out the future work that should be addressed. Also we provide a list of publications associated with the work performed in this thesis.\\ 

\part{\label{part:DDF}Driven Diffusive Fluids}
\chapter{Introduction\label{intro:DDF}}

As outlined in the general introduction (Chapter~\ref{cap0}), the first part of this thesis is devoted to the study of nonequilibrium phase transitions in \textit{Driven Diffusive Fluids}. This is a broad class of nonequilibrium systems which are characterized to be coupled to two reservoirs of energy. The one injects energy continously whereas the other substracts it in such a way that a nonequilibrium steady state is reached asymptotically. Studies of these systems have generally focused on lattice systems, i.e., simplified models based on a discretization of space and in considering interacting particles that move according to simple local rules. This is because a general formalism analogous to equilibrium statistical mechanics is lacking even for nonequilibrium steady states. The way to overcome this drawback is to develop and study simple models ---while retaining the essence of the difficulties of nonequilibrium states--- capable to capture the essential behavior.\\

\begin{figure}
\centerline{\includegraphics[width=10cm,clip=]{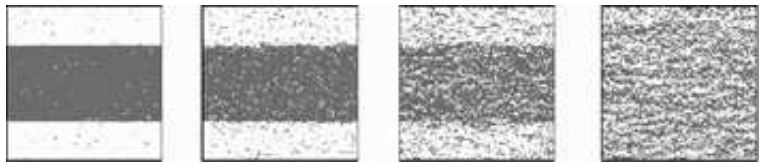}}
\caption[Typical configurations in the stationary regime of the DLG for a $128\times 128$ half-filled lattice]{\label{fig1_DDF} Typical configurations in the stationary regime of the DLG for a $128\times 128$ half-filled ($\rho=1/2$) lattice and (from left to right) $T=0.70$, $T=0.95$, $T=1.30$, and $T=1.50$. The temperatures are rescaled according with the equilibrium lattice gas critical temperature (see Appendix~\ref{apendA}): $T_{\text{Ons}}=2.269$ in energy units. The value of the field is $E=1000$, which is effectively infinite at all temperatures.}
\end{figure}

Within this context, one of most relevant models is the one devised by Katz, Lebowitz and Spohn\footnote{Sometimes is referred to as the KLS model, after the initials of its inventors.} \cite{kls}. Motivated by both the theoretical interest in nonequilibrium steady states and the physics of fast ionic conductors \cite{marro}, they introduced a minor modification to a well-known system in equilibrium statistical mechanics: the Ising model \cite{thompson,ising}. It consists of particles ---instead of spins--- diffusing on a lattice, subject to an excluded volume constraint and an attractive nearest neighbor interaction. The particles, whose total number remains conserved, are driven by an external (constant) field $E$, settled on a thermal bath at temperature $T$. We shall refer henceforth to this model as the \textit{Driven Lattice Gas} (DLG) \cite{marro,zia}. The dynamics in the DLG is according to Metropolis \cite{metropolis} transition probabilities per unit time. Consequently, for periodic boundary conditions (and for any non-zero driving field), the result is a net current of particles that generates heat which is absorbed by the thermal bath. As $E$ is increased, the system reaches saturation, i.e., particles cannot jump against the field. Eventually, anisotropic phase segregation occurs in the DLG for low enough bath temperatures: a striped liquid (rich particle) phase, which coexists with its vapor (poor particle) phase, percolates along the field direction. However, for high enough temperatures a disordered phase appears, characterized by a gas-like state. As a matter of fact, for two dimensional half-filled lattices the DLG undergoes a second-order phase transition; otherwise the transition is of first-order. Its phase behavior is best illustrated by `snapshots' of typical configurations generated by \textit{Monte Carlo} (MC) simulations \cite{mc_sims}, as in Fig.~\ref{fig1_DDF}. 

In spite of a deceptively simple definition, the steady states observed exhibit a rich and counterintuitive behavior, with many interesting features such as the violation of the fluctuation-dissipation theorem, a singular structure factor, and slow decay of spatial correlations. But perhaps its most striking, counterintuitive feature is that a strong field raises the critical temperature above the equilibrium lattice gas\footnote{See Appendix~\ref{apendA} for a quick review of the lattice gas properties.} critical temperature. We shall discuss these features in greater detail in Chapters~\ref{cap1} and \ref{cap2}. The DLG has been also useful to model, for instance, ionic currents \cite{marro,ionic}, traffic and pedestrians flows \cite{antal,helbing2}, water-in-oil microemulsions \cite{microemulsions}, electrophoresis experiments \cite{electrophoresis}, etc (see Ref.~\cite{zia} and references therein.) Moreover, the highly anisotropic configurations exhibited by the DLG resembles actual driven systems. Many natural systems exhibit spatial striped patterns on macroscopic scales \cite{cross,haken}. These are often caused by transport of matter or charge induced by a drive which leads to heterogeneous ordering. Such phenomenology occurs during segregation in driven sheared systems \cite{exp,beysens2,follow}, in flowing fluids \cite{rheology2}, and during phase separation in colloidal \cite{loewen3}, granular \cite{reis,sanchez}, and liquid--liquid \cite{liqliq} mixtures. Further examples are ripples shaped by the wind in sand deserts \cite{dunes,dunes2}, the lanes and trails formed by living organisms (animals and/or pedestrians) and vehicle traffic \cite{helbing,lanes}, and the anisotropies observed in high temperature superconductors \cite{cuprates0,cuprates1} and in two-dimensional electron gases \cite{2deg1,mosfet}. 

As a result, this model system became the theoreticians' prototype for anisotropic behavior, and is nowadays the most thoroughly studied system showing an anisotropic nonequilibrium phase transition. \\

In Chapters~\ref{cap1} and \ref{cap2} we report on novel findings on this and other related models by describing different realizations observed in (Metropolis) MC simulations. In particular, we study to what extent the DLG features are universal, or in other words, how robust is its behavior. The clue is that the DLG is, in a sense, pathological. We also devise a novel driven fluid for anisotropic phenomena, which we believe is not only of theoretical importance, but also relevant for experimentalists. In addition to MC techniques, field-theoretical Langevin-type approaches on the DLG are also developed. But before deep further on these matters, we shall describe in greater detail the computational and field-theoretical methods employed in these chapters.\\

In statistical mechanics, \textit{Monte Carlo}\footnote{More precisely, one should refer it as the \textit{Monte Carlo importance sampling} algorithm \cite{metropolis,mc_sims}. The nickname ``Monte Carlo" comes from the fact that this technique entails a large sequence of random numbers.} simulations \cite{mc_sims} play a major role on the study of phase transitions and critical phenomena. Any MC algorithm generates stochastic trajectories in the system's phase space, in such a way that the properties of the system are derived from averages over the different trajectories. To be specific, for systems both in and far from equilibrium, MC simulations involve long sequences of configurations, which evolve from one $\mathbf{c}$ to another $\mathbf{c}^{\prime}$ according to a defined transition probability per unit time (\textit{rate}), namely, $w(\mathbf{c}\rightarrow\mathbf{c}^{\prime})$. Once initial transients have decayed stationary observables can be computed as configurational averages. For systems under equilibrium conditions, such an average over configurations is referred to as the \textit{ensemble} average. In such a case, and if the \textit{ergodic hypothesis} \cite{libros_eq} is assumed, ensemble averages are equivalent to time averages\footnote{In fact, if MC simulations mimic configurational averages, Molecular Dynamics is a scheme for studying the natural time evolution (time averages) of a certain system. For thermalized (or equilibrated) systems differences between both techniques should disappear in the \textit{thermodynamic limit}. Chapters~\ref{cap3} and \ref{cap4} deal with Molecular Dynamic simulations, which are briefly described in Chapter~\ref{intro:Granos}.}. 

The acceptance rules for transitions between configurations are chosen such that these configurations occur with a frequency prescribed by the desired probability distribution. For equilibrium systems, this must be the ``canonical" stationary Gibb's distribution $\mathcal{P}_{0}=e^{-\mathcal{H}/k_{B}T}/Z$, where $H$ is the Hamiltonian, $Z$ is the partition function, $k_{B}$ is the Boltzmann's constant, and $T$ the temperature. Since our interest here is in nonequilibrium behavior, we will need to specify how a given configuration $\mathbf{c}$ evolves into a new one, $\mathbf{c}^{\prime}$, that is, $w(\mathbf{c}\rightarrow\mathbf{c}^{\prime})$. As a consequence, we must deal with a time-dependent probability distribution function $\mathcal{P}(\mathbf{c},t)$, which obeys a \textit{master equation} \cite{vkampen,gardiner}
\begin{equation}
\partial_t \mathcal{P}(\mathbf{c},t) =\sum_{\mathbf{c}^{\prime}}\left\lbrace \mathcal{P}(\mathbf{c}^{\prime},t)\,w(\mathbf{c}^{\prime}\rightarrow\mathbf{c})-\mathcal{P}(\mathbf{c},t)\,w(\mathbf{c}\rightarrow\mathbf{c}^{\prime})\right\rbrace \,.
\label{eq:master_intro_DDF}
\end{equation}
Its stationary solution, $\mathcal{P}(\mathbf{c})$, controls all the time-independent properties. To ensure that the desired equilibrium distribution $\mathcal{P}_{0}$ is reproduced, one chooses \textit{rates} which satisfy the \textit{detailed balance} condition, namely
\begin{equation}
\frac{w(\mathbf{c}^{\prime}\rightarrow\mathbf{c})}{w(\mathbf{c}\rightarrow\mathbf{c}^{\prime})}=\frac{\mathcal{P}_{0}(\mathbf{c})}{\mathcal{P}_{0}(\mathbf{c}^{\prime})}\,.
\label{eq:db_intro_DDF}
\end{equation}
This ensures that the bracket in Eq.~\eqref{eq:master_intro_DDF} vanishes. The important point here is that the ratio $\mathcal{P}_{0}(\mathbf{c})/\mathcal{P}_{0}(\mathbf{c}^{\prime})=e^{-\Delta \mathcal{H}/k_{B}T}$, where $\Delta \mathcal{H}=\mathcal{H}(\mathbf{c}^{\prime})-\mathcal{H}(\mathbf{c})$. One may thus choose rates of the form $w(\mathbf{c}\rightarrow\mathbf{c}^{\prime})=f(\Delta \mathcal{H}/k_{B}T)$, with an appropriate $f$ satisfying
\begin{equation}
f(-x)/f(x)=e^x\,.
\label{eq:db2_intro_DDF}
\end{equation}
Some choices for $f$ are the Kawasaki rate $f(x)=2\left( 1+e^x\right)^{-1}$ \cite{kawasaki_rate}, the van Beijeren \& Schulman rate $f(x)=e^{-x/2}$ \cite{vbeijeren_rate}, and the Metropolis rate $f(x)=\min\left\lbrace 1,e^{-x}\right\rbrace $ \cite{metropolis}. All of our simulations in Part~\ref{part:DDF} concern the Metropolis rate. 

The most simple way of driving a system (as in the DLG case) into a nonequilibrium steady state is to impose rates that violates detailed balance. A straightforward extension of $\Delta \mathcal{H}=\mathcal{H}(\mathbf{c}^{\prime})-\mathcal{H}(\mathbf{c})$ in Eq~\eqref{eq:db_intro_DDF} is to include the work done by the field, i.e., to define the total energy difference of the for $\Delta \mathcal{H}+Ex$, with $x$ the particle displacement in the field direction. This rate with the property Eq.~\eqref{eq:db2_intro_DDF} satisfies the detailed balance condition \textit{locally} but not globally so that microscopic reversibility of the process as in equilibrium is not guaranteed. This scheme is adopted in the DLG which (together with toroidal boundary conditions) yields the system towards a nonequilibrium steady state.

It is worth notice that there are recent computational \textit{coarse-grained} methods, e.g., \textit{Dissipative Particle Dynamics} (DPD) and \textit{Brownian Dynamics} (BD) (see for instance \cite{frenkel}), for fluid dynamics which may have several advantages over conventional computational dynamics methods. 
The DPD method ---first proposed by Hoogerbrugge and Koelman \cite{dpd} using heuristic arguments, and properly formalized by Espa\~nol and Warren \cite{espannol}--- is a useful technique when studying the mesoscopic structure of complex liquids. BD is a technique that resembles DPD but each (mesoscopic) particle feels a random force and a drag force relative to a fixed background\footnote{The important difference between BD and DPD is that BD does not satisfy Newton's third law and hence is does not conserve momentum. As a consequence, BD cannot reproduce hydrodynamic behavior.}. However, as we stressed in Chapter~\ref{cap0}, we are interested in how the macroscopic behavior emerges from \textit{microscopic} (atomistic) interactions rather than in modeling the dynamics of fluids from a \textit{mesoscopic} point of view.\\

Statistical field theories are a complementary approach to the understanding of nonequilibrium ordering in the DLG and in related models. These approaches often pose great conceptual and computational challenges in themselves, but we will focus here on critical properties and renormalization group notions through the Langevin equation \cite{vkampen,gardiner}. This is a stochastic partial differential equation, which corresponds to a mesoscopic (coarse-grained) description of the system, and it is thought to describe low-frequency, small-wave number phenomena (such as critical properties). It takes into account only the relevant symmetries for the problem under study and lets the fast degrees of freedom that we forget about in the coarse-grained description to act as noise. The Langevin equation for a field $\phi$ takes the following form \cite{vkampen,gardiner}
\begin{equation}
\frac{\partial \phi(\mathbf{r},t)}{\partial t}=F[\phi]+G[\phi]\xi(\mathbf{r},t)\,,
\label{eq:intro_DDF_langevin}
\end{equation}
where $\xi$ is a random variable which represents a uncorrelated Gaussian white noise. $F$ and $G$, which are determined in general from symmetry considerations, 
are analytic functionals of $\phi$. In fact, these functionals include every analytic term consistent with the symmetries of the microscopic system. In principle, $F$ and $G$ can be determined phenomenologically through experiments and/or symmetry arguments. Other difficult procedure which is seldom carried out is the coarse-graining operation that produces the Langevin equation from the microscopic model, which follows, say, a master equation as Eq.~\eqref{eq:master_intro_DDF}. We adopt the latter procedure (we extend the scheme devised in Ref.~\cite{paco}) in the Chapter~\ref{cap1} and in greater detail in the Appendix~\ref{apendB}.

\chapter{\label{cap1}Lennard--Jones and Lattice models of driven fluids}

The present chapter describes Monte Carlo (MC) simulations and field theoretical calculations that aim at illustrating how slight modifications of dynamics at the microscopic level may influence, even quantitatively, the resulting (nonequilibrium) steady state. With this aim, we take as a reference the \textit{driven lattice gas} (DLG), which we described qualitatively in the previous chapter. We present and study related lattice and off-lattice microscopic models in which particles, as in the DLG, interact via a local anisotropic rule. The rule induces preferential hopping along one direction, so that a net current sets in if allowed by boundary conditions. In particular, we shall discuss on the similarities and differences between the DLG and its continuous counterpart, namely, a Lennard--Jones analogue in which the particles' coordinates vary continuously. A comparison between the two models allows us to discuss some exceptional, hardly realistic features of the original discrete system ---which has been considered a prototype for nonequilibrium anisotropic phase transitions.

\section{\label{cap1:intro}Driven Lattice Gas}
The \textit{driven lattice gas} \cite{kls} is a nonequilibrium extension of the Ising model with conserved dynamics. The DLG consists of a $d$-dimensional square lattice gas in which pair of particles interact via an attractive and short--range Ising--like Hamiltonian, 
\begin{equation}
H=-4\sum_{\langle \mathbf{j},\mathbf{k} \rangle} \sigma_{\mathbf{j}} \sigma_{\mathbf{k}} \: .
\label{eq:hamilt}
\end{equation}
Here $\sigma_{\mathbf{k}}$ is the lattice occupation number at site $\mathbf{k}\in \mathbb{Z}^{d}$, and the sum runs over all the nearest-neighbor (NN) sites\footnote{For $d=2$, this corresponds with a lattice's connectivity $4$.}. Each lattice site has two possible states, namely, a particle ($\sigma_{\mathbf{k}}=1$) or a hole ($\sigma_{\mathbf{k}}=0$) may occupy each site $\mathbf{k}$. Cell dimensions for the lattice are chosen so that the NN distance (\textit{bond length}) was unity: $|\mathbf{j}-\mathbf{k}|=1$. A configuration is given by $\mathbf{\sigma}\equiv\left\lbrace \sigma_{\mathbf{k}}\,;\,\mathbf{k}\in d\text{-dimensional lattice} \right\rbrace $. We shall restrict ourselves to two-dimensional $L_{\parallel}\times L_{\perp}$ lattices, that is, $d=2$. Dynamics is induced by the competion between a heat bath at temperature $T$ and an external driving field $E$ which favors particle hops along one of the principal lattice directions, say horizontally ($\hat{x}$ direction), as if the particles were positively charged. Consequently, for periodic boundary conditions, a nontrivial nonequilibrium steady state is set in asymptotically. This is formalized through a master equation similar to Eq.~\eqref{eq:master_intro_DDF}, where the transition probabilities per unit time $w(\mathbf{\sigma}\rightarrow\mathbf{\sigma}^{\prime})$ are the Metropolis ones, namely,
\begin{equation}
w(\mathbf{\sigma}\rightarrow\mathbf{\sigma}^{\prime})=\min \left\lbrace 1,e^{-\left( \Delta H+\mathbf{E}\cdot\mathbf{\delta}\right)/(k_BT) }\right\rbrace \;.
\label{eq:metro}
\end{equation}
Where $\mathbf{E}\cdot\mathbf{\delta}$ denotes the dot product between the field $E$ (oriented along $\hat{x}$ direction) and $\mathbf{\delta}$, which is the attempted particle displacement, i.e, $\mathbf{\delta}=(\mathbf{j}-\mathbf{k})(\sigma_{\mathbf{j}} - \sigma_{\mathbf{k}})$; and $\Delta H=H(\sigma^{\prime})-H(\sigma)$. To be precise, $w(\mathbf{\sigma}\rightarrow\mathbf{\sigma}^{\prime})$ stands for the \textit{rate} for the particle-hole exchange $\sigma_{\mathbf{k}} \leftrightarrows \sigma_{\mathbf{j}}$ when the configuration is $\mathbf{\sigma}$; the particle density $\rho=N/(L_{\parallel}L_{\perp})$ remains constant during the time evolution. The exchange $\sigma_{\mathbf{k}} \leftrightarrows \sigma_{\mathbf{j}}$ corresponds to a particle jumping to a NN hole if $\sigma_{\mathbf{k}}\neq\sigma_{\mathbf{j}}$; we assume $w(\mathbf{\sigma}\rightarrow\mathbf{\sigma}^{\prime})=0$ otherwise, i.e., only NN particle-hole exchanges are allowed. This bias breaks detailed balance and establishes a nonequilibrium steady state. Typical configurations in the stationary regime of the DLG are shown in Fig.~\ref{fig1_DDF}.\\

MC simulations by the biased \textit{Metropolis} rate in Eq.~\eqref{eq:metro} reveal that, as in equilibrium, the DLG undergoes a second order phase transition. At high enough temperature, the system is in a disordered state while, below a critical point (at $T\leq T_{E}$, where $E=|\mathbf{E}|$) it orders displaying anisotropic phase segregation. That is, an striped rich--particle phase then coexists with its gas. It is also found that, for a lattice half filled of particles, the critical temperature $T_{E}$ monotonically increases with $E$ from the Onsager value\footnote{See Appendix~\ref{apendA}.} $T_{0}=T_{\text{Ons}}=2.269\,Jk_{B}^{-1}$ to $T_{\infty }\simeq 1.4\,T_{\text{Ons}}.$ This limit ($E\rightarrow \infty$) corresponds to a \textit{nonequilibrium} critical point. As a matter of fact, it was numerically shown to belong to a universality class other than the Onsager one, e.g., MC data indicates $\beta_{\mathrm{DLG}} \simeq 0.33$ (instead of the Ising value $\beta =1/8$ in two dimensions) for the order parameter critical exponent \cite{marro,beta4,beta5}. Typical configurations for the DLG in the large field limit were shown in Fig.~\ref{fig1_DDF}.

Statistical field theory is a complementary approach to the understanding of nonequilibrium ordering in the DLG. The derivation of a general mesoscopic description is still an open issue, however. Two different approaches have been proposed. The \textit{driven diffusive system} (DDS) \cite{zia,beta2,beta3}, which is a \textit{phenomenological} Langevin-type equation aimed at capturing all the relevant symmetries, predicts that the current will induce a
predominant mean--field behavior and, in particular, $\beta_{\mathrm{DDS}} =1/2.$ The \textit{anisotropic diffusive system} (ADS) \cite{paco}, which follows after a coarse graining of the master equation, rules out the relevance of the current and leads to the above--indicated MC critical exponent for $E\rightarrow \infty .$ However, the ADS approach reduces to the DDS for finite $E,$ a fact which is hard to be fitted to MC data.
In any case, field theoretical studies have constantly demanded further numerical efforts, and the DLG is nowadays the most thoroughly studied system showing an anisotropic nonequilibrium phase transitions. The topic is not exhausted, however. On the contrary, there remain unresolved matters such as the above mentioned issues concerning critical and mesoscopic behaviors, and the fact that $T_{E} $ increases with $E$, which is counterintuitive \cite{zia2,zia1}. Another significative question concerns the observation of triangular anisotropies at early times after a rapid MC quench from the homogenous state (as if $T=\infty$) to $T<T_{E}.$ The triangles happen to point against the field, which is contrary to the prediction from the DDS continuum equation \cite{triang1}.

\begin{figure}
\centerline{\includegraphics[width=10cm,clip=]{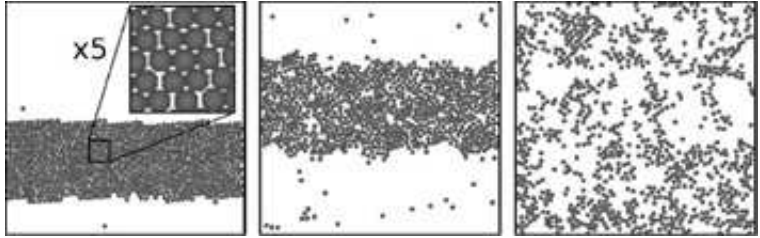}}
\caption[Typical steady state configurations of the DLJF subject to a horizontal field]{\label{fig1_1} Typical configurations during the stationary regime of the DLJF subject to a horizontal field of intensity $E^{\ast }=1.$ These graphs, which are for $N=900$ particles and density $\protect\rho ^{\ast }=0.30,$ illustrate, from left to right, \textit{(i) }coexistence of a solid and its vapor (the configuration shown is for temperature $T^{\ast }=0.20),$ \textit{(ii) }liquid--vapor coexistence $(T^{\ast }=0.35),$ and \textit{(iii) }a disordered, fluid phase 
$(T^{\ast }=0.50).$ The left--most graph shows a detail of the solid strip. The particles, which move in a square box of side $\sqrt{N/\rho^{\ast}}$, are given here an arbitrary size, namely, diameter$=1.1\protect\sigma$.}
\end{figure}

In this chapter we present a description of driven systems with continuous variation of the particles' spatial coordinates ---instead of the discrete variations in the DLG--- aiming for a new effort towards better understanding basic features of the DLG and its (and related) nonequilibrium phase transitions. This will provide a more realistic scheme for computer simulation of anisotropic fluids. Our strategy to set up the model is to follow as closely as possible the DLG. That is, we analyze an off--lattice representation of the DLG, namely, a microscopically \textit{continuum} with the same symmetries. Investigating these questions happens to clarify the puzzling situation indicated above concerning the outstanding behavior of the DLG. 

\section{\label{cap1:dljf}Driven Lennard-Jones fluid}
Consider a \textit{fluid} consisting of $N$ point-like particles in a two--dimensional $L\times L$ ($L\equiv L_{\parallel}=L_{\perp}$) box with periodic (toroidal) boundary conditions. Interactions are according to a \textit{truncated and shifted Lennard--Jones} (LJ) 12-6 potential\footnote{Often termed Weeks-Chandler-Andersen potential.} \cite{allen}:
\begin{equation}
V(r_{ij})=\left\{ 
\begin{array}{cl}
V_{LJ}(r_{ij})-V_{LJ}(r_{c}), & \text{if }r_{ij}<r_{c} \\ 
0, & \text{if }r_{ij}\geq r_{c}\:,
\end{array}%
\right. 
\label{eq:stLJP}
\end{equation}
where 
\begin{equation}
V_{LJ}(r)=4\epsilon \left[(\sigma /r)^{12}-(\sigma /r)^{6}\right]
\label{eq::fLJP}
\end{equation}
is the full LJ potential. Here, $r_{ij}=\left\vert \mathbf{r}_{i}-\mathbf{r}_{j}\right\vert $ is the relative distance between particles $i$ and $j,$ $\epsilon $ and $\sigma $ are our energy and length units, respectively, and $r_{c}$ is the \textit{cut-off} that we shall fix at $r_{c}=2.5\sigma $. The choice of this potential obeys to our strategy to set up the model following as close as possible the DLG (see Eq.~\eqref{eq:hamilt}). In fact, the potential defined in Eq.~\eqref{eq:stLJP} is only one of the multiple possibilities for an attractive short-ranged potential\footnote{It has been shown that the LJ potential provides accurate results for a variety of fluids, e.g., Argon \cite{allen}.}. The preferential hopping will be implemented as in the lattice, i.e., by adding a driving field to the potential energy. Consequently, the familiar energy balance which enters in Eq.~\eqref{eq:metro} is (assuming $k_{B}\equiv 1 $ hereafter) $H(\mathbf{c})=\sum_{i<j}V(r_{ij})$, where (\textit{continuum}) configurations $\mathbf{c}$ are here $\mathbf{c}\equiv \left\{ \mathbf{r}_{1},\ldots,\mathbf{r}_{N}\right\}$. The particle attempted displacement is therefore $\mathbf{\delta}=\mathbf{r}_{i}^{\prime }-\mathbf{r}_{i}$. Lacking a lattice, the field is the only source of anisotropy, and any trial move should only be constrained by a maximum displacement in the radial direction. That is, we take $0<|\mathbf{\delta}|<\delta_{max} ,$ where $\delta_{max} =0.5\sigma $ in our simulations. We express the relevant quantities in terms of our energy $\epsilon$ and length $\sigma$ scales, i.e., we introduce reduced units\footnote{This is very convenient in molecular simulations, and, most importantly, if we do not use reduced units, we might miss the equivalence between corresponding states (\textit{law of corresponding state}).}. From these basic units, the temperature $T$, length $L$, number density $\rho=N/L^{2}$, internal energy $U$, and field $E$ variables will be reduced respectively according to 
\begin{equation}
T^{\ast}=T/\epsilon\,,\: L^{\ast}=L/\sigma\,,\: \rho^{\ast}=\rho \sigma^2\,,\: U^{\ast}=U/\epsilon\,,\: \text{and}\: E^{\ast}=E\sigma/\epsilon \,, 
\label{eq:rescal}
\end{equation}
where $^{\ast}$ denotes reduced units. This model thus reduces for $E\rightarrow 0$ to the well-known \textit{truncated and shifted LJ fluid}, one of the most studied models in the computer simulation of fluids \cite{allen,smit}. \\

\begin{figure}
\centerline{\includegraphics[width=10cm,clip=]{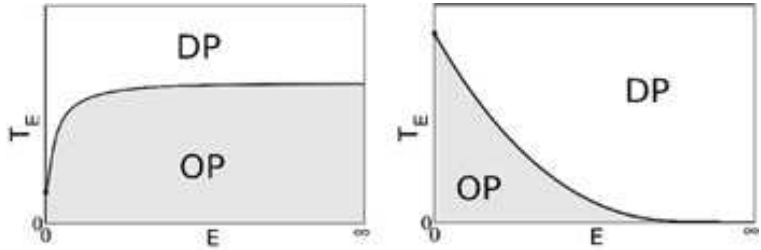}}
\caption[Schematic phase diagrams for the DLG, NDLG, and DLJF]{\label{fig2_1}Schematic phase diagrams for the three models, as defined in the main text of this Chapter, showing ordered or striped (OP) and disordered (DP) phases. The left graph is for the DLG, for which $T_{0}=T_{\text{Ons}}.$ The graph on the right is valid for both the NDLG, i.e., the DLG with next--nearest--neighbor (NNN) hops, for which $T_{0}=2.32\,T_{\text{Ons}},$ and the DLJF, for which $T_{0}^{\ast}=0.459$ \protect\cite{smit}.}
\end{figure}

We studied this \textit{driven LJ fluid} (DLJF) in the computer by the MC method using an extended \textquotedblleft canonical ensemble\textquotedblright , namely, fixed values for $N,$ $\rho^{\ast },$ $T^{\ast },$ and $E^{\ast }.$
Simulations involved up to $N=10^{4}$ particles with parameters ranging as follows: $0.5\leq E^{\ast }\leq 1.5,$ $0.20\leq \rho ^{\ast }\leq 0.60,$ $0.15\leq T^{\ast }\leq 0.55.$ The typical configurations one observes are
illustrated in Fig.~\ref{fig1_1}. As its equilibrium counterpart, the DLJF exhibits three different phases (at least): vapor, liquid, and solid (sort of close--packing phase; see the left--most graph in Fig.~\ref{fig1_1}). At
intermediate densities and low enough $T^{\ast },$ vapor and a condensed phase segregate from each other. The condensed droplet (see Fig.~\ref{fig1_1}) is not near circular as it generally occurs in equilibrium, but strip--like extending along the field direction. A detailed study of each of these phases will be reported
in Chapter~\ref{cap2} for a related system.

\begin{figure}
\centerline{\includegraphics[width=8cm,clip=]{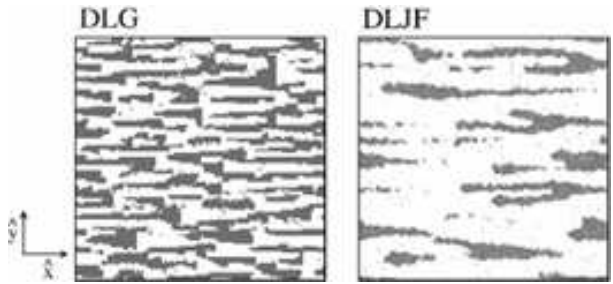}}
\caption[Triangular anisotropies observed at early times of the DLG and DLJF]{\label{fig3_1}Triangular anisotropies as observed at early times for $E=1$ in computer simulations of the lattice (left) and the off--lattice (right) models defined in the main text. The field points along the $\hat{x}$ direction. The DLG configuration is for $t=6\times 10^{4}$ MC steps in a 128$\times $128 lattice with $N=7372$ particles and $T=0.4T_{\text{Onsager}}.$ The DLJF configuration is for $t=1.5\times 10^{5}$ MC steps, $N=10^{4}$ particles, $\protect\rho^{\ast} =0.20$ and $T^{\ast}=0.23.$}
\end{figure}

A main observation is that the DLJF closely resembles the DLG in that both depict a particle current and the corresponding anisotropic interface. However, they differ in an essential feature, as illustrated by Fig.~\ref{fig2_1}. That is, contrary to the DLG, for which the critical temperature $T_{E}$ increases with $E,$ the DLJF shows a transition temperature $T_{E}$ which decreases with increasing $E.$ The latter behavior was expectable. In fact, as $E$ is
increased, the effect of the potential energy in the balance Eq.~\eqref{eq:metro} becomes weaker and, consequently, the cohesive forces between particles tend to become negligible. Therefore, unlike for the DLG, there is no phase transition for a large enough field, and $T_{E}\rightarrow 0$ for $E\rightarrow \infty $ in the DLJF. Confirming this, typical configurations in this case are fully homogeneous for any $T$ under a sufficiently large field $E.$ One may think of variations of the DLJF for which $T_{E\rightarrow \infty }=const>0$ (see Chapter~\ref{cap2}), but the present one follows more closely the DLG microscopic strategy based on Eq.~\eqref{eq:metro}.

Concerning the early process of kinetic ordering, one observes triangular anisotropies in the DLJF that point along the field direction. That is, the early--time anisotropies in the off--lattice case (right graph in Fig.~\ref{fig3_1}) are similar to the ones observed in mesoscopic approaches (as, for instance, the DDS approach), and so they point along the field, contrary to the ones observed in the discrete DLG (left graph in Fig.~\ref{fig3_1}). 

\section{\label{cap1:NDLG}Driven Lattice Gas with extended dynamics}
\begin{figure}
\centerline{\includegraphics[width=8cm,clip=]{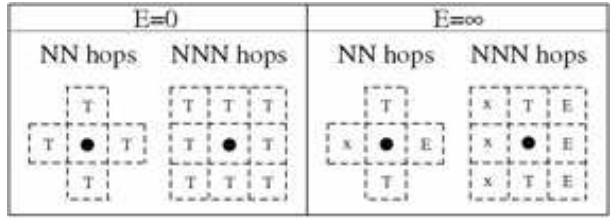}}
\caption[Schematic comparison of the sites a particle may occupy for nearest--neighbor and next--nearest--neighbor hops]{\label{fig4_1}Schematic comparison of the accessible sites a particle (at the center, marked with a dot) has
for nearest--neighbors (NN) and next--nearest--neighbors (NNN) hops at equilibrium (left) and in the presence of a large horizontal field (right). The particle--hole exchange between neighbors may be forbidden (x), depend only on the potential energy (T), or occur with probability 1 (E).}
\end{figure}

The above observations altogether suggest a unique exceptionality of the DLG behavior. This is to be associated with the fact that a driven particle is geometrically restrained in the DLG. In order to show this, we studied the lattice with an infinite drive extending the hopping to next--nearest--neighbors (NNN) \cite{valles,szabo,triang2}. That is, we extend interactions and accessible sites to the NNN. Thus, the sum in Eq.~\eqref{eq:hamilt} involves the eight next-nearest neighbors instead of the four nearest neighbors of the standard DLG. In this case, the particle-hole exchange $\sigma_{\mathbf{k}} \leftrightarrows \sigma_{\mathbf{j}}$ corresponds to a particle hop to a NNN hole if $\sigma_k\neq\sigma_j$. As illustrated in Fig.~\ref{fig4_1}, this introduces further relevant directions in the lattice, so that the resulting model, to be named here NDLG, is expected to behave closer to the DLJF. This is confirmed. For example, one observes in the discrete NDLG that, as in the continuum DLJF, $T_{E}$ decreases with increasing $E$ ---though from $T_{0}=2.32\,T_{\text{Ons}}$ in this case\footnote{For the Ising model with NNN interactions, one can easily derive theoretical estimates for the critical temperature \cite{mattis}.}---. This is illustrated in Fig.~\ref{fig2_1}. Specifically, we discuss this difference between the DLG and the NDLG in Appendix~\ref{apendC}, in where we define an intermediate model which covers both situations.

\begin{figure}
\centerline{\includegraphics[width=12cm,clip=]{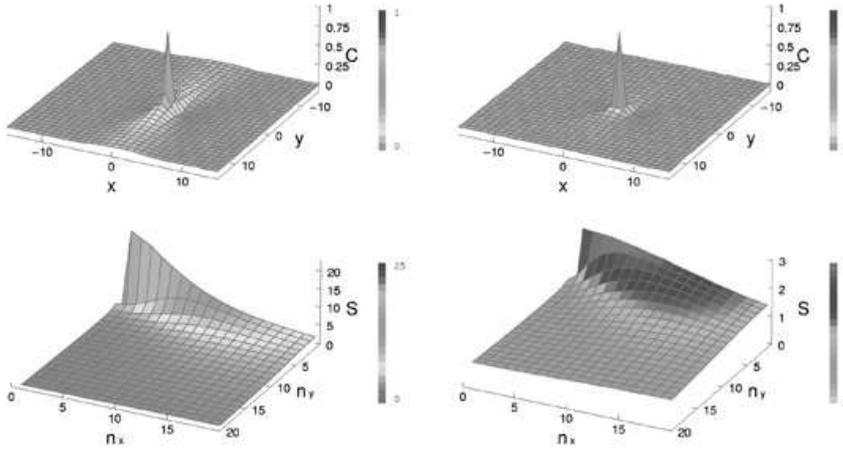}}
\caption[Surface plots of the two--point correlations and structure factor above criticality for the DLG with NN and NNN couplings]{\label{fig33_1} Surface plots of the two--point correlation (upper row) and the structure factor (lower row). These are for the DLG (left column) and NDLG (right column) on a $L_{\parallel}\times L_{\perp}=128\times 128$ lattice in the large field limit $E=\infty$. Temperatures are $T/T_{\text{Ons}}=1.58$ and $T/T_{\text{Ons}}=1.00$ for the DLG and the NDLG, respectively. Notice the change of color code with each plot.}
\end{figure}
\begin{figure}
\centerline{\includegraphics[width=8cm,clip=]{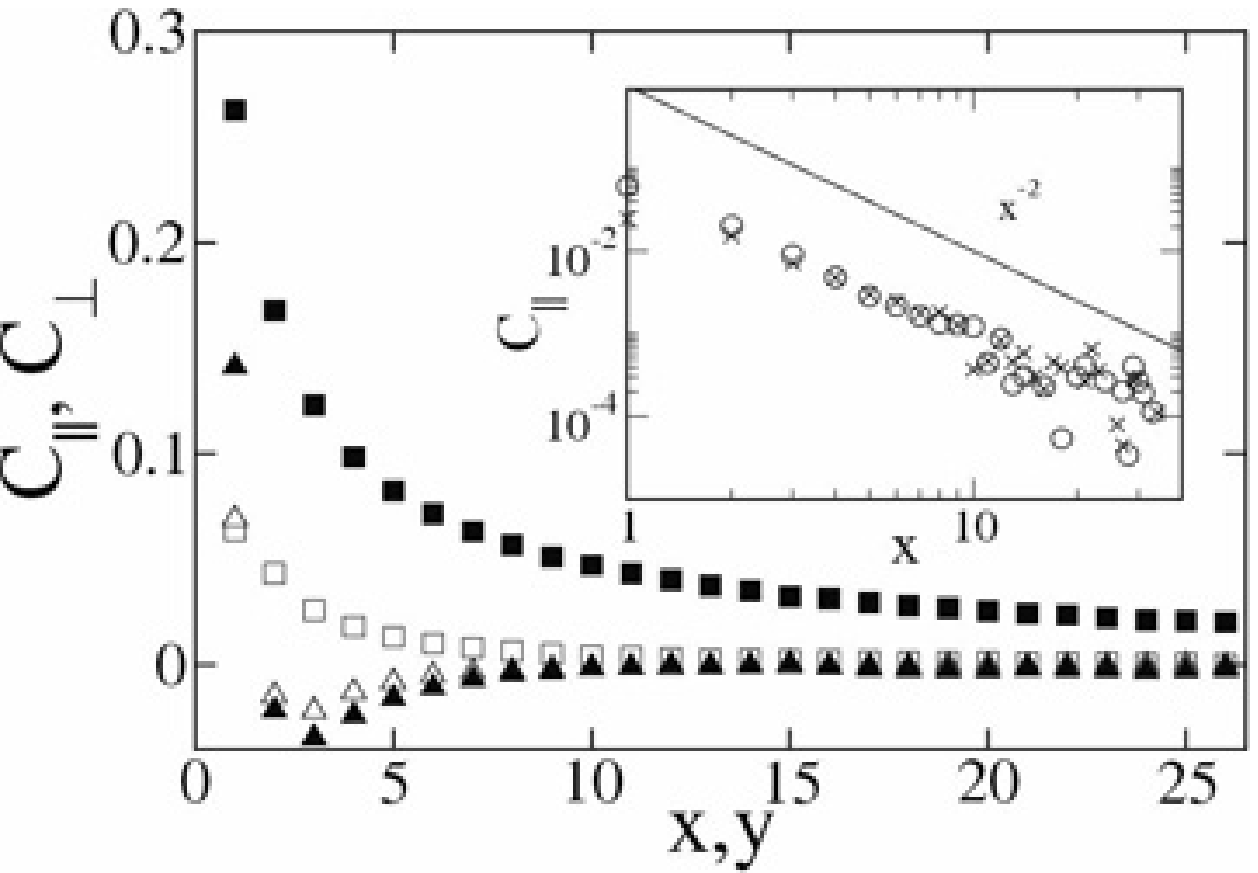}}
\caption[Parallel and transverse components of the two-point correlation function above criticality for the DLG with NN and NNN interactions]{\label{fig44_1}Parallel (squares) and transverse (triangles) components of the two-point correlation function above criticality for the DLG on a $128\times 128$ lattice with NN hops/interactions (filled symbols) at $T=1.58$ and NNN hops/interactions (empty symbols) at $T=1.$ Temperatures normalized to $T_{\text{Ons}}$. The inset shows the $x^{-2}$ power law decay in $C_{\parallel}$ for both discrete cases: DLG ($\circ$) and NDLG ($\times$).}
\end{figure}

\subsection{\label{cap1:correls}Correlations and Structure Factor}
There is also interesting information in the two--point correlation function and (equal-time) structure factor. The former, which measures the degree of order of the lattice, is defined for a half--filled lattice as 
\begin{equation}
C(\mathbf{i})=\left\langle \sigma_{\mathbf{j}}\,\sigma_{\mathbf{i}+\mathbf{j}}\right\rangle -1/4,
\label{eq:correl}
\end{equation}
where the steady average $\left\langle \cdots \right\rangle$ involves averaging over $\mathbf{j}.$ Translational invariance is assumed. The structure factor, which is the Fourier transform of the two-point correlation function, can be written in terms of the occupation variables as
\begin{equation}
S(\mathbf{\kappa})=\frac{1}{L^{2}}\left\langle \left\vert \sum_{\mathbf{r}}\sigma_{\mathbf{r}}e^{i\,\mathbf{\kappa} \cdot \mathbf{r}}\right\vert^{2}\right\rangle \,,  
\label{eq:struc}
\end{equation}
where $\mathbf{\kappa}$ is a wave vector which has components $\kappa_{x,y}=2\pi n_{x,y}/L$ with $n_{x,y}=0,1,...,L-1$. This is very useful to distinguish disordered configurations from inhomogeneous one. To avoid complications due to inhomogeneous ordered phases, we compute for both the DLG and NDLG two-body correlations and structure factors above criticality. We show in Fig.~\ref{fig33_1} the surface plots of $C$ and its Fourier transform $S$ for both dynamics. Analysis of the components (shown in Fig.~\ref{fig44_1}) along the field, $C_{\parallel}\equiv C(x,0),$ and transverse to it, $C_{\perp}\equiv C(0,y),$ shows that correlations are qualitatively similar for the DLG and the NDLG ---although somewhat weaker along the field for NNN hops. That is, allowing for a particle to surpass its forward neighbor does not modify correlations. In fact, both the DLG and the NDLG display generic slow decay of two--point correlations\footnote{This is in stark contrast with their equilibrium counterparts, the lattice gas (see Appendix~\ref{apendA}), in which correlations decay exponentially (except at the critical point). The latter is a common feature of all the systems at thermal equilibrium.} \cite{pedro}. It is generally accepted \cite{zia} that the key ingredients which yield power-law decays are \textit{(i)} a conserved dynamics, \textit{(ii)} a non-equilibrium steady state, and \textit{(iii)} spatial anisotropy associated with the dynamics. The first ingredient alone (as in equilibrium) cannot lead to generic power laws due to the validity of the \textit{fluctuation dissipation theorem}. The role of the second ingredient is to lift the theorem's constraint, so that the power laws are now generic. The role of the third one is more subtle, necessary only for producing power laws in the two-body correlations.

\begin{figure}
\centerline{\includegraphics[width=8cm,clip=]{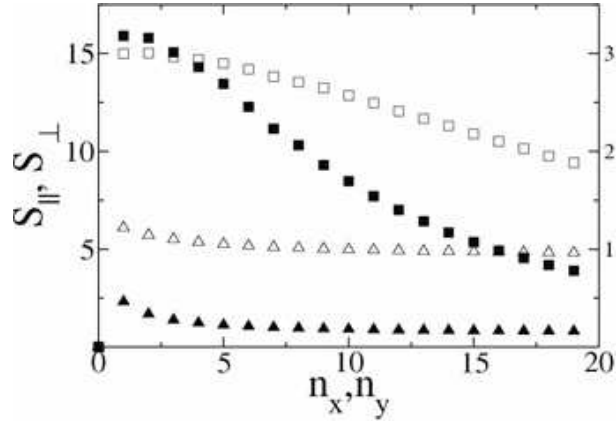}}
\caption[Parallel and transverse components of the structure factor above criticality for the DLG with NN and NNN interactions]{\label{fig5_1}Parallel (squares) and transverse (triangles) components of the structure factor above criticality for the DLG on a $128\times 128$ lattice for the DLG (filled symbols, left scale) at $T=1.58$ and for the NDLG (empty symbols, right scale) at $T=1.$ Temperatures normalized to $T_{\text{Onsager}}$. Here, $n_{x,y}=128\,\kappa_{x,y}/2\protect\pi $.}
\end{figure}

The power-law behavior\footnote{Together with the Ising symmetry violation by $E\neq0$.} translates into a discontinuity of the structure factor, namely, $\lim_{\kappa_{x}\rightarrow 0}S_{\parallel}\neq \lim_{\kappa_{y}\rightarrow 0}S_{\perp},$ where $S_{\parallel}\equiv S(\kappa_{x},0)$ and $S_{\perp}\equiv S(0,\kappa_{y})$. This is clearly confirmed in Fig.~\ref{fig5_1} for both NN and NNN hops/interactions. Notice that this singularity, which do not appear under equilibrium conditions, is an indication of the dependence of the nonequilibrium steady state on the (anisotropic) dynamic.

\subsection{\label{cap1:crit}Critical behavior}
As outlined in Section~\ref{cap1:intro}, the derivation of a general mesoscopic description is still an issue under contention. In order to shed more light on the field theoretical description of driven fluids, we try to derive a Langevin-type equation for the NDLG. To this end, we firstly compute the order parameter critical exponent by means of standard finite size scaling techniques. Secondly, we derive the Langevin equation in order to describe the NDLG critical properties, by using the ADS approach \cite{paco}. We also discuss the viability of other approaches, e.g., the DDS approach.

\subsubsection{\label{cap1:sims}MC simulations}
Assuming a half--filled lattice\footnote{By following simple symmetry arguments, the critical density is $N/L_{\perp}L_{\parallel}=1/2$.} we perform finite size scaling analysis for the NDLG by following the scheme proposed in \cite{zia} consistent with the ADS theory \cite{beta4}. The order parameter is chosen as the structure factor
\begin{equation}
m=S(0,2\pi/L_{\perp})\;,
\label{eq:orderp}
\end{equation}
(as suggested in \cite{leung,wang,praest}), which carries the intrinsic anisotropies of the system. In order to perform systematic anisotropic finite size scaling we considered system sizes $80\times 40$, $25\times 50$, $45\times 30$, and $125\times 50$. These aspect ratios satisfy $L_{\parallel}^{\nu_{\perp}/\nu_{\parallel}}=0.22\, \times \, L_{\perp}$, where $\nu_{\perp}/\nu_{\parallel}\approx 1/2$ consistent with the ADS anisotropic spatial scaling \cite{leung,wang,praest}. A strong enough field is needed to avoid crossovers from the equilibrium regime. Notice also that, as we showed above for the NDLG, a saturating field suppresses the ordering, that is, $T_{E}=0$ when $E\rightarrow \infty$. Therefore we choose an intermediate value for the field $E=12$. The corresponding critical temperature is determined by using the Binder's fourth cumulant method \cite{binder2}, which is $T_{12}=1.45\pm 0.01$. The obtained critical value $T_{12}$ was employed for the finite size scaling analysis.
 
\begin{figure}
\centerline{\includegraphics[width=8cm,clip=]{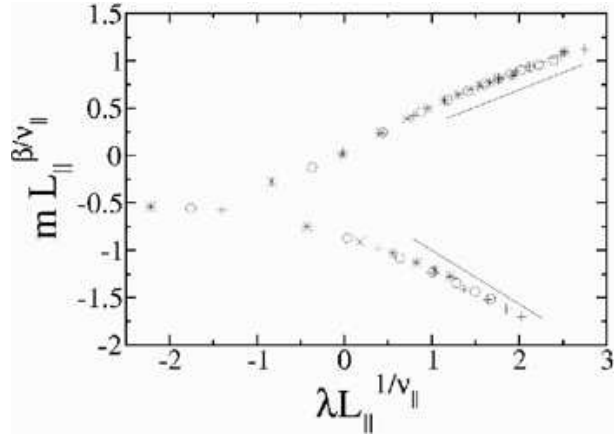}}
\caption[Log-log plot of the rescaled order parameter $m$ for different sizes]{\label{fig20_1} Log-log plot of the order parameter rescaled by $L_{\parallel}^{\beta/\nu_{\parallel}}$ vs. $\lambda L_{\parallel}^{1/\nu_{\parallel}}$, for different sizes: ($+$) $125\times 50$, ($*$) $45\times 30$, and ($\circ$) $80\times 40$. The upper straight line is for $\beta= 1/3$, whereas the lower one indicates $\gamma= 0.3$.} 
\end{figure}

Fig.~\ref{fig20_1} show the scaled order parameter $m$ upon the distance to the critical point $\lambda$, for three different sizes of $L_{\parallel}$. The former is rescaled by $L_{\parallel}^{\beta/\nu_{\parallel}}$ whereas the latter is rescaled by $L_{\parallel}^{1/\nu_{\parallel}}$. The number of MC steps considered for each temperature was $160$ millions. A perfect data collapse between the different system sizes is obtained by fixing $\nu_{\parallel}\approx 1.25$, $\gamma\approx 1.22$ and $\beta\approx 0.33$, where $\nu_{\parallel}$, $\gamma$, and $\beta$ are the corresponding critical exponents. Notice that these values are precisely the same for the DLG (with NN hops/interactions) \cite{beta4}. As shown in Fig.~\ref{fig20_1}, the asymptotic lines fit nicely with $\beta\approx 0.33$ (upper line), and $-\gamma/2 \approx -0.61$ (lower line). 

\begin{figure}
\centerline{\includegraphics[width=8cm,clip=]{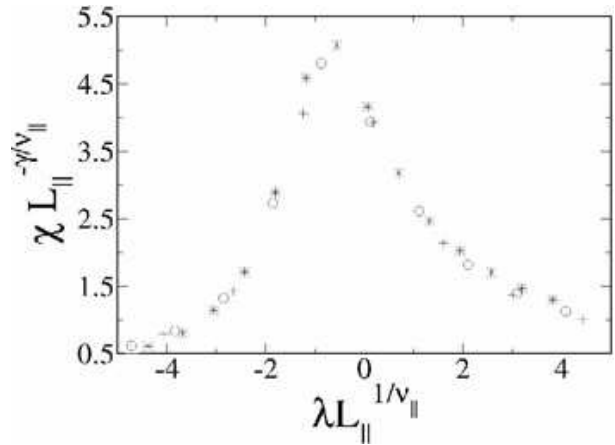}}
\caption[Log-log plot of the rescaled susceptibility $\chi$ for different sizes]{\label{fig21_1} Log-log plot of the susceptibility rescaled by $L_{\parallel}^{-\gamma/\nu_{\parallel}}$ vs. $\lambda L_{\parallel}^{1/\nu_{\parallel}}$. Symbols are as in Fig.~\ref{fig20_1}.} 
\end{figure}

\begin{figure}
\centerline{\includegraphics[width=8cm,clip=]{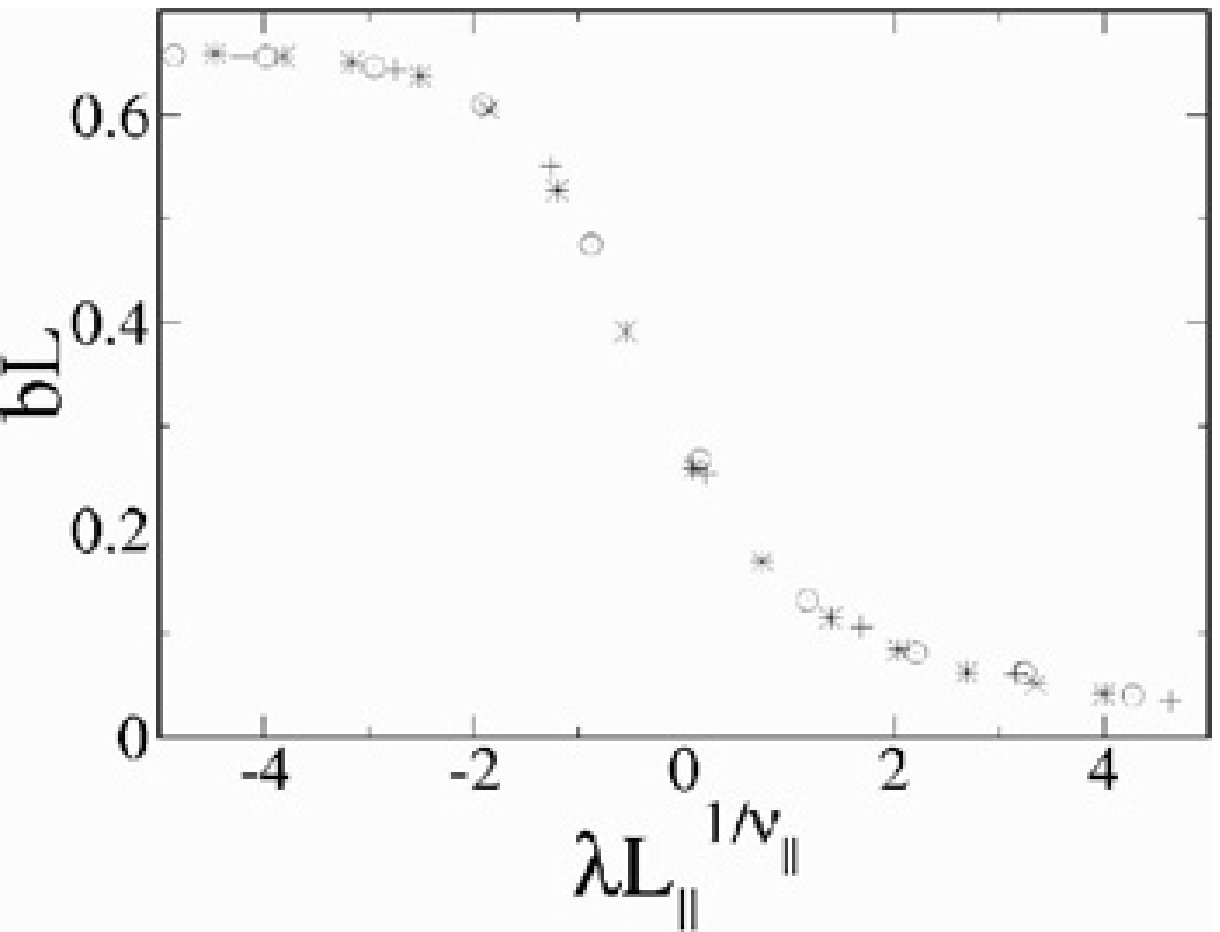}}
\caption[Scaling plot of the rescaled fourth Binder cumulant $b$ for different sizes]{\label{fig22_1} Scaling plot of the fourth Binder cumulant $b$ vs. $\lambda L_{\parallel}^{1/\nu_{\parallel}}$. Symbols are as in Fig.~\ref{fig20_1}.} 
\end{figure}

We have also computed the susceptibility, defined as the relative fluctuations of the order parameter. This reads
\begin{equation}
\chi=\frac{L_{\parallel}}{\sin{\pi/L_{\perp}}} \left( \langle m^2 \rangle - \langle m \rangle^2 \right)
\label{eq:susc}
\end{equation}
The rescaled susceptibility is shown in Fig.~\ref{fig21_1}. The best collapse is obtained with the same values as before. Plotting the dimensionless Binder cumulant $b$ as a function of the rescaled distance to the critical point with $\nu_{\parallel}\approx 1.25$, again, nearly perfect collapse is obtained for all system sizes (as depicted in Fig.~\ref{fig22_1}). 


This set of MC results support that both the DLG and the NDLG belong to the same universality class, i.e., both dynamics yield the same critical indexes. In particular, it follows that the critical properties remains unchanged when extending hops and interactions from the nearest--neighbors to the next--nearest--neighbors.

\subsubsection{\label{cap1:mesos}Mesoscopic description}
The ADS approach has been employed successfully to describe the critical behavior of the DLG \cite{paco}. The resulting Langevin--type equation, derived from a coarse--graining process from the DLG master (microscopic) equation sets up the anisotropy as the main nonequilibrium ingredient and leads to the right critical exponent in the large field limit ($\beta\approx 1/3$). The associated Langevin equation for the coarse--grained density $\phi(\textbf{r},t)$ reads
\begin{equation}
{\partial}_t \phi(\mathbf{r},t)=-\nabla_{\bot}^4 \phi+(\tau+a)\nabla_{\bot}^2 \phi +\frac{g}{3!}\nabla_{\bot}^2\phi^3+\frac{a}{2} \nabla_{\Vert}^2 \phi+\xi(\mathbf{r},t)
\label{eq:adsNN}
\end{equation}
Here, the last term $\xi$ stands for a Gaussian white noise, i.e., $\langle \xi(\mathbf{r},t)\rangle=0$ and $\langle \xi(\mathbf{r},t)\xi(\mathbf{r}^{\prime},t^{\prime})\rangle=\delta(\mathbf{r}-\mathbf{r}^{\prime})\delta(t-t^{\prime})$, representing the fast degrees of freedom, and $\tau$, $a$ and $g$ are model parameters. 

This approach is certainly valuable because it represents a detailed connection between microscopic dynamics and their mesoscopic descriptions. Therefore, this important feature suggests that its validity can be extended to describe the critical behavior of the NDLG, which was studied by computational means in the previous subsection. Such a description include the microscopic details of diagonal dynamics which, in particular, should allow us to distinguish between the DLG and the NDLG. This is in contrast of other more heuristic approaches, as the DDS \cite{zia,beta2,beta3}, which aim for the main symmetries of the DLG. The DDS approach is the natural extension of the conserved $\phi^{4}$ theory for the Ising equilibrium model \cite{hohenberg}. The most relevant prediction of the DDS is the mean field behavior $\beta=1/2$, which conflicts with MC simulations. This is not surprising, because the DLG behavior is more complicated than its equilibrium counterpart, in which the symmetries and the influence of the dynamics is well established. The same situation occurs when extending dynamics to the NNN: the DDS (heuristic) scheme leads again to the mean--field order parameter critical exponent. As a matter of fact, the DDS approach can not account for the diagonal (microscopic) degrees of freedom. Therefore, the DDS Langevin equation does not describe the NDLG critical properties. 

The question is whether the ADS approach describes properly the critical behavior observed in the NDLG. This is addressed in this section. 

Next, we outline the derivation of a Langevin equation for the NDLG by following the scheme proposed in \cite{paco}. (this derivation can be found in greater detail in Appendix~\ref{apendB}).\\

Let us define a density variable $\phi(\mathbf{r},t)$ averaged in a region of volume $v$ over which the original microscopic occupation variable ($\sigma_{\mathbf{k}}$ in Eq.~\eqref{eq:hamilt}) were averaged. The system evolves from a (coarse--grained) configuration $\mathbf{\gamma}$ to another $\mathbf{\gamma}^{\prime}$ by choosing randomly a particle at point $\mathbf{r}$ and exchanging it with one of its next nearest neighbors in the direction $\mu$ \cite{paco}. Notice that, in contrast with the previous ADS approach for the DLG with NN interactions, now $\mu$ involves the known parallel $\parallel$ and transversal $\perp$ directions as well as the two additional directions which allow for the diagonal jumps $\diagup$ and $\diagdown$. If $v$ is large enough, then $\phi$ can be considered a continuum function of $\mathbf{r}$ so we have 
\begin{equation}
\mathbf{\gamma}^{\prime}=\left\lbrace \phi+v^{-1} \, \nabla_{\mu} \delta(\mathbf{r}^{\prime}-\mathbf{r}),\; \text{where} \; \phi \in \mathbf{\gamma}        \right\rbrace 
\label{eq:conf}
\end{equation}
The distribution $P(\mathbf{\gamma},t)$ accounts for the statistical weight of each configuration $\mathbf{\gamma}$ at time $t$ and evolves according to the following Markovian master equation \cite{vkampen}
\begin{equation}
\partial_{t} P(\mathbf{\gamma},t)=\sum_{\mu=\left\lbrace \parallel, \perp, \diagup, \diagdown \right\rbrace} \int d\eta f(\eta) \int d \mathbf{r} \left[ \Omega(\mathbf{\gamma} \rightarrow \mathbf{\gamma}^{\prime})P(\mathbf{\gamma},t)-\Omega(\mathbf{\gamma}^{\prime} \rightarrow \mathbf{\gamma})P(\mathbf{\gamma}^{\prime},t) \right] 
\label{eq:master}
\end{equation}
where $\Omega(\mathbf{\gamma} \rightarrow \mathbf{\gamma}^{\prime})$ stands for the transition probability per unit time (transition rate) from $\mathbf{\gamma}$ to $\mathbf{\gamma}^{\prime}$ and $f(\eta)$ is an even function which account for the amount of mass attempted to be displaced. That is, Eq.~\eqref{eq:conf}--\eqref{eq:master} represent a proccess in which a configuration $\mathbf{\gamma}$ is exchanged with the infinitesimal neighbor of $\mathbf{r}$ ($\mathbf{\gamma}^{\prime}$) in the $\mu$ direction. As usual, the transition rates are taken to be a product of a function of the entropy times the same function of the difference between the configurations plus a term due to the effect of the driving field \cite{amit}, namely, 
\begin{equation}
\Omega(\mathbf{\gamma} \rightarrow \mathbf{\gamma}^{\prime})=D(\Delta S(\eta))\cdot D(\Delta H(\eta)+H_{\varepsilon}).
\label{eq:ads_rate}
\end{equation}
Here, $H_{\varepsilon}=\eta \, \mathbf{\mu} \cdot \mathbf{\varepsilon} \, (1-\phi^2)+\mathcal{O}(v^{-1})$, and $D$ is a function satisfying detailed balance. Importantly, in the Langevin--type equations framework, $\varepsilon$ represents a coarse--grained field acting upon a given configuration $\eta$. Indeed, the detailed dependence of the mesoscopic coefficients on the microscopic field introduced in Eq.~\eqref{eq:metro} remain still as an open issue \cite{marro}. This constraint ensures that the limiting case of $\varepsilon=0$ the steady state solution of the master equation is the canonical one, i.e. $P(\mathbf{\gamma},t=\infty)|_{\varepsilon=0}\,=\,e^{-H[\phi]}$. At a mesoscopic level, the equivalent to Eq.~\eqref{eq:hamilt} is the standard $\phi^{4}$ (Ginzburg--Landau) Hamiltonian, and following a similar procedure as here the equilibrium Langevin equation is the so called  \textit{model B} \cite{hohenberg}. The structure of the free energy consist of two contributions: a entropic and a energetic one which are given by, respectively
\begin{eqnarray} 
S(\eta)=v \int  dr \left( \frac{a}{2}\phi^{2} + \frac{g}{4!}\phi^{4}\right) \nonumber\\
H(\eta)=v \int  dr \left( \frac{1}{2} \left( \nabla \phi \right)^{2}+\frac{\tau}{2}\right).
\label{eq:ee}
\end{eqnarray}
Here $a$ and $g$ are entropic coefficients whereas the parameter $\tau$ comes from the energetic contribution. From Eq.~\eqref{eq:master} a Fokker--Planck equation is derived by means of a Kramers--Moyal \cite{vkampen} expansion. The derivation is similar to the one in Ref.~\cite{paco}, but taking into account the transversal degrees of freedom for the NDLG. After using standard techniques in the theory of stochastic processes \cite{vkampen} we derive from the Fokker--Planck equation its stochastically equivalent Langevin equation using the Ito prescription, which reads
\begin{equation}
\partial_t \phi(\mathbf{r},t)= \sum_{\mu} \nabla_{\mu} \left[ p(\lambda_{\mu}^{S},\lambda_{\mu}^{H}+\lambda_{\mu}^{\varepsilon})+q(\lambda_{\mu}^{S},\lambda_{\mu}^{H}+\lambda_{\mu}^{\varepsilon})^{1/2}\xi_{\mu}(r,t)  \right].
\label{eq:proto-L}
\end{equation}
where
\begin{eqnarray}
\lambda_{\mu}^{\varepsilon}=\mathbf{\mu} \cdot \mathbf{\varepsilon} \,(1-\phi^{2}), \nonumber\\
\lambda_{\mu}^{X}=-\left( \nabla_{\mu} \frac{\delta}{\delta \phi} \right) X, \nonumber\\
p(z_{1},z_{2})=\int dr\; \eta \,f(\eta) \,D(\eta z_{1})\,D(\eta z_{2}), \nonumber\\
q(z_{1},z_{2})=\int dr\; \eta^{2}\, f(\eta)\, D(\eta z_{1})\,D(\eta z_{2}).
\label{ea:proto-fL}
\end{eqnarray}
The time has been rescaled by $v^{-1}$ and $\xi_{\mu}(\mathbf{r},t)$ is a delta--correlated Gaussian white noise in the $\mu$ direction. After some algebra, the terms depending on both diagonal components $\diagup$ and $\diagdown$ can be split into the two parallel and transversal to the field components. We now focus on the critical region and discard irrelevant terms in the renormalization group sense by naive power counting. We perform the following anisotropic scale transformations:
\begin{equation}
\begin{split}
t\rightarrow \tau^{-z} t\,,\\
r_{\perp}\rightarrow \tau^{-1}r_{\perp}\,,\\
r_{\parallel}\rightarrow \tau^{-s}r_{\parallel}\,,\\
\phi \rightarrow \tau^{\delta}\phi\,.
\end{split}
\label{eq:adsscale}
\end{equation}
Expanding the Langevin equation Eq.~\eqref{eq:proto-L} around $\tau=0$ keeping only the more relevant terms, we found that the leading terms of the theory are $\nabla_{\perp}^{4}\phi$ and $\nabla_{\parallel}^{2}\phi$ which, together with imposing invariance of the time scale, the transverse spatial interaction, and the transverse noise, under scale transformations, lead to $s=2$, $\delta=(s+d-3)/2$, and $z=4$. According to this and after some cumbersome algebra, the Langevin equation for finite (coarse--grained) driving field contains a large bunch of field--dependent terms. Similarly to what occurs in the ADS approach for the DLG, just taking a large enough value of the field most of those terms vanish. By doing this, one is led to the following Langevin equation
\begin{equation}
{\partial}_t \phi(\mathbf{r},t)=-\nabla_{\bot}^4 \phi+(\tau+\frac{3}{2}a)\nabla_{\bot}^2 \phi + \frac{g}{4}\nabla_{\bot}^2\phi^3 +a \nabla_{\Vert}^2 \phi+\xi(\mathbf{r},t)
\label{eq:adsNNN}
\end{equation}
The resulting equation resemble the one for NN interactions (see Eq.~\eqref{eq:adsNN}), although in Eq.~\eqref{eq:adsNNN} appears additional prefactors of entropic nature. This is consistent with the fact that the model with the extended dynamics, i.e., the NDLG, possesses more neighbors than the DLG. The present equation for NNN interactions contains all the same symmetries as Eq.~\eqref{eq:adsNN} and both lead to the same critical behavior. The value of the order parameter critical exponent that follows from a renormalization group analysis of Eq.~\eqref{eq:adsNNN} is $\beta= 1/3$, which is consistent with the observed in the MC simulations described above. 

\section{\label{cap1:concl}Conclusions}
The above observations show that the nature of correlations is not enough to determine the phase diagram. There are already indications of this from the study of equilibrium systems, and also from other nonequilibrium models. That is, one may have a qualitatively different phase diagram but essentially the same two--point correlations by modifying the microscopic dynamics. It follows, in particular, that the exceptional behavior of the DLG cannot be understood just by invoking the two-point correlation function $C(\mathbf{i})$ ---or, equivalently, the structure factor $S(\mathbf{\kappa})$--- or crude arguments concerning symmetries. The fact that particles are constrained to travel precisely along the two principal lattice directions in the DLG is the cause for its singular behavior. Or in other words, the lattice geometry acts more efficiently in the DLG as an ordering agent than the field itself, which occurs rarely in actual cooperative transport in fluids. This means that the ordering in the DLG is a \textit{geometric} effect rather than \textit{dynamic}. Allowing jumping along intermediate directions, as in both the NDLG and the DLJF, modifies essentially the phase diagram but not features ---such as power--law correlations--- that seem intrinsic of the nonequilibrium nature of the phenomenon. This special arrangement of particles is also lacking, for instance, when other orientations of the driving field \cite{siders} are considered. This also would explain certain irregular configurations observed in three dimensions \cite{marro}, since this constraint is also lifted in the $d=3$ case. 

Rutenberg and Yeung \cite{triang2} also performed quenching experiments for
variations of the DLG\footnote{They allow for NNN hops, but the Hamiltonian only involves the NN interactions.}. They showed, in particular, that minor modifications in the DLG dynamics may lead to an inversion of the triangular anisotropies during the formation of clusters which finally condense into strips. Our observations above prove that such nonuniversal behavior goes beyond kinetics, namely, it also applies to the stationary state.

The unique exceptionality of the DLG has some important consequences. One is that this model involves features that are not frequent in nature. There are situations in which a drive induces stripes but not necessarily DLG behavior \cite{exp}. The fact that a particle impedes the freedom of the one behind to move along $\hat{x}$ for a large enough field may only occur very seldom in cooperative transport in fluids. The NDLG is more realistic in this sense. In any case, the great effort devoted to the DLG during two decades has revealed important properties of both nonequilibrium steady states and anisotropic phase transitions. On the other hand, it ensues that, due to its uniqueness, the DLG does not have a simple \textit{off-lattice} analog. This is because the ordering agent in the DLG is more the lattice geometry than the field itself. The fact that one needs to be very careful when modeling nonequilibrium phenomena ---one may induce both a wrong critical behavior and an spurious phase diagram--- ensues again in this example. This seems not to be so dramatic in equilibrium where, for example, the lattice gas is a useful oversimplification of a LJ fluid.

Concerning criticality, according to our observations we found that the \textit{aniso-tropic diffusive system} approach include the microscopic details of transverse dynamics which, in particular, allow one to distinguish between the DLG and the NDLG. This is in contrast with other, more heuristic approaches, e.g., the \textit{driven diffusive system}. The ADS approach lead to the right order parameter critical exponent given by MC simulations. We analyzed the order parameter, magnetization, and the fourth Binder cumulant in order to compute the critical exponents. By using anisotropic finite size scaling we have shown that, in spite of structural differences between them, both models belong to the same universality class, which is characterized by an order parameter critical exponent $\beta\approx 1/3$. 

Consequently, their critical properties do not depend on this particular extension of the dynamics. There are other nonequilibrium models in which the corresponding critical behavior changes by such an extension of the dynamic from NN to NNN. That is, for instance, the case of a closely related lattice system (with NN \textit{exclusion} under a driving field) devised by Dickman \cite{dickman}: When one extends the dynamics to NNN \cite{szabo}, the nature of the order--disorder transition which takes place in this model changes dramatically. We shall discuss on this model further in the next chapter.


Finally, we remark that in the novel fluid model introduced in this chapter, the particle `infinite freedom' is realistic, as it is also the LJ potential. It may contain some of the essential physics in a class of nonequilibrium anisotropic phenomena and phase transitions. On the other hand, the model is simple enough to be useful in computer simulations, and it is endowed of even--simpler and functional lattice analogs such as the NDLG. In the following chapter (Chapter~\ref{cap2}) we shall develop in greater detail this matter.

\chapter[Anisotropic phenomena in nonequilibrium fluids]{\label{cap2}Prototypical model for anisotropic phenomena in nonequilibrium fluids}


\begin{figure}
\centerline{\includegraphics[width=8cm,clip=]{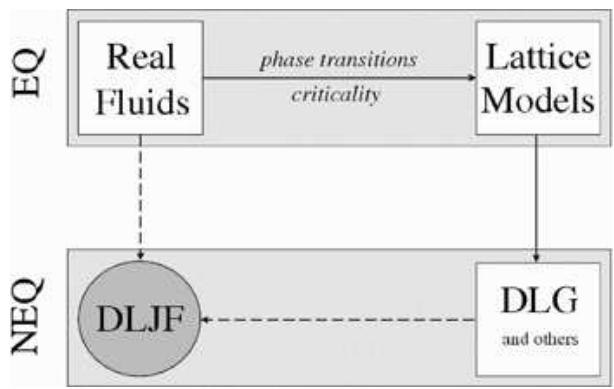}}
\caption[Off--lattice models in the realm of computer simulation of fluids]{\label{fig0_2}The place of our novel off-lattice models (as, for instance, the DLJF introduced in the previous chapter) in the realm of computer simulation of fluids.}
\end{figure}

In this chapter, we describe short--time kinetic and steady--state properties of the non--equilibrium phases, namely, (mono- and polycrystalline) solid, liquid and gas anisotropic phases in a novel driven Lennard--Jones fluid. This is a computationally--convenient model which exhibits a net current and striped structures at low temperature, thus
resembling many situations in nature. We here focus on both critical behavior and some details of the early nucleation process, by far less studied than later regimes. In spite of the anisotropy of the late--time \textquotedblleft spinodal decomposition\textquotedblright\ process, earlier nucleation seems to proceed by \textit{Smoluchowski coagulation} and \textit{Ostwald ripening},
which are known to account for nucleation in equilibrium, isotropic lattice systems and actual fluids. On the other hand, a detailed analysis of the system critical behavior rises some intriguing questions on the role of symmetries; this concerns the computer and field--theoretical modeling of nonequilibrium fluids.

\section{\label{cap2:intro}Motivation}
The microscopic mechanism of stripe formation in particular materials is still not clear\footnote{For instance, in high--temperature superconductors \cite{kivelson}, questions such as whether or not the Coulomb interactions are important in these kind of materials remain unresolved.}. Hence, lacking theory for the \textquotedblleft thermodynamic\textquotedblright\ instabilities causing the observed striped structures, one tries to link them to the microscopic dynamics of suitable model systems. As we explained in the previous chapters, the \textit{driven lattice gas} (DLG) \cite{kls}, namely, a model system in which particles diffuse under an external driving field, has been up to now a theoretical prototype of anisotropic behavior \cite{marro,zia,odor}. This model was shown in Chapter~\ref{cap1} to be unrealistic in some essential sense, however. Particle moves in the DLG are along the principal lattice directions, and any site can hold one particle at most, so that a particle impedes the one behind to jump freely along the direction which is favored to model the action of the field. Consequently, the lattice geometry acts more efficiently in the DLG as an ordering agent than the field itself, which occurs rarely ---never so dramatically--- in actual cooperative transport in fluids. In fact, actual situations may in principle be more closely modeled by means of continuum models, and this peculiarity of the DLG implies that it lacks a natural off-lattice extension.

Here we present and analyze numerically a novel nonequilibrium off-lattice, \textit{Lennard--Jones} (LJ) system which is a candidate to portray some of the anisotro-pic behavior in nature\footnote{This is in fact a related model to the \textit{driven Lennard-Jones fluid} (DLJF) introduced in the previous Chapter.}. The model, which involves a driving field of intensity $E,$ reduces to the celebrated (equilibrium) LJ \textit{fluid} \cite{allen,smit} as $E\rightarrow 0.$ For any $E>0,$ however, it exhibits currents and anisotropic phases as in many observations out of equilibrium. In particular, as the DLG, our model in two dimensions shows striped steady states below a critical point. We also observe critical behavior consistent with the equilibrium universality class. This is rather unexpected in view of the criticality reported both for the DLG. 
On the other hand, concerning the early--time relaxation before well--defined stripes form by spinodal decomposition, we first observe ---as in previous studies of relaxation towards equilibrium--- effective diffusion of small droplets, which is followed by monatomic diffusion probably competing with more complex processes. It is very likely that our observations here concerning nucleation, coexistence, criticality, and phases morphology hold also in a number of actual systems. 

Before we proceed, it is worth mentioning that the off-lattice models\footnote{Here we include the DLJF model defined in Chapter~\ref{cap1}.} devised over the course of this thesis fills a void in the physics of nonequilibrium fluids.  This is schematized in Fig.~\ref{fig0_2}. As we explained in Chapter~\ref{cap0}, lattice models are convenient, simplified models which often capture the main ingredients of equilibrium systems, particularly, close to the critical points. For instance, the \textit{lattice gas} may be a useful representation of, e.g., a Lennard-Jones fluid. However, rather surprisingly, nonequilibrium critical phenomena are by far more widely studied in nonequilibrium extensions of lattice models than in their \textit{continuum} counterparts. In this chapter, we provide a more \textit{realistic} ---without loss the efficiency of lattice models---, convenient model for the study of nonequilibrium phase transitions in fluids, and we suggest simple procedures to extend the study of actual fluids to the nonequilibrium realm.

In section~\ref{cap2:model} we define the off-lattice model, and section~\ref{cap2:results} is devoted to the main results as follows. \S\ ~\ref{cap2:kin} describes the early--time segregation process as monitored by the excess energy, which measures the droplets surface. \S\ ~\ref{cap2:struc} describes some structural properties of the steady state, namely, the radial and azimuthal distribution functions, and the degree of anisotropy. \S\ ~\ref{cap2:trans} depicts some transport properties. And \S\ ~\ref{cap2:coex} is devoted to an accurate estimate of the liquid--vapor coexistence curve and the associated critical parameters. Finally, Section~\ref{cap2:conclus} contains the main conclusions and final comments.

\section{\label{cap2:model}The model}
Consider $N$ particles of equal mass (set henceforth to unity) in a $d-$dimensional square box, $L^{d},$ with periodic boundary conditions. Interactions are via the truncated and shifted pair potential \cite{allen} (already defined in Chapter~\ref{cap1}; see Eq.~\eqref{eq:stLJP}): 
\begin{equation}
\phi (r)=\left\{ 
\begin{array}{cc}
\phi _{LJ}(r)-\phi _{LJ}(r_{c}) & \text{if }r<r_{c}\  \\ 
0 & \text{if }r\geqslant r_{c},%
\end{array}%
\right.
\end{equation}%
where 
\begin{equation}
\phi _{LJ}(r)=4\epsilon \left[ (\sigma /r)^{12}-(\sigma /r)^{6}\right]
\end{equation}%
is the full LJ potential, $r$ stands for the interparticle distance, and $r_{c}$
is a \textit{cut-off} that we set $r_{c}=2.5\sigma $. The parameters $\sigma 
$ and $\epsilon $ are, respectively, the characteristic length and energy
---that we use in the following to reduce units as in Eq.~\eqref{eq:rescal}.

Time evolution is by microscopic dynamics according to the transition
probability per unit time (\textit{rate}):
\begin{equation}
\omega ^{\left( E\right) }\left( \mathbf{\eta }\rightarrow \mathbf{\eta}^{\prime}\right) =\chi ^{(E)}\times \min \left\{ 1,e^{-\Delta \Phi/T}\right\} .  \label{rate}
\end{equation}%
Here, $\chi^{(E)}$, which is a nonequilibrium extension of the Metropolis rate \cite{metropolis}, reads
\begin{equation}
\chi ^{(E)}=\frac{1}{2}\left[ 1+\tanh \left( E\hat{x}\cdot \mathbf{\delta}\right) \right] ,  \label{bias}
\end{equation}
$E$ is the intensity of a uniform external field along the horizontal direction, say $\hat{x},$ $\mathbf{\eta }\equiv \left\{ \mathbf{r}_{1},\ldots ,\mathbf{r}_{N}\right\} $ stands for any configuration of energy 
\begin{equation}
\Phi (\mathbf{\eta })=\sum_{i<j}\phi \left( \left\vert \mathbf{r}_{i}-\mathbf{r}_{j}\right\vert \right) ,
\end{equation}
where $\mathbf{r}_{i}$ is the position of particle $i$ that can be anywhere in the $d-$torus, $\mathbf{\eta}^{\prime}$ equals $\mathbf{\eta}$ except for the displacement of a single particle by $\mathbf{\delta}=\mathbf{r}_{i}^{\prime}-\mathbf{r}_{i},$ and $\Delta \Phi\equiv \Phi (\mathbf{\eta}^{\prime})-\Phi(
\mathbf{\eta })$ is the cost of such displacement.

\begin{figure}
\centerline{\includegraphics[width=6cm,clip=]{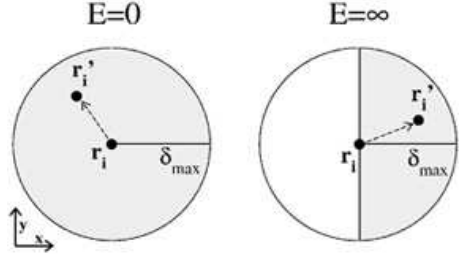}}
\caption[Schematic representation of the region which is accessible to a given particle as a consequence of a trial move]{\label{fig1_2}Schematic representation of the region (grey) which is accessible to a given particle as a consequence of a trial move for $E=0$ (left-handside) and $E=\infty $ (right-handside), assuming the \textquotedblleft infinite\textquotedblright\ field points $\hat{x},$ horizontally.}
\end{figure}

It is to be remarked that $\chi ^{(E)},$ as defined in (\ref{bias}), contains a drive bias (see Fig.~\ref{fig1_2}) such that the rate \eqref{rate} lacks invariance under the elementary transitions $\mathbf{\eta}\leftrightarrows \mathbf{\eta}^{\prime} .$ Consequently, unlike in equilibrium, there is
no detailed balance for toroidal boundary conditions if $E>0.$

Here we describe the results from a series of Monte Carlo (MC) simulations. In order to reduce unnecessary
interparticle distance calculations, we used a standard neighbor--list algorithm \cite{allen}. Our simulations concern fixed values of $N,$ with $N\leq 10^{4},$ particle density $\rho =N/L^{d}$ within
the range $\rho \in \left[ 0.2,0.6\right] ,$ and temperature $T\in \left[
10^{-2},10^{5}\right] .$ Following the fact that most studies of striped
structures, e.g., many of the ones are mentioned in the Chapter~\ref{intro:DDF}, concern two dimensions ---in particular, the DLG critical behavior is only known with some confidence for $d=2$ \cite{marro,beta4,beta5}--- we restricted ourselves to a two dimensional torus. We fixed the maximum particle displacement is $\delta _{\text{max}}=0.5$ in our simulations as shown in Fig.~\ref{fig1_2}. We deal below with steady--state averages over $10^{6}$ configurations, and kinetic or time averages over 40 or more independent runs.

The distribution of displacements $\mathbf{\delta}$ is uniform, except that the new particle position $\mathbf{r}_{i}^{\prime }$ is (most often in our simulations) sampled only from within the half--forward semi--circle of radius $\delta _{\text{max}}$ centered at $\mathbf{r}_{i},$ as illustrated on
the right-hand graph of Fig.~\ref{fig1_2}. This is because the \textit{infinite--field} limit, $E\rightarrow \infty ,$ turns out to be most relevant, and this means, in practice, that any displacement contrary to the field is forbidden. This choice eliminates from the analysis one parameter and, more importantly, it happens to match a physically relevant case. As a matter of fact, simulations reveal that any external field $E>0$ induces a
flux of particles along $\hat{x}$ ---which crosses the system with toroidal boundary conditions--- that monotonically increases with $E$, and eventually saturates to a maximum. This is a realistic stationary condition in which the thermal bath absorbs the excess of energy dissipated by the drive. Our simulations here concern only $E=\infty$ in order to maximize the nonequilibrium effects\footnote{We are interested in the far-from-equilibrium effects induced by the external driving field rather than in weak deviations from equilibrium.}.

\begin{figure}
\centerline{\includegraphics[width=8.55cm,clip=]{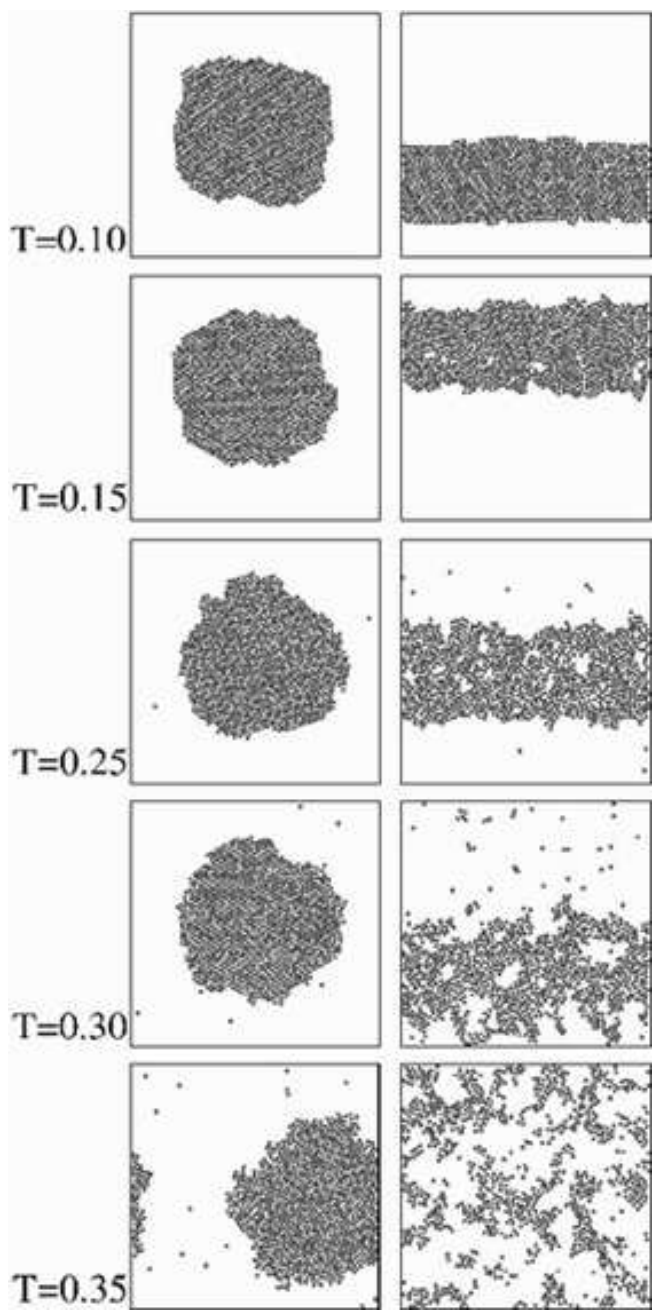}}
\caption[Typical steady--state configurations for equilibrium ($E=0$) and large field limit ($E\rightarrow \infty $)]{\label{fig2_2}Typical steady--state configurations for $E=0$ (left column) and $E\rightarrow \infty $ (right column) at different temperatures as indicated. This is for $N=1000$ and $\protect\rho =0.30.$ The field is oriented along the horizontal direction.}
\end{figure}

\section{\label{cap2:results}Main results}
Figure~\ref{fig2_2} illustrates late--time configurations, i.e., the ones that typically characterize the steady state, as the temperature $T$ is varied. These graphs already suggest that the system undergoes an
order--disorder phase transition at some temperature $T_{E}.$ This happens to be of second order for any $E>0,$ as in the equilibrium case $E=0.$ We also observe that $T_{E}$ decreases monotonically with increasing $E,$ and
that it reaches a well--defined minimum, $T_{\infty },$ as $E\rightarrow\infty .$

Figure~\ref{fig2_2} also shows that, at low enough temperature, an anisotropic interface forms between the condensed phase and its vapor; this extends along $\hat{x}$ throughout the system at intermediate densities.

\subsection{\label{cap2:kin}Phase segregation kinetics}
\begin{figure}
\centerline{\includegraphics[width=10cm,clip=]{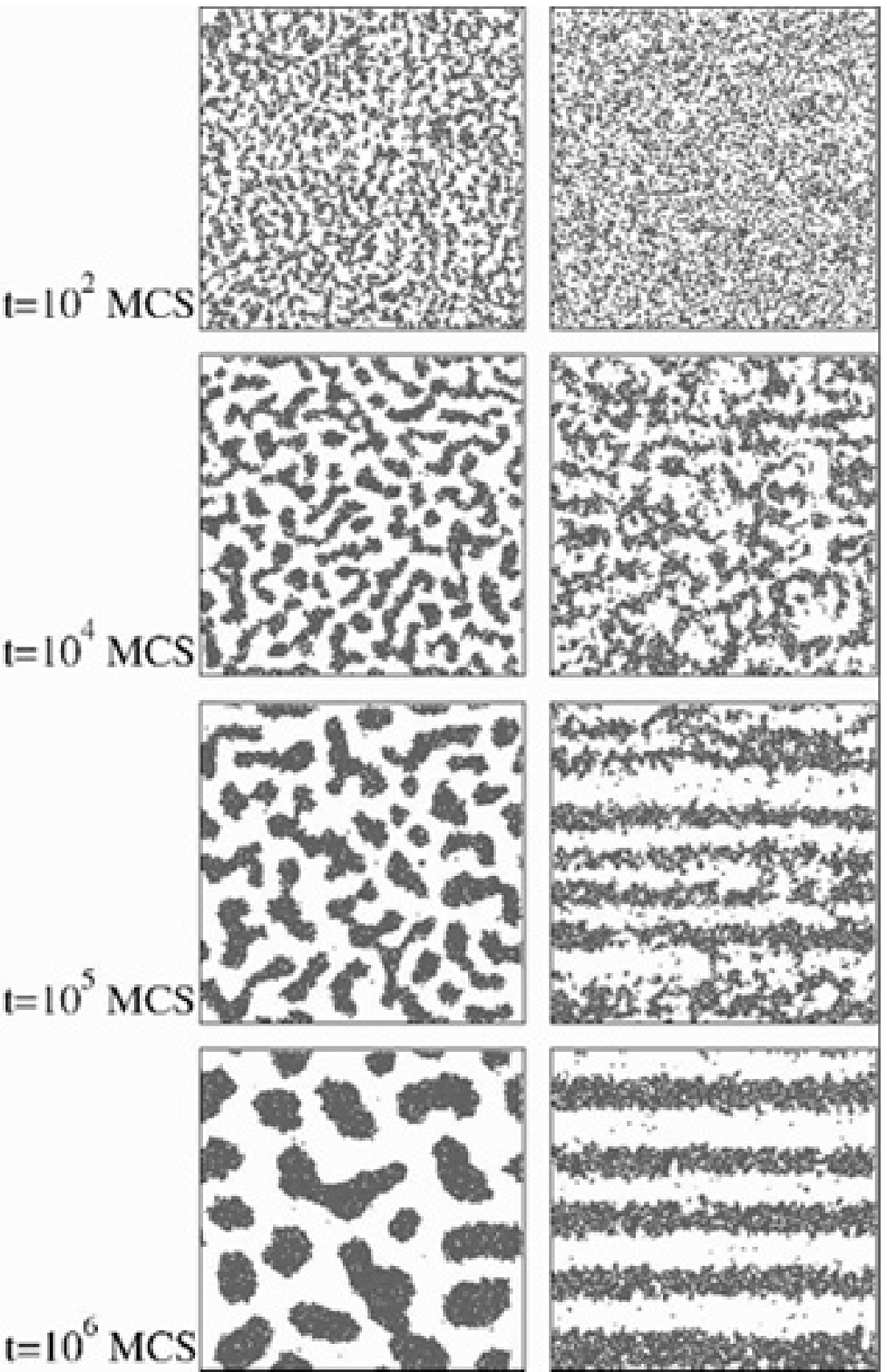}}
\caption[Typical configurations for $E=0$ and $E\rightarrow \infty$ as time proceeds during relaxation from a disordered state]{\label{fig3_2}Typical configurations for $E=0$ (top row) and $E\rightarrow \infty $ (bottom row) as time proceeds during relaxation from a disordered state (as for $T\rightarrow \infty )$ at $t=0$. The time here is proportional to the MC steps (MCS) as indicated. This is for $N=7500$, $\protect\rho =0.35,$ and $T=0.275,$ below the corresponding transition temperature.}
\end{figure}

Skipping microscopic details, the kinetics of phase segregation at late times looks qualitatively similar to the one in other nonequilibrium cases, including the DLG \cite{hurtado} and both molecular--dynamic \cite{zeng} and Cahn--Hilliard \cite{ludo} representations of sheared fluids, while it essentially differs from the one in the
corresponding equilibrium system. This is illustrated in Fig.~\ref{fig3_2}. One observes, in particular, condensation of many stripes ---as in the graph for $t=10^{5}$ in Fig.~\ref{fig3_2}--- into a single one ---as in the first three graphs at the bottom row in Fig.~\ref{fig2_2}. This process corresponds to an anisotropic version of the so--called \textit{spinodal decomposition }\cite{spinod}, which is mainly characterized by a tendency towards minimizing the interface surface as well as by the existence of a unique relevant length, e.g., the stripe width \cite{hurtado}. We focus here on a detailed study of the early-time separation process instead of the late regime, which has already been studied for both equilibrium \cite{marro2,bray} and nonequilibrium cases, including the DLG \cite{hurtado,levine}. This is because detailed descriptions of early nonequilibrium nucleation are rare as
compared to studies of the segregation process near completion. 

Following an instantaneous quench from a disordered state into $T<T_{\infty},$ one observes in our case that small clusters form, and then some grow at the expenses of the smaller ones but rather independently of the growth of other clusters of comparable size. This corresponds to times $t<10^{5}$ in Fig.~\ref{fig3_2}, i.e., before many well--defined stripes form. We monitored in this regime the \textit{excess energy} or \textit{enthalpy}, $H\left(t\right),$ measured as the difference between the averaged internal energy at time $t>0$ and its stationary value. This reflects more accurately the growth of the condensed droplets than its size or radius, which are
difficult to be estimated during the early stages \cite{toral,chinos}. In addition to its theoretical importance, the excess of energy may be relevant for experimentalists. This is because $H\left( t\right)$ may be determined in microcalorimetric experiments \cite{marro3}.

\begin{figure}
\centerline{\includegraphics[width=8cm,clip=]{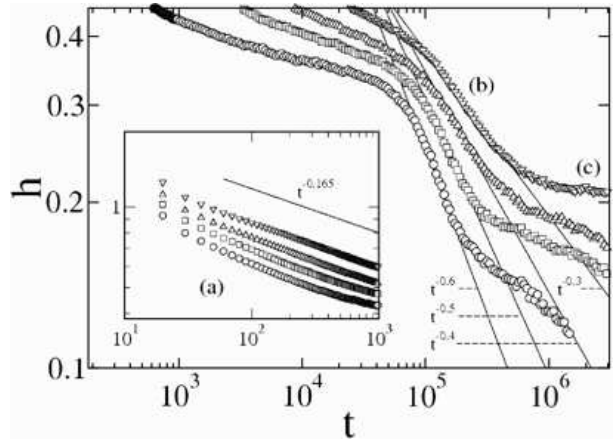}}
\caption[Time evolution of the enthalpy or excess of energy per particle for different temperatures]{\label{fig4_2}Time evolution of the enthalpy per particle for $N=7500,$ $\protect\rho =0.35$ and, from top to bottom, $T=0.200,$ $0.225$, $0.250$, and $0.275$. Straight lines are a guide to the eye; the slope of each line is indicated. The inset shows the detail at early times. (For clarity of presentation, the main graph includes a rescaling of the time corresponding to the data for $T=0.250,$ $0.225$ and $0.200$ by factors $2$, $3$ and $3$, respectively.)}
\end{figure}

The time development of the enthalpy density $h\left( t\right) =H\left(t\right) /N$ is depicted in Fig.~\ref{fig4_2}. This reveals some well--defined regimes at early times.

The first regime, (a) in the inset of Fig.~\ref{fig4_2}, follows a power law $t^{-\theta }$ with $\theta \approx 1/6$ ---which corresponds to the line shown in the graph--- independently of the temperature investigated. This is the behavior predicted by the \textit{Smoluchowski coagulation} or effective cluster diffusion \cite{binder2}. The same behavior was observed in computer simulations for $E=0$ and also reported to hold in actual experiments on binary mixtures \cite{toral,marro3}. Surprisingly, this suggests the early dominance of a rather stochastic mechanism, in which the small clusters rapidly nucleate, which is practically independent of the field, i.e., it is not affected in practice by the drive. It is also noticeable that, in contrast with simulations on lattice gases in which this regime is rather noisy, a slope $t^{-1/6}$ appears clearly in Fig.~\ref{fig4_2}. The indication of some temperature dependence in equilibrium \cite{chinos}, which is not evident here, might correspond to the distinction between \textit{deep} and \textit{shallow quenches} made in Ref.~\cite{toral} that we have not investigated out of
equilibrium.

At later times, there is a second regime, (b) in Fig.~\ref{fig4_2}, in which the anisotropic clusters merge into filaments and, finally, stripes. We observe in this regime that $\theta $ varies between $0.3$ and $0.6$ with
increasing $T$. \textit{Ostwald ripening }\cite{lifshitz}, consisting of monomers diffusion, predicts $\theta =1/3$ for fluids at thermal equilibrium. The situation here in our nonequilibrium is more complicate: it is likely that regime (b) describes a crossover from a situation which is dominated by monomers at low enough temperature to the emergence of other mechanisms \cite{bray,baum} which might be competing as $T$ is increased.

Finally, one observes a regime, (c) in Fig.~\ref{fig4_2}, which corresponds to the beginning of spinodal decomposition.

\subsection{\label{cap2:struc}Structure of the steady state}
Concerning the local structure in the stationary regime, for any $E>0,$ the anisotropic condensate changes from a solidlike hexagonal packing of particles at low temperature (e.g., $T=0.10$ in Fig.~\ref{fig2_2}), to a polycrystalline or perhaps glass--like structure with domains which show a varied morphology at, e.g., $T=0.12.$ The latter phase further transforms, with increasing temperature, into a fluid--like structure at, e.g., $T=0.30$ and, finally, into a disordered, gaseous state.

\begin{figure}
\centerline{\includegraphics[width=10cm,clip=]{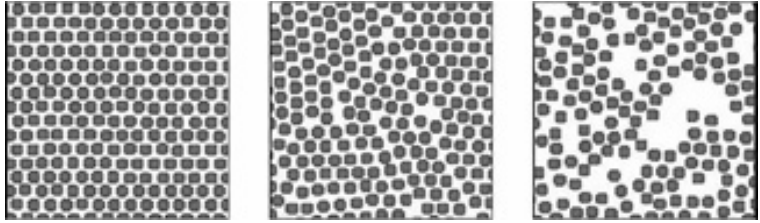}}
\caption[Details of the structure in the low$-T,$ solid phase]{\label{fig5_2}Details of the structure in the low$-T,$ solid phase as obtained by zooming into configurations such as the ones in Fig.~\ref{fig2_2}. This is for $T=0.05,$ $0.12$, and $0.25$, from left to right, respectively.}
\end{figure}

More specifically, the typical situation we observe at low temperature is illustrated in Fig.~\ref{fig5_2}. At sufficiently low temperature, $T=0.05$ in the example, the whole condensed phase orders according to a
perfect hexagon with one of its main directions along the field direction $\hat{x}.$ This is observed in approximately 90\% of the configurations that we generated at $T=0.05,$ while all the hexagon axes are slanted with
respect to $\hat{x}$ in the other 10\% cases. As the system is heated up, the stripe looks still solid at $T=0.12$ but, as illustrated by the second graph in Fig.~\ref{fig5_2}, one observes in this case several coexisting hexagonal domains with different orientations. The separation between domains is by line defects and/or vacancies. Interesting enough, as it will be shown later on, both the system energy and the particle current are practically independent of temperature up to, say $T=0.12.$ The hexagonal ordering finally disappears in the third graph of Fig.~\ref{fig5_2},
which is for $T=0.25;$ this case corresponds to a fluid phase according to the criterion below.

\begin{figure}
\centerline{\includegraphics[width=8cm,clip=]{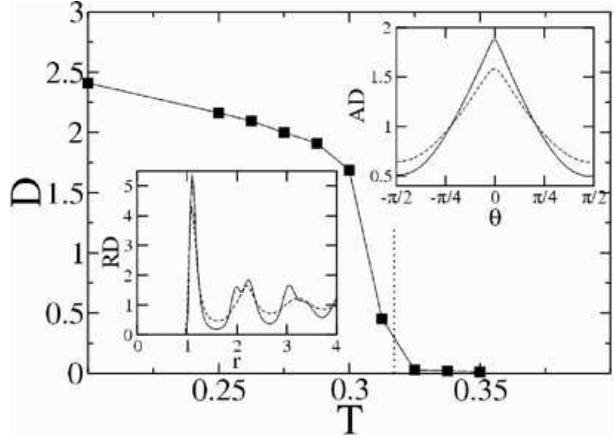}}
\caption[Degree of anisotropy (main graph), radial (lower inset) and azimuthal (upper inset) distribution functions]{\label{fig6_2}Data from simulations for $N=7000$ and $\protect\rho =0.35$. The main graph shows the degree of anisotropy, as defined in the main text, versus temperature. The vertical dotted line denotes the transition temperature. The lower (upper) inset shows the radial (azimuthal) distribution at $T=0.20,$ full line, and $T=0.30,$ dashed line.}
\end{figure}

A close look to the structure is provided by the radial distribution (RD),
\begin{equation}
g\left( r\right) =\rho ^{-2}\left\langle \sum_{i<j}\delta \left( \mathbf{r}_{i}-\mathbf{r}_{j}\right) \right\rangle ,
\end{equation}
i.e., the probability of finding a pair of particles a distance $r$ apart, relative to the case of a random spatial distribution at same density. This is shown in the lower inset of Fig.~\ref{fig6_2}. At fixed $T,$ the driven fluid is less structured than its equilibrium counterpart, suggesting that the field favors disorder. This is already evident in Fig.~\ref{fig2_2}, and it also follows from the fact that the critical temperature decreases with increasing $E.$

Further information on the essential anisotropy of the problem is beared by the azimuthal distribution (AD), which accounts for the longe range ordering. This is defined 
\begin{equation}
\alpha (\theta )=N^{-2}\left\langle \sum_{i<j}\delta \left( \theta -\theta
_{ij}\right) \right\rangle ,
\end{equation}
where $\theta _{ij}\in \lbrack 0,2\pi )$ is the angle between the line connecting particles $i$ and $j$ and the field direction $\hat{x}.$ Except at equilibrium, where this is uniform, the AD is $\pi /2-$periodic with maxima at $k\pi $ and minima at $k\pi /2,$ where $k$ is an integer. This means that the field favors long-range longitudinal ordering instead of order along other directions. This is consistent with the observations in which the orientations of the hexagons in the solidlike phase are parallel to the field direction. The AD is depicted in the upper inset of Fig.~\ref{fig6_2}. 

We also monitored the \textit{degree of anisotropy,} defined as the distance $D$
\begin{equation}
D=\int_{0}^{2\pi }d\theta\,\left\vert \alpha(\theta) -1\right\vert ,  \label{funcD}
\end{equation}%
which measures the deviation from the equilibrium, isotropic case, for which $\alpha (\theta )=1,$ independent of $\theta .$ The function (\ref{funcD}), which is depicted in the main graph of Fig.~\ref{fig6_2}, reveals the
existence of anisotropy even above the transition temperature. This shows the persistence of nontrivial two--point correlations at high temperatures which has been demonstrated for other nonequilibrium models \cite{pedro}.

\subsection{\label{cap2:trans}Transport properties}
\begin{figure}
\centerline{\includegraphics[width=8cm,clip=]{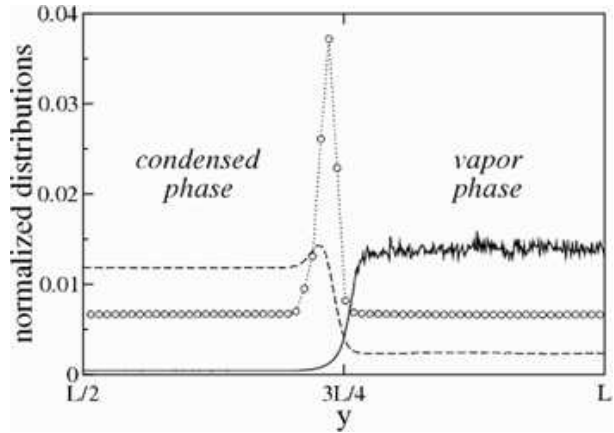}}
\caption[Transverse--to--the--field current profiles below criticality: comparison between lattice and off-lattice models]{\label{fig65_2}. Transverse--to--the--field current profiles below criticality. The shaded (full) line corresponds to the current (velocity) profile of the off--lattice model. For comparison we also show the current profile of the DLG (circle--dotted line). Since each distribution is symmetric with respect to the system center of mass (located here at $L/2$) we only show their right half parts.}
\end{figure}

The current is highly sensitive to the anisotropy. The most relevant information is carried by the transverse--to--the--field current profile $j_{\perp}$, which shows the differences between the two coexisting phases (Fig.~\ref{fig65_2}). Above criticality, where the system is homogeneous, the current profile is flat on the average. Otherwise, the condensed phase shows up a higher current (lower mean velocity) than its mirror phase, which shows up a lower current (higher mean velocity). Both the transversal current and velocity profiles are shown in Fig.~\ref{fig65_2}. The current and the density vary in a strongly correlated manner: the high current phase corresponds to the condensed (high density) phase, whereas the low current phase corresponds to the vapor (low density) phase. This is expectable due to the fact that there are many carriers in the condensed phase which allow for higher current than in the vapor phase. However, the mobility of the carriers is much larger in the vapor phase. The maximal current occurs in the interface, where there is still a considerable amount of carriers but they are less bounded than in the particles well inside the \textit{bulk} and, therefore, the field drives easily those particles. This enhanced current effect along the interface is more prominent in driven lattice models (notice the large peak in the current profile in Fig.~\ref{fig65_2}). Thus, these are clear cases of interfacially driven transport \cite{microfluidics}. Moreover in both lattice cases, DLG and NDLG (already studied in Chapter~\ref{cap1}), there is no difference between the current displayed by the coexisting phases because of the particle--hole symmetry. Such a symmetry is derived from the Ising--like Hamiltonian in Eq.~\eqref{eq:hamilt} and it is absent in the off--lattice case.

\begin{figure}
\centerline{\includegraphics[width=8.5cm,clip=]{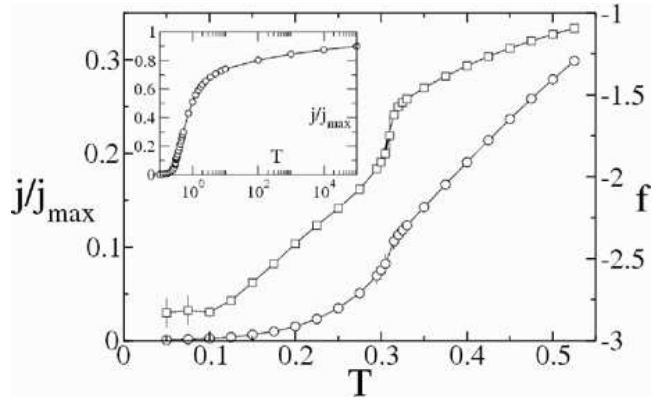}}
\caption[Temperature dependence of the mean energy and normalized net current for large field limit]{\label{fig7_2}Temperature dependence of the mean energy (squares; the scale is on the right axis) and normalized net current (circles; scale at the left) for $N=5000$ and $\protect\rho =0.30$ under \textquotedblleft infinite\textquotedblright\ field. The inset shows the $T-$dependence of the current over a wider range.}
\end{figure}

Regarding the comparison between off--lattice and lattice transport properties, Fig.~\ref{fig7_2} shows the net current $j$, defined as the mean displacement per MC step per particle, as a function of temperature. Saturation is only reached at $j_{max}=4\delta_{max}/3\pi$ when $T\rightarrow \infty$. The current approaches its maximal value logarithmically, i.e., slower than the exponential behavior predicted by the Arrhenius law. The way this limit is approached is illustrated in the inset of Fig.~\ref{fig7_2}. The sudden rising of the current as $T$ is increased can be interpreted as a transition from a poor--conductor (low--temperature) phase to a rich--conductor (high--temperature) phase, which is reminiscent of ionic currents \cite{marro}. This behavior of the current also occurs in the DLG. Revealing the persistence of correlations, the current is nonzero for any, even low $T,$ though it is small, and roughly independent of $T$, in the solid phase. 

An insight of the transition points between the different phases is obtained from the temperature dependence of the mean potential energy per particle $f=N^{-1}\left\langle \Phi (\mathbf{\eta })\right\rangle$ and from the net current $j$. In particular, one may estimate the transition point as the condensed strip changes from solid to liquid ($T\approx 0.15$) and finally changes to a fully disordered state ($T\approx 0.31$), i.e., disordered state. This is shown in Fig.~\ref{fig7_2} which exhibits well--defined changes of slope in both magnitudes when the phases transform. Notice also that the mean energy behaves linearly with temperature for $T\in \left( 0.12,0.3\right) $ in the liquid phase.


\subsection{\label{cap2:coex}Coexistence curve}
\begin{figure}
\centerline{\includegraphics[width=8cm,clip=]{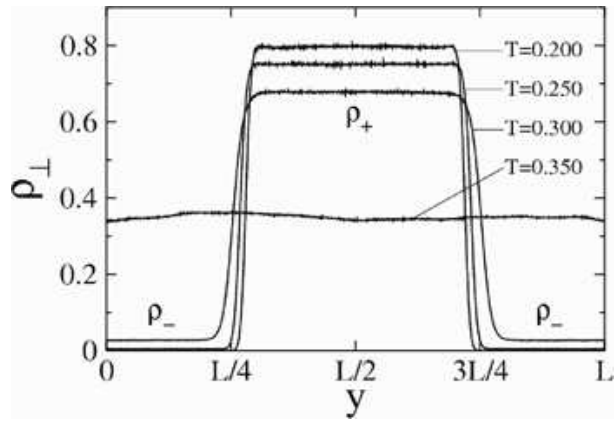}}
\caption[Density profiles transverse to the field for different temperatures]{\label{fig8_2}Density profiles transverse to the field for $N=7000,$ $\protect\rho =0.35,$ and different temperature, as indicated. The coexisting densities, $\protect\rho _{\pm },$ are indicated.}
\end{figure}

One of the main issues concerning the steady state is the liquid--vapor coexistence curve and the associated critical behavior. The (nonequilibrium) coexistence curve may be determined from the density profile transverse to
the field. This is illustrated in Fig.~\ref{fig8_2}.

At high enough temperature ---in fact, already at $T=0.35$ in this case for which the transition temperature is slightly above $0.3$--- the local density is roughly constant around the mean system density, $\rho =0.35$ in Fig.~\ref{fig8_2}. As $T$ is lowered, the profile accurately describes the existence of a single stripe of condensed phase of density $\rho _{+}$ which coexists with its vapor of density $\rho _{-}$ ($\rho _{-}\leq \rho _{+}$). The interface becomes thinner and smother, and $\rho _{+}$ increases while $\rho _{-}$ decreases, as $T$ is decreased.

\begin{figure}
\centerline{\includegraphics[width=8cm,clip=]{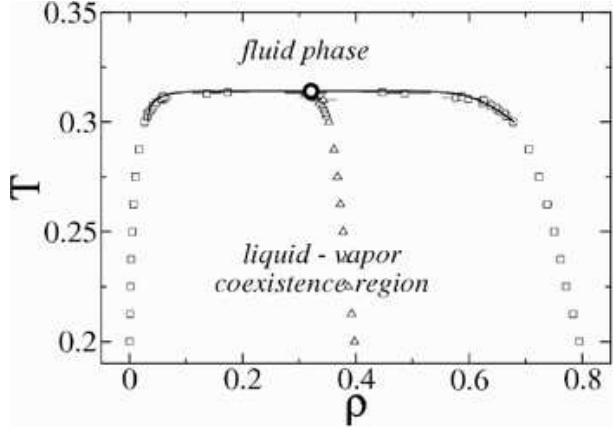}}
\caption[Liquid--vapor coexistence curve and the associated critical indexes obtained from the density profiles]{\label{fig9_2}Coexistence curve (squares) for the LJ nonequilibrium model obtained from the density profiles in Fig.~\ref{fig8_2}. The fluid phase and the coexistence region are indicated. The triangles are the arithmetic mean points, which serve to compute the critical parameters. The large circle at the top of the curve locates the critical point, and the solid line is a fit using the Wegner
expansion and the rectilinear diameter law with the critical parameters given in Table~\ref{cap2:table1}.}
\end{figure}

As in equilibrium, one may use $\rho _{+}-\rho _{-}$ as an order parameter. The result of plotting $\rho _{+}$ and $\rho _{-}$ at each temperature results in the non-symmetric liquid-vapor coexistence curve shown in Fig.~\ref{fig9_2}. The same result follows from the current, which in fact varies strongly correlated with the local density. Notice that the accuracy of our estimate of $\rho _{\pm}$ is favored by the existence of a linear interface, which enable us to get closer to the critical point than in equilibrium. Lacking a \textit{thermodynamic} theory for ``phase transitions" in nonequilibrium liquids, other approaches have to be considered in order to estimate the critical parameters. 

Consider to the rectilinear diameter law 
\begin{equation}
\frac{1}{2}(\rho _{+}+\rho _{-})=\rho _{\infty }+b_{0}(T_{\infty }-T),
\end{equation}%
which is a empirical fit \cite{biel} extensively used for fluids in equilibrium. Here $\rho_{\infty}$ and $T_{\infty}$ denotes the critical density and the critical temperature, respectively. This, in principle, has no justification out of equilibrium. However, we found that our MC data nicely fit the diameters equation. We use this fact together with a universal scaling law\footnote{The first term of a Wegner-type expansion \cite{wegner}.}
\begin{equation}
\rho _{+}-\rho _{-}=a_{0}(T_{\infty }-T)^{\beta },
\end{equation}%
to accurately estimate the critical parameters. The simulation data in Fig.~\ref{fig9_2} then yields the values in Table~\ref{cap2:table1}, which are confirmed by the familiar log--log plots. Compared to the equilibrium critical temperature reported by Smit and Frenkel \cite{smit}, one has that $T_{0}/T_{\infty }\approx 1.46,$ i.e., the change is opposite to the one for the DLG \cite{marro}. This confirms the observation above that the field acts in the nonequilibrium LJ system favoring disorder.

\begin{table}
\caption[Critical parameters]{\label{cap2:table1}Critical parameters}
\centerline{
\begin{tabular}{l|l|l}
$\rho _{\infty }$ & $T_{\infty }$ & $\beta $ \\ \hline\hline
0.321(5) & 0.314(1) & 0.10(8)
\end{tabular}}
\end{table}

The fact that the order--parameter critical exponent is relatively small may already be guessed by noticing the extremely flat coexistence curve in Fig.~\ref{fig9_2}. This is similar to the corresponding curve for the
equilibrium two--dimensional LJ fluids \cite{smit,papa1,papa2}, and it is fully consistent with the equilibrium Onsager value, $\beta =1/8.$ We therefore believe that our model belongs to the Ising universality class. In any case, although the error bar is large, one may discard with confidence both the DLG value $\beta \approx 1/3$ as well as the mean field value $\beta =1/2$ ---both cases would produce a hump visible to the naked eye in a plot such as the one in Fig.~\ref{fig9_2}. One may argue that this result is counterintuitive, as our model \textit{apparently} has the short--range interactions and symmetries that are believed to characterize the DLG. Further understanding for this difference will perhaps come from the statistical field theory.

\section{\label{cap2:conclus}Conclusions}
In summary, the present (non--equilibrium) Lennard--Jones system, in which particles are subject to a constant driving field, has two main general features. On the one hand, this case is more convenient for computational purposes, than others such as, for instance, standard molecular--dynamics realizations of driven fluid systems. On the other hand, it seems to contain the necessary essential (microscopic) physics to be useful as a prototypical model for anisotropic behavior in nature.

This model is the natural extension to nonequilibrium anisotropic cases of the Lennard-Jones model which has played an important role in analysing equilibrium fluids. For zero field, it reduces to the familiar LJ fluid. Otherwise, it exhibits some arresting behavior, including currents and striped patterns. We have identified a process which seem to dominate early nucleation before anisotropic spinodal decomposition sets in. Interesting enough, they seem to be identical to the ones characterizing a similar situation in equilibrium. Surprisingly, we have also found that the model critical behavior is consistent with the Ising one for $d=2$, i.e., $\beta\approx 1/8$, but not with the critical behavior of the \textit{driven lattice gas}, i.e., $\beta\approx 1/3$. This is puzzling. Moreover, this result is quantitatively similar to the one in a related driven lattice system. Szolnoki and Szab\'o \cite{szabo} showed that assuming nearest-neighbor \textit{repulsion} under a driving field \cite{dickman} and extending the dynamics to next-nearest-neighbor hops, the order--disorder transition which the model undergoes becomes second order for all drive, and belongs to the Ising universality class. Furthermore, it is straightforward to see that if one allows particles' spatial coordinates to vary continuously then the model reduces to a \textit{hard-disks} gas under a driving field, and the second order phase transition vanishes. The transition becomes similar to the melting transition for systems in equilibrium, which is described by the KTHNY scenario \cite{melting}. The fact that in both nonequilibrium diffusive models ---which share similar dynamics--- the critical behavior is consistent with Ising is rather surprising. For instance, using bare arguments of statistical field theory, symmetries seem to bring our system closer to the nonequilibrium lattice model (the DLG) than to the corresponding equilibrium case (the LJ fluid). 
 
Therefore, a principal conclusion is that ---regarding the modeling of complex systems--- spatial discretization may change significantly not only morphological properties, but also critical properties. This is in contrast with the concept of universality in equilibrium systems, where critical properties are independent of dynamic details. The main reason for this disagreement might be the particle--hole symmetry violation in the driven \textit{Lennard--Jones} fluid. However, to determine exactly this statement will require further study. 

In any case, the additional freedom of the present, off--lattice system, which in particular implies that the particle--hole symmetry which occurs in lattice models is violated ---which induces the coexistence-curve asymmetry in Fig.~\ref{fig9_2} in accordance with actual systems--- are likely to matter more than suggested by some naive intuition. Indeed, the question of what are the most relevant ingredients and symmetries which determine unambiguously the universal properties in driven diffusive fluids is still open. Nevertheless, the above important difference between the lattice and the off--lattice cases results most interesting as an unquestionable nonequilibrium effect; as it is well known, such microscopic detail is irrelevant to universality concerning equilibrium critical phenomena.

%


Future research should address the development of a field-theoretical description aiming at shedding light on the critical behavior of these models. Off--lattice models pose a new challenge for any available Langevin-type approach in which to test their viability. In principle, the \textit{anisotropic diffusive system} approach described in the previous chapter appears to be more appropriate because it allows the consideration of the microscopic dynamical details. In fact, it is expected that a careful analysis in the derivation process from the master equation to the final Langevin equation should revail the mechanisms that lead to Ising behavior.

Further study of the present nonequilibrium LJ system and its possible variations is suggested. A principal issue to be investigated is the apparent fact that the full nonequilibrium situations of interest can be described by some rather straightforward extension of equilibrium theory. We here deal with some indications of this concerning early nucleation and properties of the coexistence curve. No doubt it would be interesting to compare more systematically the behavior of models against the varied phenomenology which was already reported for anisotropic fluids. This should also help a better understanding of nonequilibrium critical phenomena.\\

Finally, one of our main aims at proposing this model has been motivate further research, both experimental and theoretical: we try to motivate new experiments on anisotropic decomposition besides theoretical work. In fact, our observations on the early-time separation process and structural properties are easily accessible by micro calorimetric and spectroscopy experiments, for instance.

\part{\label{part:Granos}Driven Granular Gases}
\chapter{Introduction\label{intro:Granos}}

The second part of this thesis focuses on \textit{rapid granular flows}, also referred as \textit{granular gases} \cite{campbell,kadanoff,thorsten1,thorsten2,goldhirsch1}. They are granular systems fluidized by the application of an external driving, e.g., by mechanical vibration \cite{umbanhowar,olafsen}, shear \cite{shear}, electrostatically \cite{aranson}, or others means \cite{aransonRMP,li}. In these systems, collisions between particles are considered nearly instantaneous compared with the mean free time. In fact, it is assumed (by definition) that the flow is governed by binary collisions ---as in molecular gases---, whereas three--body (or more) collisions are neglected. Instead of moving in many--particle blocks, each particle moves freely and independently of even its nearest neighbors. Furthermore, granular gases exhibit many states and instabilities, e.g., convection \cite{convect1,convect2,khain1}, that characterize molecular gases. In spite of these similarities with classical molecular fluids, there are significant differences between both which make granular gases strikingly different. The origin of differences lies, as noted in the general introduction (Chapter~\ref{cap0}), in 
the inelastic character of collisions between grains, and in the irrelevance of the thermal energy $k_{B}T$. 

One important coarse-grained (continuum) approach to modeling granular gases is hydrodynamics. Granular hydrodynamics has a great predictive power and looks ideally suitable for a description of large--scale ---long-wavelength and long-time--- patterns observed in rapid granular flows as, for instance, a plethora of clustering phenomena \cite{goldhirsch2,olafsen,kudrolli,baruch3}, vortices \cite{jaeger,wildman,forterre}, oscillons \cite{umbanhowar,tsimring_oscillon_theory}, and shocks \cite{shocks}. This approach consists on a set of partial differential equations for the velocity $\mathbf{v}(\mathbf{r},t)$, temperature $T(\mathbf{r},t)$ (average fluctuation kinetic energy), pressure $p(\mathbf{r},t)$, and density (or number density $n(\mathbf{r},t)$) $\rho(\mathbf{r},t)$ fields. Hydrodynamic equations are analogous to the Navier Stokes equations for molecular fluids, but include a sink term in the energy equation which accounts for dissipation in collisions. For two-dimensional granular fluids, these read
\begin{equation}
\begin{split}
\frac{\partial n}{\partial t}+ \nabla \left( n \mathbf{v}\right)=0\qquad \text{mass conservation,}\\
n\left( \frac{\partial \mathbf{v}}{\partial t}+\left( \mathbf{v}\cdot \nabla \right) \mathbf{v}\right)=\nabla \cdot p+n\mathbf{g} \qquad \text{momentum conservation,}\\
n\left( \frac{\partial T}{\partial t}+\left( \mathbf{v}\cdot \nabla \right) T\right) =-\nabla \cdot \mathbf{w}+p:\nabla \mathbf{v}-I \qquad \text{energy conservation}.
\end{split}
\label{eq:hydrodyn}
\end{equation}
Here, $p$ is the stress tensor, $\mathbf{w}$ is the heat flux, $I$ is the rate of energy loss by collisions per unit area per unit time, and $\mathbf{g}$ is a external force acting on the particles. Under some assumptions, more fundamental \textit{kinetic theory} validates these equations and supplies constitutive relations\footnote{These are the explicit expressions of $p$, $\mathbf{w}$, and $I$ in terms of the basic fields $n$, $T$, and $\mathbf{v}$.} \cite{jenkins,kin,breycubero}. However, the exact criteria for the validity of (granular) kinetic theory and Navier--Stokes hydrodynamics \cite{jenkins,haff,tan} is still lacking. This is because first--principle derivations of universally applicable \textit{continuum} theory of granular gas is not a simple task, even in the dilute limit. Unfortunately, the drastic simplification, provided by the binary collisions assumption, is far from sufficient for a derivation of hydrodynamics, and additional hypothesis are needed. Assuming a hard-disks system, and that the random motion of even neighboring particles are uncorrelated (\textit{molecular chaos} assumption or \textit{stosszahlansatz}) allows for derivations from the Boltzmann or Chapman-Enskog kinetic equation, properly generalized to account for the inelasticity of the particle collisions. An important additional assumption, made in the process of the derivation of hydrodynamics from the Chapman--Enskog equation, is scale separation \cite{goldhirsch1}. Hydrodynamics demands (for inhomogeneous and/or unsteady) flows that the mean free path of the particles is much less than any characteristic length scale, and the mean free time between two consecutive collisions be much less than any characteristic time scale, aimed to be described hydrodynamically. All of these hypothesis are justified for moderate densities and for not too large inelasticity. These should be verified, in every specific system, after the hydrodynamic problem is solved and the characteristic length and time scales determined. In such case, systematic derivations of the constitutive relations ---the equation of state and transport coefficients--- of granular hydrodynamics from the Boltzmann or Enskog equation \cite{sela,brey1,lutsko} use an expansion in the Knudsen number and, in some versions of theory, in inelasticity \cite{sela}. 

A finite inelasticity immediately brings complications \cite{goldhirsch1}. Correlations between particles, developing already at a moderate inelasticity, may invalidate the molecular chaos hypothesis \cite{correlations}. The correlations become stronger as the inelasticity of the collisions increases. The normal stress difference and deviations of the particle velocity distribution from the Maxwell distribution also become important for inelastic collisions. As a result, the Navier-Stokes granular hydrodynamics should not be expected to be quantitatively accurate beyond the limit of small inelasticity. In any case, despite the severe limitations intrinsic to it, the nearly elastic case is conceptually important, just because hydrodynamics is supposed to give here a quantitatively accurate leading order theory. This is the regime (nearly elastic collisions) considered in the following Chapters~\ref{cap3} and \ref{cap4}.

A further set of difficulties for hydrodynamics arise at large densities. In mono-disperse systems strong correlations appear, already for \textit{elastic} hard spheres, at the disorder-order transition. As multiple thermodynamic phases may coexist there \cite{chaikin}, a general continuum theory of hard sphere fluids, which has yet to be developed, must include, in addition to the hydrodynamic fields, an order-parameter field. Furthermore, even on a specified branch of the thermodynamic phase diagram, we do not have first--principle constitutive relations. Last but not least, kinetic theory of, for instance, a finite--density gas of elastic hard spheres in two dimensions is plagued by non--existence of transport coefficients because of the long-time tail in the velocity pair autocorrelation function \cite{divergence,rojo}.

Another potentially important limitation of the validity of granular hydrodynamics (or, rather, of any continuum approach to rapid granular flows) is due to the noise caused by the discrete nature of particles (discrete particle noise). This may play a dramatic role in rapid granular flow \cite{barrat,swinney,baruch2}.

Noise is stronger here than in classical molecular fluids simply because the number of particles is much smaller. In addition, noise can be amplified at thresholds of hydrodynamic instabilities as found, for example, in Rayleigh--B\'enard convection of classical fluids \cite{RBconvection} or in granular gases \cite{baruch2}. A new challenge for theory is a quantitative account of this noise. A promising approach at small or moderate densities is ``Fluctuating Granular Hydrodynamics": a Langevin--type theory that takes into account the discrete particle noise by adding a delta--correlated (gaussian) noise terms in the momentum and energy equations, in the spirit of the Fluctuating Hydrodynamics by Landau and Lifshitz \cite{ll,brey3}. At high densities, however, such a theory is again beyond reach.\\

An alternative approach to granular flows, besides experiments, are simulations. Computer simulations have a valuable role in providing essentially exact results for certain problems which would otherwise be intractable. In addition, simulations help us to obtain the intuition we need in order to understand the complex behavior exhibited by granular material. Here (Chapters~\ref{cap3} and \ref{cap4}), we test the predictions from the hydrodynamic equations by comparing with \textit{Molecular Dynamics} simulations \cite{thorsten3,allen,rapaport}. Molecular Dynamics is a technique for computing the steady state and transport properties of a many--body system. This is in many respects very similar to real experiments. The general idea is simple: integrate numerically the microscopic equations of motion (Newton's equations\footnote{Typically, because of the macroscopic size of grains, quantum effects can be discarded with confidence.}) simultaneously for all particles ($1\leq i \leq N$)
\begin{equation}
m_i\;\frac{d^{2}\mathbf{r}_i}{dt^{2}}=\mathbf{F}_{i}(\mathbf{r}_{1},\cdots,\mathbf{r}_{N},\mathbf{v}_{1},\cdots,\mathbf{v}_{N})
\label{eq:gMD}
\end{equation}
This type of Molecular Dynamics is called \textit{force--based}. However, there are systems in which this scheme is very inefficient (although, of course, still applicable) \cite{thorsten3}. In systems where most of the time each of particles propagates along a ballistic trajectory, the typical duration of a collision is much shorter than the mean time between successive collisions of a particle, i.e., binary collisions can be assumed. This is, in fact, what occurs in rapid granular flows. In such a case, if the post--collisional velocities are obtained from the pre--collisional velocities a \textit{event--driven} Molecular Dynamics simulation can be set up\footnote{There are previous works in which event--driven algorithms has been even applied to dense granular systems \cite{baruch3}.}. As a result, event--driven algorithms are much more efficient, in the sense of simulation speed, than force--based algorithms (integrating directly Eq.~\ref{eq:gMD}). We will employ event-driven molecular dynamic simulations to test the validity of hydrodynamic prediction in the following chapters.\\

A standard prototypical model for driven granular gases, employed in many theories and simulations, as well as in experiments, is that of a monodisperse collection of uniform frictionless hard spheres (hard disks in two--dimensions) whose collisions are characterized by constant normal restitution coefficient $\mu_{\perp}$\footnote{The restitution coefficient typically depends weakly on the velocity, but using a constant simplifies the theoretical derivations greatly \cite{thorsten3}.}. Usually, this model is employed to study dense or diluted flows, with or without gravity, and with or without a interstitial fluid. Here the tangential degrees of freedom are neglected, i.e., $\mu_{\parallel}=1$, and only the normal restitution coefficient $\mu_{\perp}$ appears. However, it is worth mentioning that these two assumptions are ill-suited (see a nice discussion in \cite{thorsten3}). This simplification is owed exclusively to the huge theoretical complication which comes along with the incorporation of the correct (velocity dependent) tangential restitution coefficient. More complex models may involve nonspherical particles, in general, irregularly shaped particles, and are also polydisperse. In addition, except for astrophysical studies \cite{thorsten1} or experiments performed in vacuum \cite{umbanhowar} the grains are embedded in an ambient fluid. Nonetheless, the effect of a interstitial fluid, as air or water, can be sometimes neglected in determining many properties of the system \cite{campbell}. This is particularly true when grain--grain interactions are stronger than fluid or grain--fluid interactions. But as the level of detail increases so does the computational effort. In spite of its shortcomings as a practical tool, the hard--spheres prototype enables the elucidation of many essential features which are the responsible for the observed macroscopic behavior.\\

Regarding the mechanism which drives the particles, in Chapters~\ref{cap3} and \ref{cap4} we consider rapidly vibrating boundaries. We will assume that the characteristic frequency of vibration of the wall is much larger than the collision rate of the grains. This guarantees the absence of correlations between successive collisions of particles with the vibrating wall. In addition, it will be assumed that the amplitude of the vibration is much smaller than the mean free path of the particles, so collective motions in the system are avoided. Therefore, the vibrated wall can be represented as a hydrodynamic heat flux, i.e., as a ``\textit{thermal}" wall. From a computational point of view, the thermal wall is implemented as follows: whenever a particle collides with the wall, it forgets its velocity and chooses a new one from a Maxwell distribution according to a certain, previously prescribed temperature $T$. It is important to notice that in order to avoid that the gas approaches the incorrect final temperature, the random normal velocities have to be chosen according to the probability distribution
\begin{equation}
N(v_{n})=\frac{m}{T}v_{n} \mathit{exp}\left[ {-\frac{m v_{n}^{2}}{2T}}\right]  \,,
\label{eq:norma_distrib}
\end{equation}
and the tangential velocities have to be determined from 
\begin{equation}
N(v_{t})=\sqrt{\frac{m}{2\pi T}} \mathit{exp}\left[ {-\frac{m v_{t}^{2}}{2T}}\right]  \,.
\label{eq:tang_distrib}
\end{equation}
See \textit{e.g}. Ref.~\cite{thorsten3}, pages 173-177, for detail.\\

The first simulation of a system of particles colliding ineslatically and driven by a side themal wall was considered, in one dimension, by Du, Li, and Kadanoff \cite{li_kadanoff_one_dimension}. For typical initial conditions, the system in Ref.~\cite{li_kadanoff_one_dimension} evolves to a state where the particles are separated into two groups. Almost all particles form a low--temperature cluster in a small region, while a few remaining particles move with high velocities. Clearly, this steady state cannot be described by granular hydrodynamics. The results in Ref.~\cite{li_kadanoff_one_dimension} brought into question the validity of granular hydrodynamics in general. Nevertheless, some results \cite{esipov,grossman} have showed that this ``anomaly" seems to vanish in higher dimensions (our studies developed in Chapters~\ref{cap3} and \ref{cap4} concern two-dimensional systems). Therefore, the validity of hydrodynamics merits further studies.

\chapter{\label{cap3}Phase separation of a driven granular gas in annular geometry}

In this chapter we address phase separation of a monodisperse gas of inelastically
colliding hard disks confined in a two-dimensional annulus, the inner circle of
which represents a ``thermal wall". When described by granular hydrodynamic
equations, the basic steady state of this system is an azimuthally symmetric
state of increased particle density at the exterior circle of the annulus. When
the inelastic energy loss is sufficiently large, hydrodynamics predicts
spontaneous symmetry breaking of the annular state, analogous to the van der
Waals-like phase separation phenomenon previously found in a driven granular gas
in rectangular geometry. At a fixed aspect ratio of the annulus, the phase
separation involves a ``spinodal interval" of particle area fractions, where the
gas has negative compressibility in the azimuthal direction. The heat conduction
in the azimuthal direction tends to suppress the instability, as corroborated by
a marginal stability analysis of the basic steady state with respect to small
perturbations. To test and complement our theoretical predictions we performed
event-driven molecular dynamics simulations of this system. We clearly
identify the transition to phase separated states in the simulations, despite
large fluctuations present, by measuring the probability distribution of the
amplitude of the fundamental Fourier mode of the azimuthal spectrum of the
particle density. We find that the instability region, predicted from
hydrodynamics, is always located within the phase separation region observed in
the simulations. This implies the presence of a binodal (coexistence) region,
where the annular state is metastable.

\section{\label{cap3:intro}Introduction}
This chapter deal with a simple model of \textit{rapid} granular flows, also referred to as granular gases: large assemblies of inelastically colliding hard spheres 
\cite{campbell,kadanoff,thorsten1,thorsten2,goldhirsch1,brilliantov}. In the
simplest version of this model the only dissipative effect taken into account is
a reduction in the relative normal velocity of the two colliding particles,
modeled by the coefficient of normal restitution, see below. 

As we explained in the previous chapter, under some assumptions a hydrodynamic description of granular gases becomes
possible. In short, the molecular chaos assumption allows for a description in terms of the Boltzmann or Enskog equations, properly generalized to account for the inelasticity of particle collisions, followed by a systematic derivation of hydrodynamic equations \cite{sela,brey1,lutsko}, and for inhomogeneous flows hydrodynamics demands also scale separation. The implications of these conditions can be usually seen only \textit{a posteriori}, after the
hydrodynamic problem in question is solved, and the hydrodynamic length/time scales are determined.  

We will restrict ourselves in this chapter (and also in Chapter~\ref{cap4}) to nearly elastic collisions and moderate gas densities where, based on previous studies, hydrodynamics is expected to be an accurate leading order theory. Though
quite restrictive, these assumptions allow for a detailed quantitative study
(and, quite often, prediction) of a variety of pattern formation phenomena in
granular gases. One of these phenomena is the phase separation instability
\cite{khain1,baruch2,argentina,livne2,brey2,livne1,khain2} that was first
predicted from hydrodynamics and then observed in molecular dynamic simulations.
This instability arises already in a very simple, indeed prototypical setting: a
monodisperse granular gas at zero gravity confined in a rectangular box, one of
the walls of which is a ``thermal" wall. The basic state of this system is the
stripe state. In the hydrodynamic language it represents a laterally uniform
stripe of increased particle density  at the wall opposite to the driving wall.
The stripe state was observed in experiment \cite{kudrolli}, and this and
similar settings have served for testing the validity of quantitative modeling
\cite{li_kadanoff_one_dimension,esipov,grossman}. It turns out that (i) within a ``spinodal"
interval of area fractions and (ii) if the system is sufficiently wide in the
lateral direction, the stripe state is unstable with respect to small density
perturbations in the lateral direction \cite{khain1,brey2,livne1}. Within a
broader ``binodal" (or coexistence) interval the stripe state is stable to small
perturbations, but unstable to sufficiently large ones \cite{argentina,khain2}.
In both cases the stripe gives way, usually via a coarsening process, to
coexistence of dense and dilute regions of the granulate (granular ``droplets"
and ``bubbles") along the wall opposite to the driving wall
\cite{argentina,livne2,khain2}. This far-from-equilibrium phase separation
phenomenon is strikingly similar to a gas-liquid transition as described by the
classical van der Waals model, except for large fluctuations observed in a broad
region of aspect ratios around the instability threshold \cite{baruch2}. The
large fluctuations have not yet received a theoretical explanation.

This chapter addresses a phase separation process in a different geometry. We will
deal here with an assembly of hard disks at zero gravity, colliding
inelastically inside a two-dimensional annulus. The interior wall of the annulus
drives the granulate into a non-equilibrium steady state with a
(hydrodynamically) zero mean flow. Particle collisions with the exterior wall
are assumed elastic. The basic steady state of this system, as predicted by
hydrodynamics, is the \textit{annular} state: an azimuthally symmetric state of
increased particle density at the exterior wall. The phase separation
instability manifests itself here in the appearance of dense clusters with
broken azimuthal symmetry along the exterior wall. Our main objectives are to characterize the instability and compute the phase diagram by using granular hydrodynamics
(or, more precisely, granular hydrostatics, see below) and event driven
molecular dynamics simulations. By focusing on the annular geometry, we hope to
motivate experimental studies of the granular phase separation which may be
advantageous in this geometry. The annular setting avoids lateral side walls
(with an unnecessary/unaccounted for energy loss of the particles). Furthermore,
driving can be implemented here by a rapid rotation of the (slightly eccentric
and possibly rough)
interior circle\footnote{It is worth notice that the posed setting resembles morphology and dynamical processes
in planetary rings \cite{brahic}, where clustering, spontaneous symmetry
breaking, and oscillatory instabilities, between many others, also may occur.}.

We organized the chapter as follows. Section~\ref{cap3:hydro} deals with a hydrodynamic
description of the annular state of the gas, which is completely
described by three dimensionless parameters: the grain area fraction, the
inelastic heat loss parameter, and the aspect ratio. As we will be dealing only with
states with a zero mean flow, we will call the respective equations hydrostatic.

A marginal stability analysis predicts a spontaneous symmetry breaking of the
annular state. We compute the marginal stability curves and compare them to the
borders of the spinodal (negative compressibility) interval of the system. In
Section~\ref{cap3:sims} we report event-driven molecular dynamics simulations
of this system and compare the simulation results with the hydrostatic theory. In Section~\ref{cap3:conclus} we summarize the main results.

\section{\label{cap3:hydro}Particles in an annulus and granular hydrostatics}
\subsection{The density equation}
Let $N$ hard disks of diameter $d$ and mass $m=1$
move, at zero gravity, inside an annulus of aspect ratio $\Omega=
R_{\text{ext}}/R_{\text{int}}$, where $R_{\text{ext}}$ is the exterior radius
and $R_{\text{int}}$ is the interior one. The disks undergo inelastic collisions
with a constant coefficient of normal restitution
$\mu$. For simplicity, we neglect the rotational degree of freedom of the particles. In each
inter-particle collision the kinetic energy is continuously transferred into
heat, while the momentum is conserved.
The (driving) interior wall is modeled by a thermal wall kept at temperature
$T_{0}$, whereas particle collisions with the exterior wall are considered
elastic. The energy transferred from the thermal wall to the granulate
dissipates in the particle inelastic collisions, and we assume that the system
reaches a (non-equilibrium) steady state with a zero mean flow. We restrict
ourselves to the nearly elastic limit by assuming a restitution coefficient
close to, but less than, unity: $1-\mu \ll 1$. This allows us to safely use
granular hydrodynamics \cite{goldhirsch1}.  For a zero mean flow steady state
the continuity equation is obeyed trivially, while the momentum
and energy equations yield two \textit{hydrostatic} relations:
\begin{equation}
\nabla \cdot \mathbf{w} (\mathbf{r})+I=0\,, \; \; \; p=const\,,
\label{en_balance_prev}
\end{equation}
where $\mathbf{w}$ is the local heat flux, $I$ is the energy loss term due to
inelastic collisions, and $P=P(n,T)$ is the gas pressure that depends on the
number density $n(\mathbf{r})$ and granular temperature $T(\mathbf{r})$. Despite the ineslatic collisions, if the macroscopic field gradients are not very large, it seems natural to postulate linear relation between fluxes and ``thermodynamic" forces. Then, we
adopt the classical Fourier relation for the heat flux
$\mathbf{w}(\mathbf{r})=-\kappa \nabla T(\mathbf{r})$ (where $\kappa$ is the
thermal conductivity), omitting a density gradient term $\gamma\, \nabla n(\mathbf{r})$. In the dilute limit
this term was derived in Ref.~\cite{brey1}. It can be neglected in the nearly
elastic limit which is assumed throughout the second part of this thesis.

The momentum and energy balance equations read
\begin{equation}
\nabla \cdot \left[ \kappa \nabla T(\mathbf{r}) \right] =I\,, \; \; \;
p=const\,, \label{en_balance}
\end{equation}
To get a closed formulation, we need constitutive relations for $p(n,T)$,
$\kappa(n,T)$ and $I(n,T)$. We will employ the widely used semi-empiric
transport coefficients derived by Jenkins and Richman \cite{jenkins} from the Enskog kinetic equation, generalized to account for inelastic collision losses, for moderate densities:
\begin{equation}
\begin{split}
\kappa=\frac{2d n T^{1/2} \tilde{G}}{\pi^{1/2}}
\left[ 1+\frac{9\pi}{16} \left( 1+\frac{2}{3\tilde{G}}\right)^{2}\right],\\
I=\frac{8(1-\mu)n T^{3/2}\tilde{G}}{d\sqrt{\pi}}\,,
\end{split}
\label{JR}
\end{equation}
and the equation of state first proposed by Carnahan and Starling \cite{carnahan}
\begin{equation}
p=n T(1+2\tilde{G})\,,
\label{CS}
\end{equation}
where $\tilde{G}=\nu (1-\frac{7\nu}{16})/(1-\nu)^{2}$ and $\nu=n \left( \pi
d^{2}/4 \right) $ is the solid fraction. 
Let us rescale the radial coordinate by $R_{\text{int}}$ and introduce the
rescaled inverse density $Z(r,\theta)=n_{c}/n(r,\theta)$, where $n_{c}=2/\left(
\sqrt{3}d^2 \right)$ is the hexagonal close packing density. The rescaled radial
coordinate $r$ now changes between $1$ and $\Omega\equiv
R_{\text{ext}}/R_{\text{int}}$, the aspect ratio of the annulus. As in the
previous work \cite{khain1}, Eqs. \eqref{en_balance}, \eqref{CS} and \eqref{JR} can be
transformed into a single equation for the inverse density $Z(r)$:
\begin{equation}
\nabla \cdot \left[ \mathcal{F}(Z)\nabla Z\right] =\Lambda \mathcal{Q}(Z)\,,
\label{eq:density_eq}
\end{equation}%
where
\begin{equation}
\begin{split}
\mathcal{F}(Z)=\mathcal{F}_{1}(Z)\mathcal{F}_{2}(Z),\\
\mathcal{Q}(Z)=\frac{6}{\pi} \frac{Z^{1/2}\mathcal{G}}{(1+2\mathcal{G})^{3/2}},\\
\mathcal{F}_{1}(Z)=\frac{\mathcal{G}(Z)\left[ 1+{\frac{9\pi}{16}}
\left( 1+\frac{2}{3\mathcal{G}}\right)^2\right]}{Z^{1/2}(1+2\mathcal{G})^{5/2}},\\
\mathcal{F}_{2}(Z)=1+2\mathcal{G}+ \frac{\pi}{\sqrt{3}}\frac{Z\left(
Z+\frac{\pi}{16\sqrt{3}}\right) }
{\left( Z-\frac{\pi}{2\sqrt{3}}\right)^3},\\
\mathcal{G}(Z)=\frac{\pi}{2\sqrt{3}}\frac{\left(
Z-\frac{7\pi}{32\sqrt{3}}\right)} {\left( Z-\frac{\pi}{2\sqrt{3}}\right) ^2}.
\end{split}
\label{funct}
\end{equation}%
The dimensionless parameter $\Lambda \equiv (2\pi/3) (1-\mu)
\left(R_{\text{int}}/d\right)^2$ is the hydrodynamic inelastic loss parameter.
The boundary conditions for Eq.~\eqref{eq:density_eq} are
\begin{equation}
\partial Z(1,\theta)/\partial\theta=0 \quad \text{and} \quad \nabla_{n} Z(\Omega,\theta)=0\,,
\label{eq:boundary}
\end{equation}
The first of these follows from the constancy of the temperature at the
(thermal) interior wall which, in view of the constancy of the pressure in a
steady state, becomes constancy of the density. The second condition demands a
zero normal component of the heat flux at the elastic wall.
Notice that neither the pressure nor the
temperature at the thermal wall enter in
Eqs.~\eqref{eq:density_eq}-\eqref{eq:boundary} \cite{livne1,khain1}. This is a
consequence of the fact that the hard-core interactions between particles does
not introduce a characteristic energy scale. The temperature only sets the
temperature scale in the system and affects the pressure, which is constant in
the steady state problem.
Finally, working with a fixed number of particles, we demand the normalization
condition
\begin{equation}
\int_{0}^{2\pi} \int_{1}^{\Omega}  Z^{-1} (r,\theta) r dr d\theta  = \pi f
(\Omega^2-1)\,, \label{eq:norma}
\end{equation}
where $$f=\frac{N}{\pi n_{c}R_{\text{int}}^{2}(\Omega^2-1)}$$ is the area
fraction of the grains in the annulus. Equations
(\ref{eq:density_eq})-(\ref{eq:norma}) determine all possible steady state
density profiles, governed by three dimensionless parameters: $f$, $\Lambda$,
and $\Omega$.

\begin{figure}
\centerline{\includegraphics[width=8cm,clip=]{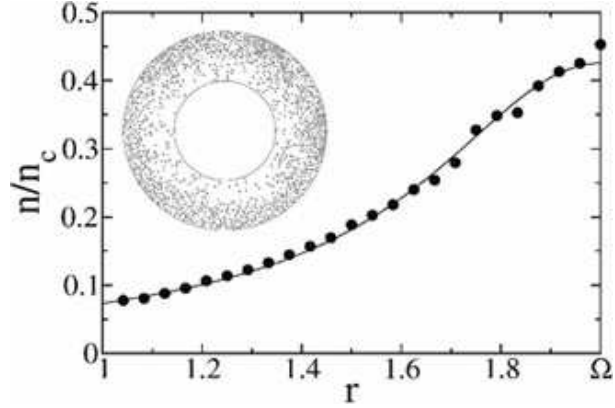}}
\caption[The normalized density profiles obtained from simulations and hydrostatics]{\label{fig1_3} The normalized density profiles obtained from the molecular dynamics
simulations (the dots) and hydrostatics (the line) for $\Omega=2$,
$\Lambda=81.09$, and $f=0.356$ (equivalently, $z_{\Omega}=2.351$). The
simulations were carried out with $N=1250$ particles, $\mu=0.92$, and
$R_{\text{int}}=22.0$. Also shown is a typical snapshot of the system at the
steady state as observed in the simulation.}
\end{figure}

\subsection{Annular state}
The simplest solution of the density equation
(\ref{eq:density_eq}) is azimuthally symmetric ($\theta$-independent): $Z=
z(r)$. Henceforth we refer to this basic state of the system as the
\textit{annular state}. It is determined by the following equations:
\begin{equation}
\begin{split}
\left[ r \mathcal{F}(z) z^{\prime} \right]^{\prime} =
r \Lambda \mathcal{Q}(z),\; z^{\prime}(\Omega)=0, \; \text{and} \\
\int_{1}^{\Omega} z^{-1} r dr=(\Omega^{2}-1)f/2 \,,
\end{split}
\label{eq:annular}
\end{equation}
where the primes denote $r$-derivatives. In order to solve the second order
equation (\ref{eq:annular}) numerically, one can prescribe the inverse density
at the elastic wall $z_{\Omega}\equiv z(\Omega)$. Combined with the no-flux condition at
$r=\Omega$, this condition define a Cauchy problem for $z(r)$
\cite{livne2,khain1}. Solving the Cauchy problem, one can compute the respective
value of $f$ from the normalization condition in Eq.~\eqref{eq:annular}. At
fixed $\Lambda$ and $\Omega$, 
there is a one-to-one relation between $z_{\Omega}$ and $f$. Therefore, an alternative parameterization of the
annular state is given by the scaled numbers $z_{\Omega}$, $\Lambda$, and
$\Omega$. The same is true for the marginal stability analysis performed in the
next subsection.

Figure~\ref{fig1_3} depicts an example of annular state that we found numerically.
One can see that the gas density increases with the radial coordinate\footnote{See Refs.~\cite{macnamara,goldhirsch2,livne2,hopkins} for clustering instability.}, as
expected from the temperature decrease via inelastic losses (the velocities of the particles decrease), combined with the constancy of the (hydrodynamic) pressure throughout the system. The hydrodynamic density
profile agrees well with the one found in our simulations, see below.

\begin{figure}
\centerline{\includegraphics[width=8cm,clip=]{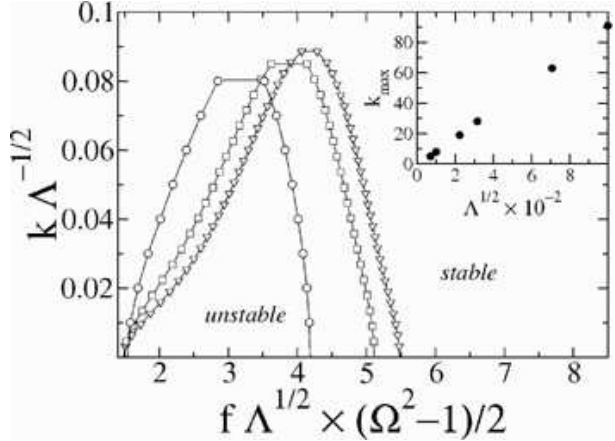}}
\caption[Main Graph: The marginal stability curves $k=k(f)$ at fixed aspect ratios $\Omega$ for different values of the heat loss parameter $\Lambda$. Inset: The dependence of $k_{max}$ on $\Lambda^{1/2}$]{\label{fig2_3}  The main graph: the marginal stability curves $k=k(f)$
(where $k$ is an integer) for $\Omega=1.5$ and $\Lambda=10^4$ (circles),
$\Lambda=5\times 10^4$ (squares), and $\Lambda=10^5$ (triangles). For a given
$\Lambda$ the annular state is stable above the respective curve and unstable
below it, as indicated for $\Lambda=10^4$. As $\Lambda$ increases the
marginal stability interval shrinks. The lines connecting dots serve to guide the eye.
The inset: the dependence of $k_{max}$ on $\Lambda^{1/2}$.  The straight line
shows that, at large $\Lambda$, $k_{max}\propto \Lambda^{1/2}$.}
\end{figure}

\subsection{Phase separation}
Mathematically, phase separation manifests itself in
the existence of \textit{additional} solutions to
Eqs.~\eqref{eq:density_eq}-\eqref{eq:norma} in some region of the parameter
space $f$, $\Lambda$, and $\Omega$. These additional solutions are \textit{not}
azimuthally symmetric\footnote{As we shall show in Section~\ref{cap3:sims}, such
instability also is observed in simulations.}. 
Solving Eqs.~\eqref{eq:density_eq}-\eqref{eq:norma} for  fully two-dimensional
solutions is not easy \cite{livne1}. One class of such solutions, however,
bifurcate continuously from the annular state, so they can be found by
linearizing Eq.~\eqref{eq:density_eq}, as in rectangular geometry
\cite{livne1,khain1}. In the framework of a time-dependent hydrodynamic
formulation, this analysis corresponds to a \textit{marginal stability} analysis
which involves a small perturbation to the annular state. For a single azimuthal
mode $\sim \sin(k\theta)$ (where $k$ is integer) we can write
$Z(r,\theta)=z(r)+\varepsilon\; \Xi(r) \sin(k\theta)$, where $\Xi(r)$ is a
smooth function, and $\varepsilon \ll 1$ a small parameter. Substituting this
into Eq.~\eqref{eq:density_eq} and linearizing the resulting equation yields a
$k$-dependent second order differential equation for the function
$\Gamma(r)\equiv\mathcal{F}[Z(r)]\,\Xi(r)$:
\begin{equation}
\Gamma^{\prime\prime}_k+\frac{1}{r}\Gamma^{\prime}_k- \left(
\frac{k^{2}}{r^{2}}+\frac{\Lambda \mathcal{Q}^{\prime}(Z)}
{\mathcal{F}(Z)}\right) \Gamma_{k}=0\,. \label{eq:MSA}
\end{equation}%
This equation is complemented by the boundary conditions
\begin{equation}
\Gamma(1)=0  \;\;\; \mbox{and} \;\;\; \Gamma^{\prime}(\Omega)=0\,.
\label{eq:MSA_BC}
\end{equation}%
For fixed values of the scaled parameters $f$, $\Lambda$, and $\Omega$,
Eqs.~\eqref{eq:MSA} and \eqref{eq:MSA_BC} determine a linear eigenvalue problem
for $k$. Solving this eigenvalue problem numerically, one obtains the marginal
stability hypersurface  $k=k(f,\Lambda,\Omega)$. For fixed $\Lambda$ and
$\Omega$, we obtain a marginal stability curve $k=k(f)$. Examples of such
curves, for a fixed $\Omega$ and three different $\Lambda$ are shown in
Fig.~\ref{fig2_3}. Each $k=k(f)$ curve has a maximum $k_{max}$, so that a density
modulation with the azimuthal wavenumber larger than $k_{max}$ is stable. As
expected, the instability interval is the largest for the fundamental mode
$k=1$. The inset in Fig.~\ref{fig2_3} shows the dependence of $k_{max}$ on
$\Lambda^{1/2}$.  The straight line shows that, at large $\Lambda$,
$k_{max}\propto \Lambda^{1/2}$, as in rectangular geometry \cite{khain1}.

\begin{figure}
\centerline{\includegraphics[width=8cm,clip=]{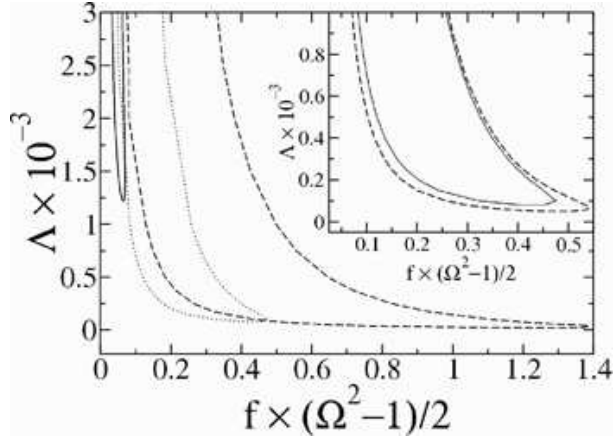}}
\caption[Two-dimensional projections of the phase diagram: Marginal stability and negative compressibility curves]{\label{fig3_3} Two-dimensional projections on the $\Lambda$-$f$ plane of
the phase diagram at $\Omega=1.5$ (the solid line), $\Omega=3$ (the dotted
line), and $\Omega=5$ (the dashed line). The inset shows more clearly, for
$\Omega=3$, that the marginal stability curve (the solid line) lies within the
negative compressibility region (depicted by the dashed line).}
\end{figure}

Two-dimensional projections of the ($f$, $\Lambda$, $\Omega$)-phase diagram at
three different $\Omega$ are shown in Fig.~\ref{fig3_3} for the fundamental mode.
The annular state is unstable in the region bounded by the marginal stability
curve and stable elsewhere. Therefore, the marginal stability analysis predicts
loss of stability of the annular state within a finite interval of $f$, that is
at $f_{min}(\Lambda,\Omega)<f<f_{max}(\Lambda,\Omega)$.

The physical mechanism of this phase separation \textit{instability} is the
negative compressibility of the granular gas in the azimuthal direction, caused
by the inelastic energy loss. To clarify this point, let us compute the
\textit{pressure} of the annular state, given by Eq.~\eqref{CS}. First we
introduce a rescaled pressure $P=p/(n_{c}T_{0})$ and, in view of the pressure
constancy in the annular state, compute it at the thermal wall, where $T=T_{0}$
is prescribed and $z(1)$ is known from our numerical solution for the annular
state. We obtain
$$
P(f,\Lambda,\Omega)=\frac{1+2\mathcal{G}(z(1))}{z(1)}\,.
$$
The spinodal (negative compressibility) region is determined by the necessary
condition for the \textit{instability}: $\left(
\partial P/\partial f\right)_{\Lambda,\Omega}<0$, whereas the borders of the
spinodal region are defined by $\left( \partial P/\partial
f\right)_{\Lambda,\Omega}=0$. That is, computing the derivative we arrive at $\partial P/\partial f=
(\partial P/\partial z(1))( \partial z(1)/\partial z(\Omega))
( \partial z(\Omega)/\partial f)$. One can easily check that the first
and third multipliers on the righthand side of this relation are
always negative, thus the sign of $\left( \partial P/\partial f \right)_{\Lambda,\Omega}$ is
determined by the sign of $\left( \partial z(1)/\partial z(\Omega) \right)_{\Lambda,\Omega}$.

\begin{figure}
\centerline{\includegraphics[width=8cm,clip=]{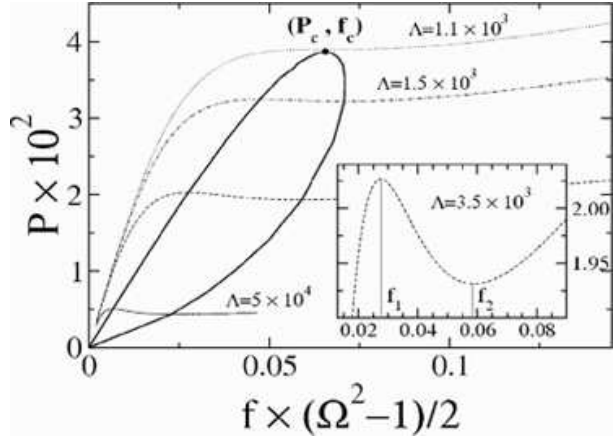}}
\caption[The scaled steady state granular pressure $P$ versus the
grain area fraction $f$ at fixed $\Omega$, and the spinodal interval]{\label{fig4_3} The scaled steady state granular pressure $P$ versus the
grain area fraction $f$ for $\Omega=1.5$ and $\Lambda=1.1\times 10^{3}$ (the
dotted line), $\Lambda=1.5\times 10^{3}$ (the dash-dotted line),
$\Lambda=3.5\times 10^{3}$ (dashed line), and $\Lambda=5\times 10^{4}$ (the
solid line). The inset shows a zoom-in for $\Lambda=3.5\times 10^{3}$. The
borders $f_1$ and $f_2$ of the spinodal interval are determined from the
condition $\left(
\partial P/\partial f \right)_{\Lambda,\Omega}=0$. The thick solid line
encloses the spinodal balloon where the effective azimuthal compressibility of
the gas is negative.}
\end{figure}

Typical $P(f)$ curves for a fixed $\Omega$ and
several different $\Lambda$ are shown in Fig.~\ref{fig4_3}.
One can see that, 
at sufficiently large $\Lambda$, the rescaled pressure $P$ goes down with an
increase of $f$ at an interval $f_{1}<f<f_{2}$. That is, the effective
compressibility of the gas with respect to a redistribution of the material in
the azimuthal direction is negative on this interval of area fractions. By
joining the spinodal points $f_1$ and (separately) $f_2$ at different $\Lambda$,
we can draw the spinodal line for a fixed $\Omega$. As $\Lambda$ goes down, the
spinodal interval shrinks and eventually becomes a point at a critical point
$(P_{c},f_c)$, or $(\Lambda_{c},f_c)$ (where all the critical values are
$\Omega$-dependent). For $\Lambda<\Lambda_c$ $P(f)$ monotonically increases and
there is no instability.

What is the relation between the spinodal interval $(f_{1},f_{2})$ and the
marginal stability interval $(f_{min},f_{max})$? These intervals would coincide
were the azimuthal  wavelength of the perturbation infinite (or, equivalently,
$k\rightarrow 0$), so that the azimuthal heat conduction would vanish. Of
course, this is not possible in annular geometry, where $k\ge 1$. As a result,
the negative compressibility interval must include in itself the marginal
stability interval $(f_{min},f_{max})$. This is what our calculations indeed
show, see the inset of Fig.~\ref{fig3_3}. That is, a negative compressibility is
necessary, but not sufficient, for instability, similarly
to what was found in rectangular geometry \cite{khain1}. 

Importantly, the instability region of the parameter space is by no means not
the \textit{whole} region the region where phase separation is expected to
occur. Indeed, in analogy to what happens in rectangular geometry
\cite{argentina,khain2}, phase separation is also expected in a \textit{binodal}
(or coexistence) region of the area fractions, where the annular state is stable
to small perturbations, but unstable to sufficiently large ones. The whole
region of phase separation should be larger than the instability region, and it
should of course \textit{include} the instability region. Though we did not
attempt to determine the binodal region of the system from the hydrostatic
equations (this task has not been accomplished yet even for rectangular
geometry, except in the close vicinity of the critical point \cite{khain2}), we
determined the binodal region from our molecular dynamics simulations reported in the next
section.

\section{\label{cap3:sims}Molecular Dynamics Simulations}
\subsection{Method}
We performed a series of event-driven molecular dynamics simulations of this
system using an algorithm described by P\"{o}schel and Schwager
\cite{thorsten3} and sketched in the previous chapter. Simulations involved $N$ hard disks of diameter $d=1$ and mass
$m=1$. After each collision of particle $i$ with particle $j$, their relative
velocity is updated according to
\begin{equation}
\vec{v}_{ij}^{\,\prime}=\vec{v}_{ij} - \left( 1+\mu \right) \left(
\vec{v}_{ij}\cdot \hat{r}_{ij}\right)\hat{r}_{ij}\,, \label{eq:velocs}
\end{equation}
where $\vec{v}_{ij}$ is the precollisional relative velocity, and
$\hat{r}_{ij}\equiv \vec{r}_{ij}/\left|\vec{r}_{ij}\right|$ is a unit vector
connecting the centers of the two particles. Particle collisions with the
exterior wall $r=R_{\text{ext}}$ are assumed elastic. The interior wall is kept
at constant temperature $T_{0}$ that we set to unity. This is implemented as
explained in Chapter~\ref{intro:Granos}. When a particle collides with the wall it forgets its velocity and
picks up a new one from a proper Maxwellian distribution with temperature
$T_{0}$.
The time scale is therefore $d(m/T_{0})^{1/2}=1$. The initial condition is a
uniform distribution of non-overlapping particles inside the annular box. Their
initial velocities are taken randomly from a Maxwellian distribution at
temperature $T_{0}=1$. In all simulations the coefficient of normal restitution
$\mu=0.92$ and the interior radius $R_{\text{int}}/d=22.0$ were fixed, whereas
the the number of particles $527\leq N\leq 7800$ and the aspect ratio
$1.5\leq\Omega\leq 6$ were varied. In terms of the three scaled hydrodynamic
parameters the heat loss parameter $\Lambda=81.09$ was fixed whereas $f$ and
$\Omega$ varied.

To compare the simulation results with predictions of our hydrostatic theory,
all the measurements were performed once the system reached a steady state. This
was monitored by the evolution of the total kinetic energy $(1/2)
\sum_{i=1}^{N}\vec{v}_{i}^{\;2}$, which first decays and then, on the
average, stays constant.

\begin{figure}
\centerline{\includegraphics[width=11cm,clip=]{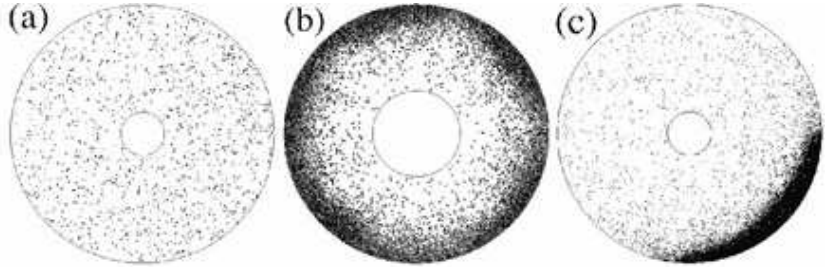}}
\caption[Typical steady state snapshots]{\label{fig5_3} Typical steady state snapshots for $N=1250$ and $\Omega=6$
(a); $N=5267$ and $\Omega=3$ (b), and $N=6320$ and $\Omega=6$ (c). Panels (a)
and (b) correspond to annular states of the hydrostatic theory, whereas panel
(c) shows a broken-symmetry (phase separated) state.}
\end{figure}

\subsection{Steady States}
Typical steady state snapshots of the system, observed
in our molecular dynamics simulation, are displayed in Fig.~\ref{fig5_3}. Panel (a) shows a dilute
state where the radial density inhomogeneity, though actually present, is not
visible by naked eye. Panels (b) and (c) do exhibit a pronounced radial density
inhomogeneity. Apart from visible density fluctuations, panels (a) and (b)
correspond to annular states. Panel (c) depicts a broken-symmetry (phase
separated) state. When an annular state is observed, its density profile agrees
well with the solution of the hydrostatic equations
(\ref{eq:density_eq})-(\ref{eq:norma}). A typical
example of such a comparison is shown in Fig.~\ref{fig1_3}. 

\begin{figure}
\centerline{\includegraphics[width=10cm,clip=]{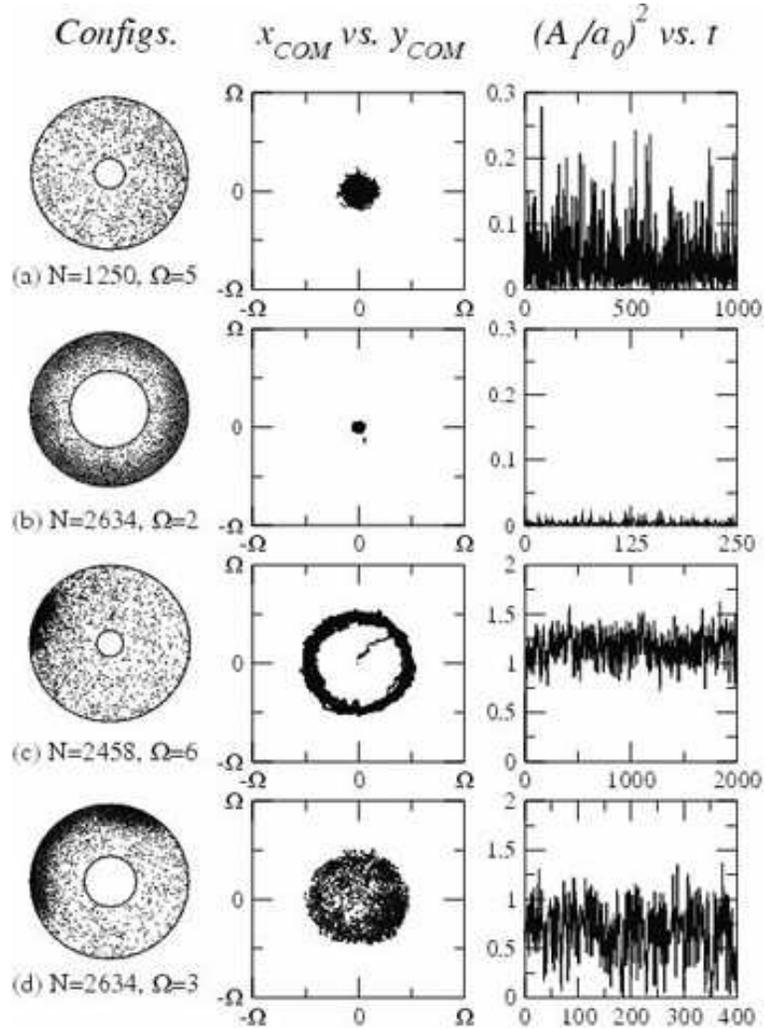}}
\caption[Typical steady state snapshots and the temporal evolution of the COM and of the squared amplitude
of the fundamental Fourier mode]{\label{fig6_3} Typical steady state snapshots (the left column; particle size looks larger than it is) and the
temporal evolution of the COM (the middle column) and of the squared amplitude
of the fundamental Fourier mode (the right column). The temporal data are
sampled each $150$ collisions per particle. Each row corresponds to one
simulation with the indicated parameters. The vertical scale of panels (a) and (b)
was stretched for clarity.}
\end{figure}

Let us fix the aspect ratio $\Omega$ of the annulus at not too a small value and
vary the number of particles $N$. First, what happens on a qualitative level?
The simulations show that, at small $N$, dilute annular states, similar to
snapshot (a) in Fig.~\ref{fig5_3}, are observed. As $N$ increases,
broken-symmetric states start to appear. Well within the unstable region, found
from hydrodynamics, a high density cluster appears, like the one shown in
Fig.~\ref{fig5_3}c, and performs an erratic motion along the exterior wall. As $N$
is increased still further, well beyond the high-$f$ branch of the unstable
region, an \textit{annular state} reappears, as in Fig.~\ref{fig5_3}b. This time,
however, the annular state is denser, while its local structure varies from a
solid-like (with imperfections such as voids and line defects) to a liquid-like.

To characterize the spatio-temporal behavior of the granulate at a steady state,
we followed the position of the center of mass (COM) of the granulate. Several
examples of the COM trajectories are displayed in Fig.~\ref{fig6_3}. Here cases (a)
and (b) correspond, in the hydrodynamic language, to annular  states. There are,
however, significant fluctuations of the COM around the center of the annulus.
These fluctuations are of course not accounted for by hydrodynamic theory. In
case (b), where the dense cluster develops, the fluctuations are much weaker that
in case (a). More interesting are cases (c) and (d). They correspond to
broken-symmetry states: well within the phase separation region of the parameter
space (case c) and close to the phase separation border (case d). The COM
trajectory in case (c) shows that the granular ``droplet" performs random motion
in the azimuthal direction, staying close to the exterior wall. This is in
contrast with case (d), where fluctuations are strong both in the azimuthal and in
the radial directions. Following the actual snapshots of the simulation, one
observes here a very complicated motion of the ``droplet", as well as its
dissolution into more ``droplets", mergers of the droplets \textit{etc}. 
Therefore, as in the case of granular phase separation in rectangular geometry
\cite{baruch2}, the granular phase separation in annular geometry is accompanied
by considerable spatio-temporal fluctuations. In this situation a clear
distinction between a phase-separated state and an annular state, and a
comparison between the simulations and hydrodynamic theory, demand proper
diagnostics. We found that such diagnostics are provided by the azimuthal
spectrum of the particle density and its probability distribution.

\begin{figure}
\centerline{\includegraphics[width=10cm,clip=]{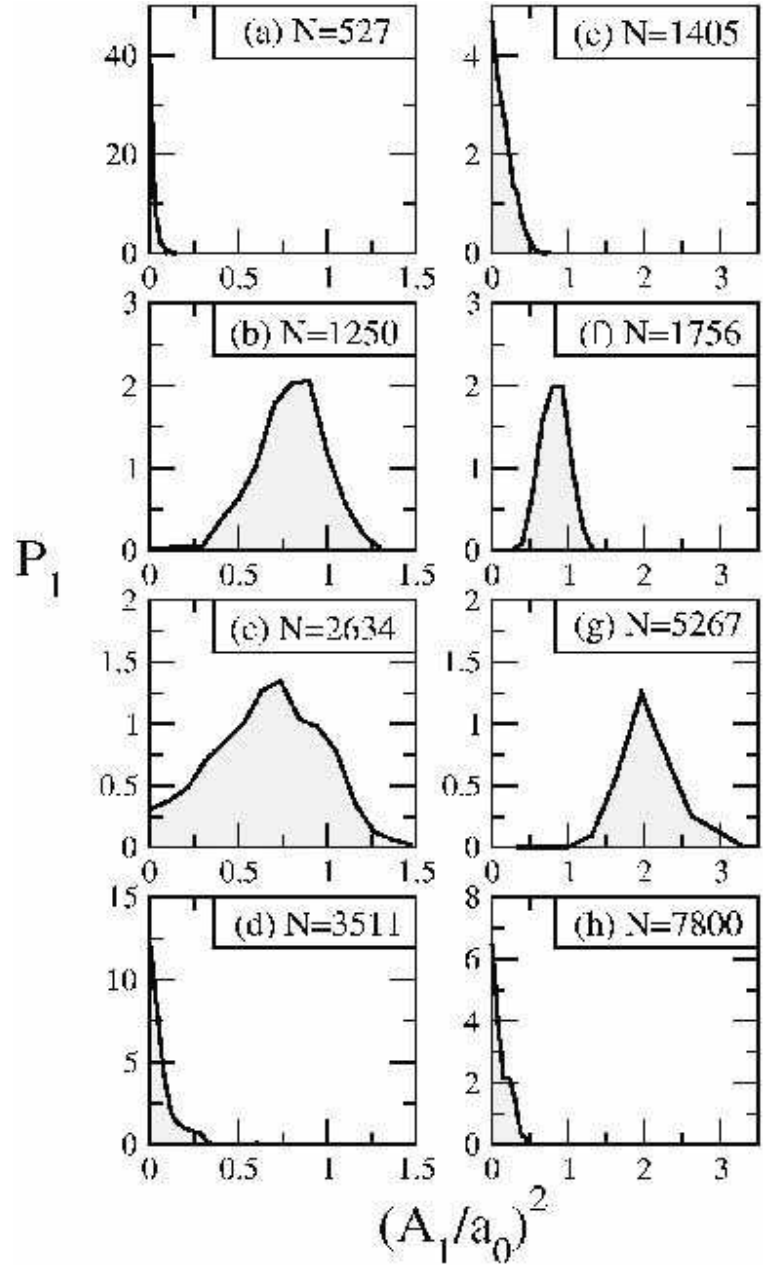}}
\caption[The normalized probability distribution functions at a different number of particles]{\label{fig7_3} The normalized probability distribution functions
$P_{1}\left(A_{1}^{2}/a_{0}^{2} \right)$ for $\Omega=3$ (the left column) and
$\Omega=5$ (the right column) at a different number of particles.} 
\end{figure}

\subsection{Azimuthal Density Spectrum}
Let us consider the (time-dependent)
rescaled density field $\nu(r,\theta,t)=n(r,\theta,t)/n_c $ (where $r$ is
rescaled to the interior wall radius as before), and introduce the integrated
field $\hat{\nu}(\theta, t)$:
\begin{equation}
\hat{\nu}(\theta,t) = \int_{1}^{\Omega}\,\nu(r,\theta,t)\, r \, dr \,.
\end{equation}
In a system of $N$ particles, $\hat{\nu}(\theta,t)$ is normalized so that
\begin{equation}\label{nunorm}
    \int_{0}^{2 \pi} \hat{\nu} (\theta,t)\, d\theta=\frac{N}{n_c\,R^2_{\text{int}}}\,.
\end{equation}
Because of the periodicity in $\theta$ the function $\hat{\nu}(\theta,t)$ can be
expanded in a Fourier series:
\begin{equation}
\hat{\nu}(\theta,t) = a_0+\sum_{k=1}^{\infty} \left[ a_k (t) \cos (k\theta) +
b_k(t) \sin (k\theta)\right] \,,
\end{equation}
where $a_0$ is independent of time because of the normalization condition
(\ref{nunorm}). We will work with the quantities
\begin{equation}
A_k^{2}(t)\equiv a_k^{2}(t) + b_k^{2}(t) \,, \;\;\;k\ge1\,. \label{eq:spectral}
\end{equation}
For the (deterministic) annular state one has $A_k=0$ for all $k\ge 1$, while
for a symmetry-broken state $A_k>0$. The relative quantities $A_k^2(t)/a_{0}^2$
can serve as measures of the azimuthal symmetry breaking. As is shown in
Table~\ref{tab:Ak}, $A_1^2(t)$ is usually much larger (on the average) that the
rest of $A_{k}^{2}(t)$. Therefore, the quantity $A_1^2(t)/a_{0}^2$ is sufficient
for our purposes.

\begin{table}
\caption[The averaged squared relative amplitudes $\langle A_{k}^2(t)\rangle/a_{0}^2$ for the first three modes $k=1,2$ and $3$]{\label{tab:Ak}The averaged squared relative amplitudes $\langle
A_{k}^2(t)\rangle/a_{0}^2$ for the first three modes $k=1,2$ and $3$. (a)
$N=2634$, $\Omega=3$, (b) $N=5267$, $\Omega=4$, (c) $N=1000$, $\Omega=2.25$, and
(d) $N=1250$, $\Omega=3$.}
\centerline{
\begin{tabular}{l|l|l|l|l}
$k$&(a)&(b)&(c)&(d)\\
\hline\hline
$1$ & $0.66\pm 0.05$ & $0.39\pm 0.04$ & $0.30\pm 0.08$ & $0.77\pm 0.05$\\
$2$ & $0.04\pm 0.02$ & $0.05\pm 0.02$ & $0.07\pm 0.01$ & $0.28\pm 0.09$\\
$3$ & $0.03\pm 0.02$ & $0.03\pm 0.03$ & $0.02\pm 0.02$ & $0.11\pm 0.08$\\
\end{tabular}}
\end{table}

Once the system relaxed to a steady state, we followed the temporal evolution of
the quantity $A_{1}^{2}/a_0^2$. Typical results are shown in the right column of
Fig.~\ref{fig6_3}. One observes that, for annular states, this quantity is usually
small, as is the cases (a) and (b) in Fig.~\ref{fig6_3}. For broken-symmetry states
$A_{1}^{2}$ is larger, and it increases as one moves deeper into the phase
separation region. (Notice that the averaged value of $A_{1}^{2}/a_0^2$ in (c) is
larger than in (d), which means that (c) is deeper in the phase separation
region.) Another characteristics of $A_{1}^{2}(t)/a_0^2$ is the magnitude of
fluctuations. One can notice that, in the vicinity of the phase separation
border the fluctuations are stronger (as in case (d) in Fig.~\ref{fig6_3}).

All these properties are encoded in the \textit{probability distribution}
$P_{1}$ of the values of $\left(A_{1}/a_{0}\right)^{2}$: the ultimate tool of
our diagnostics. Figure~\ref{fig7_3} shows two series of measurements of this
quantity at different $N$: for $\Omega=3$ and $\Omega=5$. By following the
position of the maximum of $P_{1}$ we were able to to sharply discriminate
between the annular states and phase separated states and therefore to locate
the phase separation border. When the maximum of $P_1$ occurs at the zero value
of $\left(A_{1}/a_{0}\right)^{2}$ (as in cases (a) and (d) and, respectively, (e) and
(h) in Fig.~\ref{fig7_3}), an annular state is observed. On the contrary, when the
maximum of $P_1$ occurs at a non-zero value of $\left(A_{1}/a_{0}\right)^{2}$
(as in cases (b) and (c) and, respectively, (f) and (g) in Fig.~\ref{fig7_3}), a phase
separated state is observed. In each case, the width of the probability
distribution (measured, for example, at the half-maximum) yields a direct
measure of the magnitude of fluctuations. Near the phase separation border,
strong fluctuations (that is, broader distributions) are observed, as in case (c)
of Fig.~\ref{fig7_3}.

\begin{figure}
\centerline{\includegraphics[width=8cm,clip=]{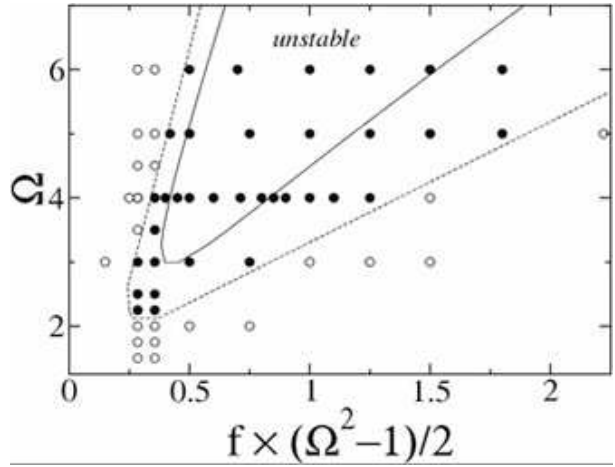}}
\caption[The $\Omega$-$f$ phase diagram for $\Lambda=81.09$ obtained from molecular dynamics simulations and hydrodynamics]{\label{fig8_3} The $\Omega$-$f$ phase diagram for $\Lambda=81.09$. The
solid curve is given by the granular hydrostatics: it shows the borders of the
region where the annular state is unstable with respect to small perturbations.
The filled symbols depict the parameters in which phase separated states are
observed, whereas the hollow symbols show the parameters at which annular states
are observed. The dashed line is an estimated binodal line of the system.}
\end{figure}

Using the position of the maximum of $P_1$ as a criterion for phase separation,
we show, in Fig.~\ref{fig8_3}, the $\Omega-f$ diagram obtained from the molecular dynamics
simulations. The same figure also depicts the hydrostatic prediction of the
\textit{instability} region. One can see that the instability region is located
within the phase separation region, as expected.

\section{\label{cap3:conclus}Conclusions}
We combined equations of granular hydrostatics and event-driven molecular dynamics simulations
to investigate spontaneous phase separation of a monodisperse gas of
inelastically colliding hard disks in a two-dimensional annulus, the inner
circle of which serves as a ``thermal wall". A marginal stability analysis
yields a region of the parameter space where the annular state ---the basic,
azimuthally symmetric steady state of the system--- is unstable with respect to
small perturbations which break the azimuthal symmetry. The physics of the
instability is negative effective compressibility of the gas in the azimuthal
direction, which results from the inelastic energy loss. Simulations of this
system show phase separation, but it is masked by large spatio-temporal
fluctuations. By measuring the probability distribution of the amplitude of the
fundamental Fourier mode of the azimuthal spectrum of the particle density we
were able to clearly identify the transition to phase separated states in the simulations. We found that the instability region of the parameter space,
predicted from hydrostatics, is located within the phase separation region
observed in the molecular dynamics simulations. This implies the presence of a binodal
(coexistence) region, where the annular state is \textit{metastable}, similar to
what has been found in rectangular geometry \cite{argentina,khain2}. We hope our
results will stimulate experimental work on the phase separation instability.

\begin{figure}
\centerline{\includegraphics[width=8cm,clip=]{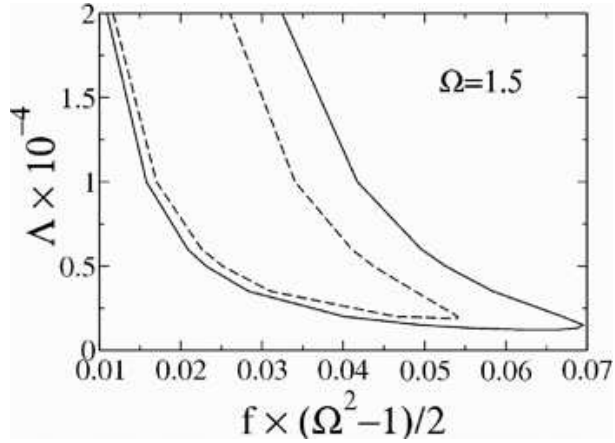}}
\caption[The marginal stability lines for our main setting and for an alternative setting]{\label{fig9_3} The marginal stability lines for our main setting (the
solid line) and for an alternative setting in which the thermal wall is at
$r=R_{\textit{ext}}$ and the elastic wall is at $r=R_{\textit{int}}$ (the dashed
line).}
\end{figure}

Finally, we have also investigated an alternative setting in which the exterior
wall is the driving wall, while the interior wall is elastic. The corresponding
hydrostatic problem is determined by the same three scaled parameters $f$,
$\Lambda$ and $\Omega$, but the boundary conditions must be changed accordingly.
Here azimuthally symmetric clusters appear near the (elastic) interior wall.
Symmetry breaking instability occurs here as well. We found very similar
marginal stability curves here, but they are narrower (as shown in
Fig.~\ref{fig9_3}) than those obtained for our main setting. 


\chapter[Close--packed granular clusters]{Close--packed granular clusters}
\label{cap4}

Dense granular clusters often behave like macro-particles. We address this interesting phenomenon in a model system of inelastically colliding hard disks inside a circular box, driven by a thermal wall at zero gravity. Molecular dynamics simulations show a close-packed cluster of an almost circular shape, weakly fluctuating in space and isolated from the driving wall by a low-density gas. The density profile of the system agrees well with the azimuthally symmetric solution of granular hydrostatic equations employing constitutive relations by Grossman \textit{et al}, whereas the widely used Enskog-type constitutive relations show
poor accuracy. We find that fluctuations of the center of mass of the system are Gaussian. This suggests an effective Langevin description in terms of a macro-particle, confined by a harmonic potential and driven by a delta-correlated noise. Surprisingly, the fluctuations persist when increasing the number of particles in the system.

\section{Introduction\label{cap4:intro}}
This chapter addresses granular hydrodynamics and fluctuations in a simple two--dimensional granular system under conditions when existing hydrodynamic descriptions \cite{sela,brey1,lutsko} break down because of large density, \textit{not} large inelasticity. In view of the difficulties mentioned in Chapters~\ref{intro:Granos} and \ref{cap3}, attempts of a first--principle description of hydrodynamics and fluctuations should give way here to more practical, empiric or semi--empiric, approaches. One such approach to a hydrodynamic (or, rather, hydrostatic) description was suggested by Grossman \textit{et al.} \cite{grossman}. In the present chapter, we put it into a test in an extreme case when macro--particles (granular clusters with the maximum density close to the hexagonal close packing)
form. The model system we are dealing with was first introduced by Esipov and P\"{o}schel \cite{esipov}. It is an assembly of $N \gg 1$ identical disks of mass $m$, diameter $d$ and coefficient of normal restitution $\mu$, placed inside a circular box of radius $R$ at zero gravity (typical steady state configurations are shown in Fig.~\ref{fig1_4}). The circular wall of the box, which supply energy continuously to the granulate, is kept at constant temperature $T_0$, i.e., the vibrating boundary is represented as a thermal wall (under conditions specified in Chapter~\ref{intro:Granos}). There are not additional external forces acting on the system.

We used a model of inelastically colliding hard disks to develop kinetic and hydrodynamic descriptions. 
We measure, using event--driven molecular dynamics simulations, the radial density profiles of the system, including the close--packed part. Furthermore, we solve numerically a set of granular hydrostatic equations which employ the constitutive relations by Grossman \textit{et al.} and show that, in a wide range of parameters, there is good agreement between the two. A marginal stability analysis will show that there are no steady--state solutions with broken azimuthal symmetry which would bifurcate from the azimuthally symmetric solution. We also show that, for the same setting, the Enskog-type CRs \cite{jenkins} perform poorly. Finally, we investigate in some detail, for the first time, fluctuations of the macro--particles, by measuring the radial probability distribution function (PDF) of the center of mass of the system. These fluctuations turn out to be Gaussian, which suggests an effective Langevin description of the system in terms of a macro--particle, confined by a harmonic potential and driven by a white noise. Surprisingly, the fluctuations persist as the number of particles in the system is increased.

\section{\label{cap4:sims}Event--driven molecular dynamics simulations}
As we outlined in Chapter~\ref{intro:DDF}, we employed a standard event--driven molecular dynamics algorithm \cite{thorsten3}. One of the simplest models one can employ to study flows dominated by binary particle collisions \cite{goldhirsch1,brilliantov} consists of $N$ inelastically colliding hard disks of mass $m$ and diameter $d$. Rotational degree of freedrom is neglected, and thus inelasticity is modeled by the coefficient of normal restitution $\mu < 1$, which is considered constant (velocity independent). When two particles collide (say $1\leq i,j\leq N$), their tangential velocities are unchanged, whereas the relative normal velocity is decreased. Using momentum conservation, the post--collisional velocities in terms of the pre--collisional read
\begin{equation}
\begin{split}
\vec{v}_i^{\,\prime} = & \vec{v}_i - \frac{1+\mu}{2}\left[\left(\vec{v}_i-\vec{v}_j\right)\cdot \hat{e}_{ij}\right]\hat{e}_{ij}\,,\\
\vec{v}_j^{\,\prime} = & \vec{v}_j + \frac{1+\mu}{2}\left[\left(\vec{v}_i-\vec{v}_j\right)\cdot
\hat{e}_{ij}\right]\hat{e}_{ij}\,,
\end{split}
\label{eq:velocs_bis}
\end{equation}
where primed quantities stand for post--collisional velocities, and\\
$\hat{e}_{ij}\equiv \left(\vec{r}_i-\vec{r}_j\right)/\left|\vec{r}_i-\vec{r}_j\right|$ is unitary vector oriented along the direction connecting their centers\footnote{Notice that Eq.~\eqref{eq:velocs_bis} is equivalent to Eq.~\eqref{eq:velocs}, but expressed in a more detailed way.}.
The circular boundary supply continuously energy to the system as if were driven by a vibrating wall. If the frequency of the oscillating wall is large enough, the energy supply may be modelled by a heated wall. The thermal wall implementation \cite{thorsten3} is simple: whenever a particle touches the wall its velocity is chosen from a Maxwell distribution according to $T_{0}$ (which is measure in units of energy), that we set to unity. We put $m=1$ and fixed $R/d=100$ and $\mu=0.888$, while the total number of particles $N$ served as the control parameter in the simulations. So the scaled time unit is $d\left( m/T_{0}\right)^{1/2}=1 $.

\begin{figure}
\centerline{\includegraphics[width=12cm,clip=]{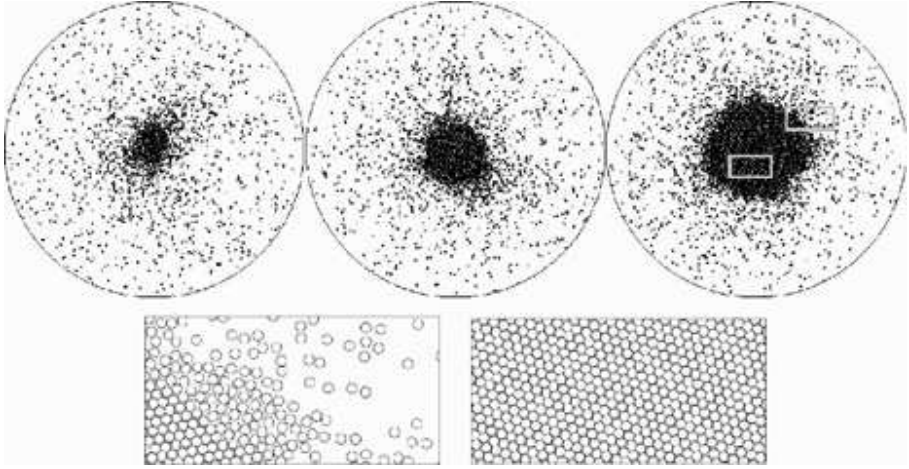}}
\caption[Snapshots and magnifications of the close--packed cluster]{Snapshots of the system (top, from left to right) for $N=1716$, $2761$ and $N=5055$. Also shown are
magnifications (bottom) of the indicated areas in the rightmost snapshot. Notice the coexistence of the different kinds of packings.} 
\label{fig1_4}
\end{figure}

We were mostly interested in a hydrodynamic (low Knudsen number) regime, when the mean free path is small compared
to the system size and the mean free time is small compared with any hydrodynamic time scale. This requires $N \gg R/d =100$. For $N$ in the range of a few hundred, we observed a dilute granular gas with an increased density in the center of the box (although not observable to naked eye). The clustering in the center becomes more pronounced as $N$ grows. The clustering can be easily explained in the hydrodynamics language: because of the inelastic collisions the granular temperature goes down as one moves away from the wall toward the center of the box. This, combined with the constancy of the pressure throughout the system, causes an increased particle density at the center. As $N$ increases further, the particle density in the center approaches the hexagonal close packing value $n_c = 2/(\sqrt{3} d^2)$. Figure \ref{fig1_4} shows snapshots of the system for three different, but sufficiently large, values of $N$. A perfect hexagonal packing is apparent nearby the center. Movies of these simulations show that the macro--particle (close--packed cluster) position fluctuates around the center of the box, while the cluster shape fluctuates around a circular shape. Notice that both kind of fluctuations, in particular the off--center location if the cluster, can not be explained hydrodynamically. 

Our diagnostics was focused on two radial distributions: the number density of the particles $n(r)$ and the PDF of a radial position of the center of mass of the system $P(r_m)$, defined below. To measure $n(r)$ (likewise in Chapter~\ref{cap3}), we recorded all particle coordinates in intervals of $100,000$ particle collisions. For each snapshot we introduced bins in the form of concentric circular rings of width $0.5$, centered in the center of mass of the system. For each particle we determined the fraction of its area falling in each of the rings, summed the contributions of all particles to each bin, and divided the result by the bin area. The resulting radial profile was averaged over many snapshots (typically 1000). Similar method is employed to measure $P(r_m)$. As we are interested in steady--state distributions, we disregarded initial transients. This was monitored by the time--dependence of the average kinetic energy of the particles (the first decayed and then approached an almost constant value) and the time--dependence of the center of mass itself.

\section{\label{cap4:hydro}Hydrostatic theory}
As the fluctuations are relatively weak, it is natural to start with a purely hydrodynamic description. For a zero mean flow this is a \textit{hydrostatic} theory, as it operates only with (time--independent) granular density $n(\mathbf{r})$, temperature $T(\mathbf{r})$ and pressure $p(\mathbf{r})$. The energy input at the thermal wall is balanced by dissipation due to inter--particle collisions, and one can employ the momentum and energy balance equations:
\begin{equation}
\begin{split}
p={\rm const}\,,\\
\nabla \cdot (\kappa \, \nabla T) = I \,.
\end{split}
\label{eq:energy1}
\end{equation}
Here $\kappa$ is the thermal conductivity and $I$ is the rate of energy loss by collisions. Note that the heat flux, entering the thermal balance in Eq.~\eqref{eq:energy1}, does not include an inelastic term, proportional to the \textit{density} gradient \cite{sela,brey1,lutsko}. In the nearly elastic limit $1-\mu \ll 1$, that we are interested in, this term can be neglected. The boundary condition at the thermal wall is $T(r=R,\theta)=T_0$, where $r$ and $\theta$ are polar coordinates with the origin at the center of the box.

To proceed from here and compute numerical factors, we need constitutive relations: an equation of state
$p=p\,(n,T)$ and relations for $\kappa$ and $I$ in terms of $n$ and $T$. As we attempt to describe close--packed clusters, the first--principle standard techniques, based on the Boltzmann or Enskog equations,
are inapplicable. Grossman \textit{et al.} \cite{grossman} suggested a set of semi--empiric relations in two dimensions, that are valid for all densities, all the way to hexagonal close packing. Their approach ignores possible coexistence beyond the disorder--order transition and assumes that the whole system is on the thermodynamic branch extending to the hexagonal close packing. The delicate issue of phase coexistence (liquid--like phase, close--packed phase with multiple domains) occur here is close analogy to the system of \textit{elastic} hard spheres \cite{thorsten1,chaikin,luding}. Grossman \textit{et al.} employed free volume arguments in the vicinity of the close packing, and suggested an interpolation between the hexagonal--packing limit and the well--known low--density relations. The resulting \textit{global} equation of state and constitutive relations \cite{grossman} read
\begin{equation}
\begin{split}
p= n\,T\, \frac{n_c+n}{n_c-n}\,,\\
\kappa=\frac{\sigma\,n\,(\alpha l + d)^2 T^{1/2}}{l} \,,\\
I= (\sigma/\gamma l)\, (1-\mu^2) \,n \, T^{3/2}\,.
\end{split}
\label{state}
\end{equation}
 Here $l$ is the mean free path, which is given by an interpolation formula \cite{grossman}
\begin{equation}
l=\frac{1}{\sqrt{8}n d}\, \frac{n_c-n}{n_c-an}\,, 
\label{meanfree}
\end{equation}
and $a= 1-(3/8)^{1/2}$. These relations include three dimensionless numerical factors of order unity: $\alpha$,
$\gamma$ and $\sigma$, where the latter drops out from the steady--state problem. They can be calculated exactly from the velocity distribution function. However, this distribution is poorly understood \cite{goldhirsch1}. Grossman \textit{et al.} determined the optimum values $\alpha=1.15$ and $\gamma=2.26$, verified by a detailed comparison with molecular dynamics simulations of a system of inelastic hard disks in a rectangular box without gravity, driven by a thermal wall, and numerical solutions of the hydrostatic equations~\eqref{eq:energy1} in rectangular geometry. We adopted the same values of $\alpha$\footnote{The value of $\alpha$ differs in systems with gravity ($\alpha\approx 0.6$) \cite{baruch3}.} and $\gamma$ in our calculations for the circular geometry.

\begin{figure}
\centerline{\includegraphics[width=10.0cm,clip=]{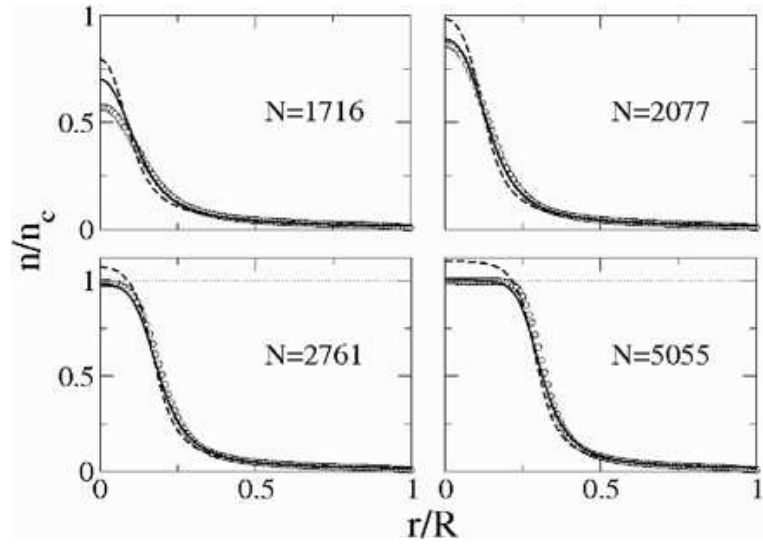}}
\caption[Scaled density $n/n_c$: Comparison between hydrostatics and event--driven simulations]{Scaled density $n/n_c$ versus scaled radius $r/R$ as
observed in simulations (circles) and predicted by the hydrostatic theory
with the constitutive relations by Grossman \textit{et al.} \cite{grossman} (solid
lines) and with the Enskog-type relations \cite{jenkins} (dashed lines) for
four different values of the total number of particles $N$. For the
rest of parameters please see the text. The dotted lines indicate
$n/n_c=1$.}
\label{fig2_4}
\end{figure}

Employing Eqs.~\eqref{state} and \eqref{meanfree} we can reduce Eqs.~\eqref{eq:energy1}
to a single equation for the rescaled inverse density $z(r,\theta)\equiv
n_c/n(r,\theta)$. In the rescaled coordinates $\mathbf{r}/R \to \mathbf{r}$ the
circle's radius is $1$, and the governing equation becomes

\begin{equation}
\frac{1}{r} \frac{\partial}{\partial r} \left[ r F(z)\frac{\partial z}{\partial
r}\right] +\frac{1}{r^2}\frac{\partial}{\partial \theta}\left[F(z) \frac{\partial
z}{\partial \theta}\right]=\Lambda Q(z)\,, 
\label{govern0full}
\end{equation}

where
\begin{equation}
  \begin{split}
\mathcal{F}(z)=&\frac{(z^2+2z-1)\left[\alpha z(z-1)+\sqrt{32/3}(z-a)\right]^2}
{(z-a)(z-1)^{1/2}z^{3/2}(z+1)^{5/2}}\\
\mathcal{Q}(z)=&\frac{(z-a) (z-1)^{1/2}}{(z+1)^{3/2} z^{1/2}}\,,
  \end{split}
\label{Q}
\end{equation}
and ${\Lambda} = (32/3\gamma)\,(R/d)^2\,(1-\mu^2)$ is the hydrodynamic inelasticity parameter, first introduced in Ref.~\cite{livne1} in the context of rectangular geometry \footnote{Notice that, no matter how small (but finite) the inelasticity is, the parameter $\Lambda$ goes to infinity in the thermodynamic limit $R \to \infty$.}. As the
total number of particles $N$ is fixed, $z^{-1}(r,\theta)$ satisfies a
normalization condition:
\begin{equation}
  \int_0^{2 \pi} \,d\theta \int_0^{1} \,dr\,r\, z^{-1}(r,\theta) = \pi f\,,
  \label{normfull}
\end{equation}
where
\begin{equation}
f=\frac{\sqrt{3}}{2 \pi}N\left(\frac{d}{R}\right)^2
\end{equation}
is the average area fraction of the particles.

Most interesting state among the one--dimensional states is the azimuthally symmetric state ($\theta$--independent), which is indeed one of the possible solutions of Eq.~\eqref{eq:energy1}. In this case, assuming azimuthal symmetry, we rewrite Eq.~\eqref{eq:energy1} as
\begin{equation}
\begin{split}
p(n,T) = {\rm const}\,,\\
\frac{1}{r} \frac{d}{dr} \left[r\,\kappa(n,T)\, \frac{dT}{dr} \right] = I(n,T)\,. 
\end{split}
\label{eq:govern1}
\end{equation}
Equations~\eqref{eq:govern1} can be reduced to a single equation for the scaled inverse density $z(r)\equiv n_c/n(r)$. In the rescaled coordinate $r$, Eqs.~\eqref{eq:govern1} reduces to
\begin{equation}
\frac{1}{r}\frac{d}{dr}\left[r
\mathcal{F}(z)\,\frac{dz}{dr}\right]=\Lambda\, \mathcal{Q}(z)\,. 
\label{ode}
\end{equation}

The fixed total number of particles yields a normalization condition, which read
\begin{equation}
  \int_0^1 \, z^{-1}(r)\,r \,dr = f/2\,.
  \label{normalization}
\end{equation}
Equations~\eqref{ode} and \eqref{normalization}, together with the obvious boundary condition $dz/dr|_{r=0} = 0$ (this is a consequence of the fact that the temperature gradient vanish at the center of the system), form a complete set. The hydrostatic problem is completely determined by two scaled parameters $\Lambda$ and $f$. This can be solved numerically as explained in Chapter~\ref{cap3} for the annular geometry. That is, in order to solve Eq.~\eqref{ode}, instead of consider the area fraction $f$ one can prescribe the inverse density at the center of the system $z_{0}$. This condition, combined with the no--flux condition at $r=0$ define a Cauchy problem for $z^{-1}(r)$ \cite{khain1}. Solving the Cauchy problem one can compute the respective value for $f$ from Eq.~\eqref{normalization}. At fixed $\Lambda$, $z_{0}$ is a strictly monotonically decreasing function of $f$. Therefore, an alternative parameterization of the azimuthally symmetric state is given by the scaled numbers $\Lambda$ and $z_{0}$. The same considerations keep for the marginal stability analysis (see below).\\

\begin{figure}
\centerline{\includegraphics[width=8cm,clip=]{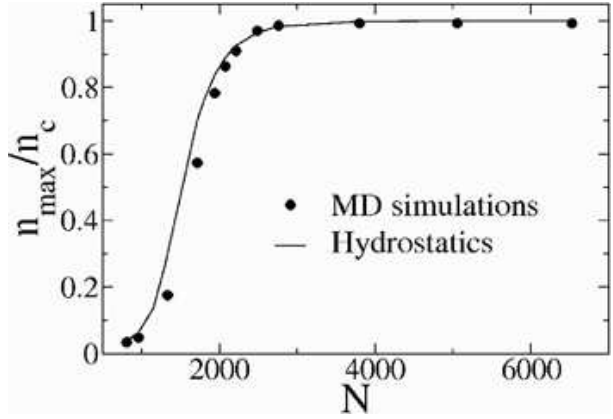}}
\caption[Maximum scaled density $n_{max}/n_c$: Comparison between hydrostatics and event--driven simulations]{Maximum scaled density $n_{max}/n_c$ as a function of $N$ as predicted by the hydrostatic theory and observed in simulations.}
\label{fig3_4}
\end{figure}

In our event--driven simulations we varied the number of particles $N$. Accordingly, the scaled parameter $f$ is varied, while $\Lambda=9980$ was constant. Figure~\ref{fig2_4} shows a comparison of the hydrostatic radial density profiles with the radial density profiles obtained in simulations, for four different values of $N$. The simulations profiles were averaged over 1000 uncorrelated configurations. It can be seen from Fig.~\ref{fig2_4} that for $N=1716$ the theory overestimates the density in the center of the box. This is expected as, for relatively small $N$, the maximum density is considerably less than the close--packing density, and the accuracy of the constitutive relations by Grossman \textit{et al}. is not as good. For larger $N$ the agreement rapidly improves. As Fig.~\ref{fig3_4} shows, the agreement between the hydrostatic theory and the simulations persists for the \textit{maximum} radial densities at different number of particles $N$. The agreement is excellent for the densities approaching the close packing density, whereas for small $N$, there is small deviations. In absolute scale, however, the agreement is very good. We also computed the density profiles using another set of semi-empiric constitutive relations\footnote{These were already employed in Chapter~\ref{cap3}.}, those obtained in the spirit of Enskog theory \cite{jenkins}. These read,
\begin{equation}
\begin{split}
\kappa=\frac{2d n T^{1/2} G}{\pi^{1/2}} \left[ 1+\frac{9\pi}{16} \left( 1+\frac{2}{3G}\right)^{2}\right],\\ 
I=\frac{8(1-\mu)n T^{3/2}G}{d\sqrt{\pi}}\,,
\end{split}
\label{JR4cap4}
\end{equation}
and the equation of state first proposed by Carnahan and Starling \cite{carnahan}
\begin{equation}
P=n T(1+2G)\,,
\end{equation}
where $G=\nu (1-\frac{7\nu}{16})/(1-\nu)^{2}$ and $\nu=n \left( \pi d^{2}/4 \right) $ is the solid fraction. One can clearly see from Fig.~\ref{fig2_4} that the Enskog-type constitutive relations predict unphysically high densities in the cluster.\\

\begin{figure}
\centerline{\includegraphics[width=8cm,clip=]{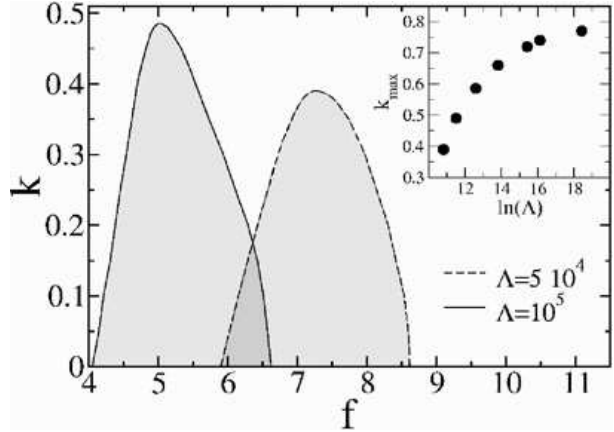}}
\caption[(Main graph) The marginal stability curves $k=k(f)$ for two different values of $\Lambda$. (Inset) $k_{max}$ as a function of $\ln (\Lambda)$]{The
marginal stability curves $k=k(f)$ for $\Lambda=5\times 10^{4}$ and
$\Lambda=10^{5}$, as indicated. The shaded area denotes in both cases the
(unphysical) instability region that is obtained if one ignores the quantization
of $k$: $k=1,2, \dots$. The inset shows $k_{max}$ as a function of $\ln
(\Lambda)$.}
\label{fig4_4}
\end{figure}

Are the azimuthally symmetric states stable with respect to small perturbations?
We performed marginal stability analysis to find out whether there are
steady-state solutions with broken azimuthal symmetry, $\hat{z}(r,\theta)$ that
bifurcate continuously from an azimuthally symmetric solution $z(r)$. Under
additional assumption that the possible \textit{instability} of the azimuthally
symmetric state is purely growing (that is, \textit{not} oscillatory), the
marginal stability analysis yields the instability borders. The marginal
stability analysis goes along the same lines as that developed for the
rectangular geometry \cite{khain1,khain2,livne1,livne2,baruch2}. In the framework of time--dependent hydrodynamics, this corresponds to marginal stability analysis of the azimuthally symmetric solution with respect to small perturbations along the azimuthal coordinate $\theta$. Let us search a steady-state close to an azymuthally-symmetric state:

\begin{equation}
\hat{z}(r,\theta)=z(r)+\varepsilon \; \phi_{k}(r) \, \sin(k\, \theta) \;,
\label{eq:linear}
\end{equation}
As $\hat{z}(r,\phi+2\pi) = \hat{z}(r,\phi)$, $k$ must be an integer which can be chosen to
be non-negative. Substituting Eq.~\eqref{eq:linear} into Eq.~\eqref{govern0full} and linearizing around the azimuthally symmetric state $z(r)$ with respect to the small correction $\varepsilon \ll 1$, we obtain a linear
eigenvalue problem, where $k=k(f,\Lambda)$ plays the role of the eigenvalue:
\begin{equation}
\zeta_{k}^{\prime\prime}+\frac{1}{r}\zeta_{k}^{\prime}-\left[ \frac{k^2}{r^2}+\frac{\Lambda \mathcal{Q}^{\prime}(z)}{\mathcal{F}(z)}\right] \zeta_{k}=0\;,
\label{eq:eigen}
\end{equation}
where $\zeta_{k}(r)=\phi_{k}(r)\, \mathcal{F}(z)$. This equation is complemented by the boundary conditions
\begin{equation}
\zeta_{k}(0)=0\qquad \zeta_{k}(1)=0\;,
\label{eq:eigen_bound}
\end{equation}
and can be solved numerically. Let us ignore for a moment the quantization of
the eigenvalue $k$ and, while looking for $k$, assume that it is a (positive)
\textit{real} number. In that case, a numerical solution yields, for a fixed
$\Lambda$, a curve $k=k(f)$, see Fig.~\ref{fig4_4} for two examples. At a fixed
$k$, the azimuthally symmetric state is unstable within an interval of area
fractions. The foots of one such curve corresponds to the (hypothetical) case
when $k$ tends to zero (that is, the azimuthal wavelength tends to infinity).
The instability interval becomes narrower when $k$ is increased, and it shrinks
to a point at a maximum $k_{max}$, signalling that a density modulation with a
sufficiently short azimuthal wavelength should be stable for all $f$. For
example, when $\Lambda=5\times 10^4$, the marginal stability curve $k=k(f)$ has
its maximum at $k_{max}\approx 0.38$ (see Fig.~\ref{fig4_4}) that is less than unity. Going back to the
physical case, where $k$ is a positive integer, we see that all the values for
$k$, determined from the eigenvalue problem, are unphysical. That is, there are
no solutions to the eigenvalue problem that would satisfy the boundary
conditions and the quantization condition $k=1,2,3 \dots$. We observed a similar
behavior for values of $\Lambda$ up to $10^8$. Though $k_{max}$ increases with
$\Lambda$, the increase is extremely slow: slower than logarithmical (see the
inset of Fig.~\ref{fig4_4}).  These numerical results strongly indicate that the
azimuthally symmetric states are stable with respect to small perturbations.
This is in marked contrast with the presence of bifurcating states with broken
symmetry in similar settings of a granular gas driven by a thermal wall, but in
rectangular \cite{khain1,khain2,livne1,livne2,baruch2} and annular (Chapter~\ref{cap3}) geometries. The
absence of the bifurcating states with broken symmetry in the circular geometry
gives a natural explanation to the persistence of circle-shaped cluster shapes
as observed in our MD simulations. This prediction is very robust, as it is independent of the values $\alpha$ and $\gamma$, and even of the constitutive relations employed. For instance, the same result is found by employing Enskog--type relations \cite{jenkins} (see Eq.~\eqref{JR4cap4}).

\begin{figure}
\centerline{\includegraphics[width=10.0cm,clip=]{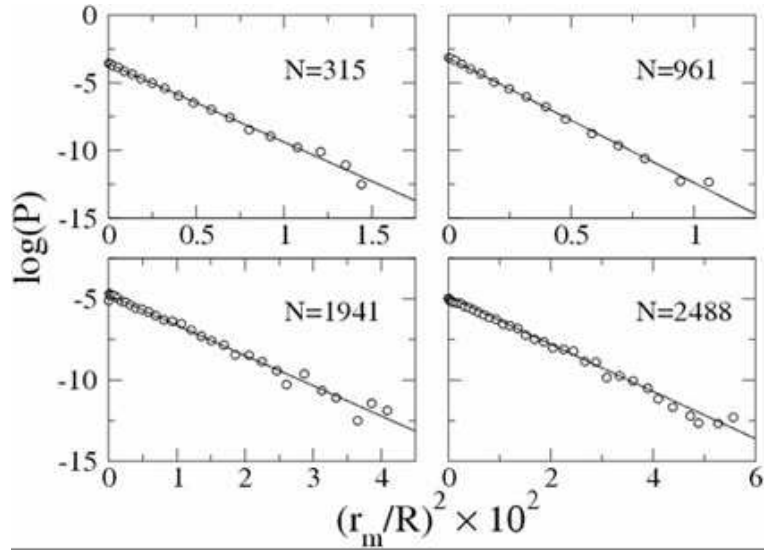}}
\caption[Probability distribution functions of the center of mass of the macro--particle]{The logarithm of the radial probability distribution function $P(r_m)$ of the center of mass position versus $(r_m/R)^2$, for four values of $N$.}
\label{fig5_4}
\end{figure}

\section{\label{cap4:fluc}Macro--particle fluctuations}
Now we turn to stationary fluctuations, that is, after transients die out. To better characterize the fluctuation--dominated behavior of this system, we compute the radial coordinate $r_m(t)$ of the center of mass of the system. The radial probability distribution function $P(r_m, t)$ is normalized by the condition $2\pi \int_0^R P(r_m, t)\, r_m dr_m =1$, where we have returned to the dimensional coordinate. Typical molecular dynamics results are presented in Fig.~\ref{fig5_4}, which shows $\log P(r_m)$ versus $(r_m/R)^2$, for four different values of $N$. The observed straight lines clearly indicate a Gaussian distribution, both for small and large $N$ values.

\begin{figure}
\centerline{\includegraphics[width=8cm,clip=]{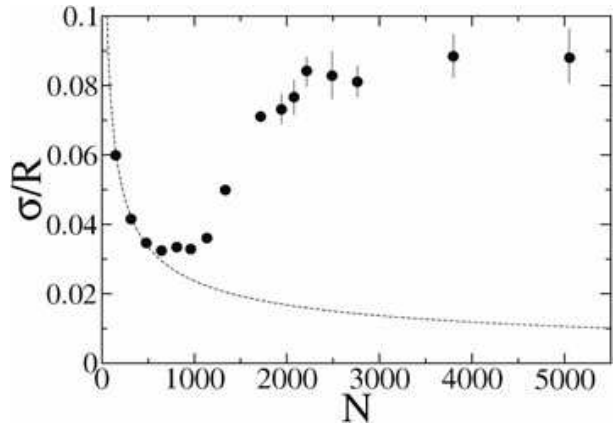}}
\caption[Scaled standard deviation of the center of mass position]{Scaled standard deviation of the center of mass position as a function of $N$, obtained in molecular dynamics simulations (dots). The dashed line shows is for $\sigma/R=0.75\, N^{-1/2}$. This power law characterizes the fluctuations of an ideal (elastic) gas in equilibrium.}
\label{fig6_4}
\end{figure}

This finding strongly suggests a Langevin description of the macro--particle. One can consider the macro--particle performing an over--damped motion in a (central--symmetric) confining harmonic potential\footnote{The systematic restoring force which confines the macro--particle has a potential $U(r_m)$. Not far from the center, $U(r_m)$ can be approximated by a harmonic potential.} $U(r_m)=k r_m^2/2$ and driven by a (delta--correlated) discrete--particle noise $\eta(t)$. The Langevin equation for this problem reads
\begin{equation}
h \dot{r}_m+k r_m=\eta(t)\,,
\label{Langevin}
\end{equation}
where $h$ is the damping rate, $\langle \eta(t) \eta(t^{\prime})\rangle=2h \Gamma \,\delta(t-t^{\prime})$, and $\Gamma$ is an effective magnitude of the discrete--particle noise. In the simplest approach the close--packed cluster and the discrete--particle noise can be characterized, for each $N$, respectively, by a point--like mass $M$ and mobility $\chi$, and by $T_{0}$, the temperature of the circular wall. A stochastic equivalent description is provided in terms of the Fokker-Planck equation for $P(r_m,t)$ \cite{vkampen,gardiner}. In the limit $\Gamma \to 0$, the (deterministic) steady-state solution of Eq.~\eqref{Langevin} is $r_m=\dot{r}_m=0$: the macro--particle at rest, located at the center of the box. At $\Gamma>0$, the steady state distribution function is 
given by the steady state solution of the Fokker-Planck equation \cite{vkampen,gardiner} which is $P(r_m)= (\pi \,\sigma^2)^{-1}\,\exp (- r_m^2/\sigma^2)$, where $\sigma^2 = \Gamma/k$, and the normalization constant is computed
for $\sigma \ll R$. The variance $\sigma^2$ is the ratio of $\Gamma$ (a characteristic of the discrete noise) and $k$ (a macroscopic quantity). 
An important insight 
can be achieved from the $N$--dependence of $\sigma$, obtained by molecular dynamics (see Section \ref{cap4:sims}). In analogy with equilibrium systems, one might expect the relative magnitude of fluctuations to decrease with increasing $N$. Surprisingly, this is not what we observed, see Fig.~\ref{fig6_4}. One can see that $\sigma(N)$ approaches a plateau, that is fluctuations persist at large $N$. The small--$N$ behavior (the first three data points: $N=150$, $315$ and $480$) in Fig.~\ref{fig6_4} agrees with the dependence $\sigma/R = {\cal O}(N^{-1/2})$, expected for an ideal gas in equilibrium. Not surprisingly, for these relatively small $N$ the clustering effect is small. On the other hand, at very large $N$, when the whole system approaches close packing (in our event--driven simulations, this corresponds to $N = 36 275$), $\sigma/R$ must go to zero. We could not probe this regime, however, as simulations became prohibitively long already at $N \sim 15 000$. An apparently related phenomenon has been recently reported for rectangular geometry, in the context of a van der Waals--like phase separation in a granular gas driven by a thermal wall \cite{baruch2}.

\section{\label{cap4:conclus}Conclusions}
We employed a simple model system \cite{esipov} to investigate the structure and fluctuations of dense clusters emerging in granular gases. The density profiles, obtained in event--driven molecular dynamics simulations, are well described by hydrostatic equations which employ the constitutive relations suggested by Grossman et al. \cite{grossman}. This agrees with similar results in rectangular geometry, with and without gravity \cite{baruch3,grossman,livne1}. We also extended their validity for lower restitution coefficient, and for the circular geometry. The agreement is particularly good for the large density region (close packing). The azimuthally symmetric state is predicted to be stable, fact which is apparent with our simulations. The observed Gaussian fluctuations of the center of mass, which are too large to be explained by diffusion, suggest an effective Langevin description in terms of a macro--particle in a confining potential of hydrodynamic nature, driven by discrete particle noise. Surprisingly, the fluctuations persist as the number of particles in the system is increased. This unexpected behavior is stark contrast with the shown by elastic gases in thermal equilibrium.\\

Other important point is the transition from a "smooth state" (with enhanced density in the middle) to a "cluster state" with a pronounced and sharp--edged cluster in the center, as a function of particle number density. It has not been stated before, that the process of cluster formation requires a certain particle number and the transition between a homogeneous (or irregular) state to a regularly clustered state (with a persisting cluster in the middle) is rather sharp. In our case it is at $N=1700\pm 300$.\\

It is worth mention that the global constitutive relations of Grossman \textit{et al.} \cite{grossman} completely ignore the issue of coexistence of different phases of the granulate: the liquid--like phase, the random close--packed phase, etc. However, the vibrofluidized steady state, considered in this chapter, has a zero mean flow. Therefore the viscosity terms in the hydrodynamic equations vanish. This fact is not merely a technical simplification. The shear viscosity of granular flow is finite in the liquid--like phase, and infinite in the (multiple) domains of the random close--packed phase. The effective total viscosity of the system is expected to diverge when the coarse--grained density slightly exceeds the freezing density \cite{divergence,rojo}. This invalidates any hydrodynamic description for sufficiently dense flows, and requires the introduction of an order parameter and a different type of the stress--strain relation into the theory (cf. Ref.~\cite{esipov}). Luckily, these complications do not appear for a zero--mean--flow state. Indeed, the equation of state, heat conductivity, and the inelastic heat loss rate do no exhibit any singularity around the freezing point, and all the way to the hexagonal close packing. Therefore, the hydrostatic description remains reasonably accurate far beyond the freezing point.\\

Future work should focus on dense clusters (more precisely, on their ``core" regions) and characterize the positional and orientational ordering at different $N$. This can be achieved by measuring two correlation function: the positional correlation function and the orientational correlation function. The motivation is the following. There are classical papers of Kosterlitz and Thouless, Young and Nelson, and Halperin, from the 70-ies ---these yielded the KTHNY theory of melting, reviewed by Strandburg in Ref.~\cite{melting}---, which suggested that melting in assemblies of \textit{elastic} hard disks occurs (as the density decreases) via two continuous phase transition: the first one from the perfect hexagonally packed phase to the so called \textit{hexatic}\footnote{This phase can exist between solid and liquid phases. The system in the hexatic phase has no long-range translational order but has quasi-long-range orientational order.} phase, and the second one from the hexatic phase to a liquid--like phase.  
A major question which remains to be answered concerning our inelastic system is whether there is any signature of additional instabilities, as the density in the central region increases.

\part{\label{part:apend}Appendices}
\appendix
\chapter{\label{apendA}Equilibrium properties of the Lattice Gas}
The \textit{driven lattice gas} studied in Chapter~\ref{cap1} reduces to the (equilibrium) \textit{lattice gas} \cite{yang_lee_lg} for zero field. This is a model for density fluctuations and liquid-gas phase transformations. As its nonequilibrium counterpart, this is defined in a square lattice with $N=L_{\parallel}L_{\perp}$ sites\footnote{We consider here only two-dimensional lattices.}. Thus, each lattice site $i$ can exist in two states, occupied by a particle or empty, labelled by an occupation variable $n_{i}=1$ or $n_{i}=0$, respectively. The \textit{lattice gas} can be mapped to the \textit{Ising} model with \textit{ferromagnetic} couplings \footnote{Historically, the Ising model \cite{ising} was introduced before the lattice model \cite{yang_lee_lg}. Likewise, the lattice gas as and the \textit{binary alloy} model \cite{thompson} are equivalent.}. The Ising model is perhaps the simplest system that undergoes a nontrivial phase transition in two or more dimensions. The model was initially proposed to describe ferromagnetism ---the presence of spontaneous magnetization in metals such as Fe and Ni below a critical temperature $T_{c}$---. The relation between the lattice gas and the Ising model is set by simple change of variables, namely, $s_{i}=(2n_{i}-1)/2$, where $s_{i}$ are spin variables capable of two orientations, ``up"  and ``down". The spin up $s_i=1/2$ and the spin down $s_i=-1/2$ states correspond to occupied ($n_i=1$) and unoccupied ($n_{i}=0$) cells, respectively. Each spin $i$ interacts with one another via an exchange interaction $J_{ij}>0$. Assuming ferromagnetic coupling and isotropic interactions ($J_{ij}\equiv J>0$), the Ising Hamiltonian is
\begin{equation}
H=-J\sum_{\langle ij\rangle}s_i s_j -B\sum_{i=0}^{N}s_i \;,
\label{eq:ising}
\end{equation}
where $\langle ij\rangle$ denotes neighbor interactions and $B$ is the external magnetic field, which plays the role of the chemical potential of the lattice gas. In fact, the \textit{canonical} partition function for the ferromagnetic model is the \textit{grandcanonical} partition function \cite{libros_eq} for the lattice gas model. Therefore the lattice model is isomorphic with the Ising model. Thermodynamically the descriptions of the ising model and the lattice gas are equivalent, but there are simplifying features for the Ising magnet. Henceforth we shall refer to them without distinction.

Lars Onsager \cite{onsager,thompson} gave in 1944 the (exact) partition function for the two-dimensional model, in the absence of external magnetic field. The Onsager solution showed the existence of a second order phase transition at $T_{c}=2/ln\left( 1+\sqrt{2}\right)\,Jk_{B}^{-1}$, referred to as the Onsager temperature $T_{\text{Ons}}$. The spontaneous magnetization
\begin{equation}
\left|M\right|=\left|\frac{1}{N}\sum_{i=1}^{N}s_i\right|
\label{eq:magnet}
\end{equation}
\begin{figure}
\centerline{\includegraphics[width=10cm,clip=]{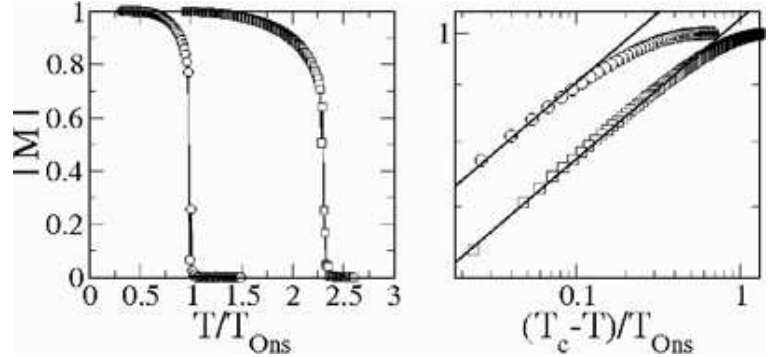}}
\caption[Spontaneous magnetization and order parameter critical exponent for the Ising model with NN and NNN couplings]{\label{fig1_A} Phase transition in the Ising model with NN (circles) and NNN couplings (squares). Left panel: Spontaneous magnetization $M$ as a function of the temperature. The magnetization vanishes abruptly at $T_{c}/T_{\text{Ons}}=1$ for NN couplings, and at $T_{c}/T_{\text{Ons}}=2.32$ for the NNN case. Otherwise is zero for $T>T_{c}$ in both cases. Right panel: As expected in equilibrium systems \cite{renormalization} the order parameter critical exponent do not depend on the dynamics details. The solid lines represent slopes of $\beta_{\text{Ising}}=1/8$.}
\end{figure}
is the order parameter for this transition. Above the critical temperature the (average) spontaneous magnetization is zero, whereas below criticality $M$ is non-zero, thus one finds the system in a ferromagnetic state. Due to the up-down symmetry of the Ising system, the spontaneous magnetization is doubly degenerate, with an ``up" phase and a ``down" phase. In the terminology of the lattice gas, there should be a dense (liquid) phase coexisting with a dilute (gas) phase below the critical temperature. Otherwise, typical configurations are disordered. The left panel in Fig.~\ref{fig1_A} shows the second order phase transition for nearest-neighbor (NN) couplings and next-nearest-neighbor (NNN) couplings, as schematized in Fig.~\ref{fig4_1} in Chapter~\ref{cap1}. For NN couplings, the first sum in Eq.~\eqref{eq:ising} runs over the four NN sites, and for the NNN case it runs over the eight NNN sites. As is shown in Fig.~\ref{fig1_A}, the phase transition occurs exactly at $T_{\text{Ons}}$ for NN couplings. The immediate effect of considering additional neighbors, as in the NNN case, is the rising of the critical temperature. For NNN couplings, the critical temperature raises to $T_{c}=2.32\,T_{\text{Ons}}$ \cite{mattis}. In the neighborhood of the critical temperature the spontaneous magnetization behaves like $M\approx \left( 1-T/T_c\right)^{\beta_{\text{\begin{tiny}Ising\end{tiny}}}}$ as $T\rightarrow T_c\;-$. Here, $\beta_{\text{\begin{tiny}Ising\end{tiny}}}$ is the order parameter critical exponent. The right panel of Fig.~\ref{fig1_A} shows the behavior of $M$ near $T_c$ for NN and NNN couplings. In both cases, $\beta_{\text{\begin{tiny}Ising\end{tiny}}}=1/8$. This result is consistent with (equilibrium) renormalization group ideas \cite{renormalization}, which set up that the critical exponents are universal. It means that there are only a few parameters, such as dimensionality and underlying symmetries which are relevant near the critical point, therefore critical exponents do not depend on the particular system, either magnetic or fluid. 
\begin{figure}
\centerline{\includegraphics[width=12cm,clip=]{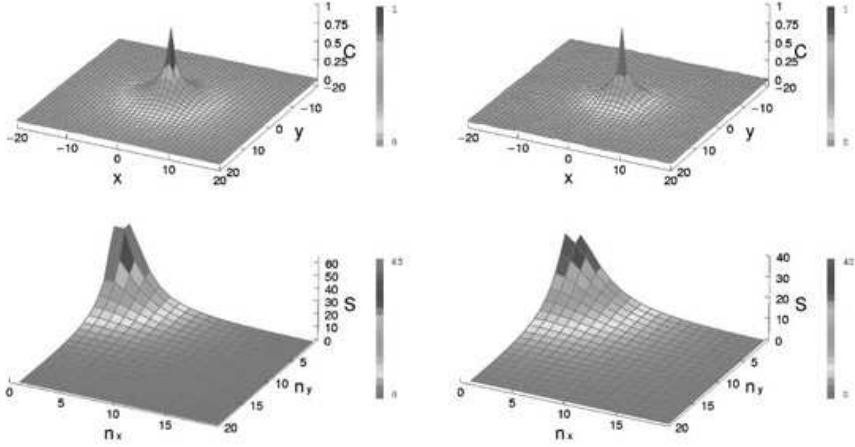}}
\caption[Surface plots of the two--point correlations and structure factor above criticality for the equilibrium lattice gas with NN and NNN interactions]{\label{fig2_A} Surface plots of the two--point correlation (upper row) and the structure factor (lower row). These are for the Ising magnet, or equivalently the lattice gas, on a $L_{\parallel}\times L_{\perp}=128\times 128$ lattice with NN (left column) and NNN interactions (right column). To avoid complications due to inhomogeneous ordered phases, we choose temperatures above criticality. Temperatures are $T/T_{\text{Ons}}=1.09$  for the NN and $T/T_{\text{Ons}}=2.61$ for the NNN case. Notice the change of color code with each plot.}
\end{figure}

\begin{figure}
\centerline{\includegraphics[width=8cm,clip=]{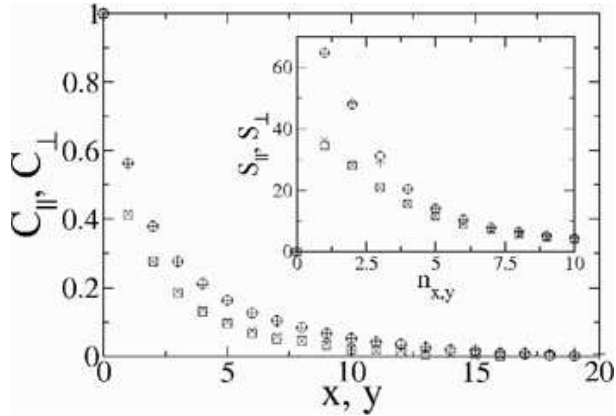}}
\caption[Projections of the two--point correlations and structure factor above criticality for the equilibrium lattice gas]{\label{fig3_A} Main graph: Parallel $C_{\parallel}$ and transverse $C_{\perp}$ projections of the two--point correlation function. Symbols $\circ$ (parallel) and $+$ (transverse) are for NN interactions, whereas $\square$ (parallel) and $\times$ (transverse) correspond to the NNN case. Inset: Parallel $S_{\parallel}$ and transverse $S_{\perp}$ projections of the structure factor. Here, $n_{x,y}=128\,\kappa_{x,y}/2\protect\pi $ are integers. Symbols are as in the main graph. System size and temperatures are as in Fig.~\ref{fig2_A}.}
\end{figure}

In zero magnetic field the pair-correlation function is of particular interest since it in a sense measures the ``degree of order" of the lattice. It is defined in general from the partition function as $C=\left\langle s_{i}s_{j} \right\rangle$, where $\left\langle\cdots\right\rangle$ (see also Eq.~\eqref{eq:correl} in Chapter~\ref{cap1}) denotes the proper ensemble average. Two-point correlations are shown in Fig.~\ref{fig2_A}. Correlations in the equilibrium Ising model are short ranged, except at the critical point where the free energy shows a nonanalytic point, controlled by a correlation length $\xi$. Specifically, they decay exponentially (in Fig.~\ref{fig2_A} or more clearly in Fig.~\ref{fig3_A}) when $r$, a typical inter-particle separation, becomes large compared to $\xi$. Translated into the momentum space, the pair correlation function lead to the structure factor $S(\kappa_{x},\kappa_{y})$ (see Eq.~\eqref{eq:struc} in Chapter~\ref{cap1}), i.e., structure factor is the Fourier transform of the two-point correlation function. Structure factors are shown in Figs.~\ref{fig2_A} and \ref{fig3_A}. Exponential decay in correlations, takes the form of ``analyticity at the origin" in the momentum space. Notice also that this analyticity in the origin do not depend on whether one takes isotropic or anisotropic couplings. As indicated above, we have assumed isotropic couplings between spins, and therefore both $C$ and $S$ are isotropic (more clearly in Fig.~\ref{fig3_A}). In Chapter~\ref{cap1} we show that the \textit{driven lattice gas} behaves quite differently, displaying generic power law decay in two-point correlations, not only at the critical point, and discontinuity singularity at the origin in the structure factor.

\chapter[Interfacial stability of Driven Lattice Gases for $E\rightarrow \infty$]{\label{apendC}Interfacial stability of Driven Lattice Gases under saturating field}

In Chapter~\ref{cap1} we pointed out one of the most striking, even counterintuitive, features of the \textit{Driven Lattice Gas} (DLG) with nearest-neighbor (NN) interactions: Monte Carlo (MC) simulations with Metropolis rates show that the critical temperature $T_{E}$ monotonically increases with the external driving field from the Onsager value $T_{0}=T_{\text{Ons}}=2.269\,Jk_{B}^{-1}$ to $T_{\infty }\simeq 1.4\,T_{\text{Ons}}$. As we concluded, this is associated with the fact that a driven particle is geometrically restrained in the DLG. In fact, allowing hops and interactions to the next-nearest-neighbors (NNN) one observes just the opposite behavior: the critical temperature decreases as the field increases. Moreover, there is no phase transition for a large enough field, i.e.,  $T_{E}\rightarrow 0$ as $E\rightarrow \infty $. Or in other words, in the DLG with NNN interactions a strong field prevents transition to an ordered low-temperature phase. This is better illustrated in Fig.~\ref{fig2_1} in Chapter~\ref{cap1}. Qualitatively, this difference of behavior in the infinite field limit between the two models may be understood in terms of stability of the liquid-gas interface. That is, in contrast with the NN case, a (well-ordered) interface cannot be stable under a strong field in the NNN case. This suggest a balance between thermal effects, which favor the stability (the order) of the interface \cite{siders}, and driving effects, which would drive particles out of the interface favoring disorder. 

In order to study this balance, the disparate behavior between both dynamics, and characterize the two competing tendencies ---i.e, the field vs. temperature balance---  we propose a generalized driven lattice gas model. This is an intermediate model between the two cases above mentioned (the NN and the NNN case) which captures the behavior in the large field limit observed in each system. It is defined on a two-dimensional lattice $L_{\parallel}\times L_{\perp}$ with periodic boundary conditions in which particles interact via the usual Ising Hamiltonian,
\begin{equation}
H=-4\sum_{\langle j,k \rangle} \sigma_j \sigma_k \: .
\label{eq:ham_alpha}
\end{equation}
Each site $i\in \mathbb{Z}^{2}$ is either occupied by a particle or empty, which we denote by a lattice occupation number $\sigma_i$ taking two values: $\sigma_i=1$ (occupied) or $\sigma_i=0$ (empty). The sum in the Hamiltonian runs over all the NNN sites. As in the standard DLG, dynamics is induced by the competion between a heat bath at temperature $T$ and an external driving field $E$, which break the \textit{detailed balance} condition. The field is assumed pointing along one of the principal lattice directions, say horizontal. Time evolution is then by microscopic dynamics according to the \textit{Metropolis} transition probability per unit time $w$ (this was previously defined in Eq.~\eqref{eq:metro} in Chapter~\ref{cap1}):
\begin{equation}
w(\mathbf{\sigma}\rightarrow\mathbf{\sigma}^{\prime})=\min\left\lbrace 1,e^{\left( \Delta H+\vec{E}\cdot \vec{x}\right)/k_BT} \right\rbrace \;.
\label{eq:rate_alpha}
\end{equation}
Here $\Delta H$ is the energy difference of a particle-hole exchange and $\vec{E}\cdot \vec{x}$ denotes the dot product between the field and the displacement of the exchange. $\mathbf{\sigma}\equiv\left\lbrace \sigma_{i}\,;\,i\in \text{two-dimensional lattice} \right\rbrace $ stands for a configuration.

\begin{figure}
\centerline{\includegraphics[width=3cm,clip=]{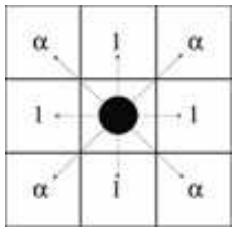}}
\caption[Schematic diagram of the accessible sites for a particle in the $\alpha$DLG model]{\label{fig1_C}Schematic diagram of the accessible sites a particle (at the center, marked with a dot) has for the $\alpha$DLG. Attempts (particle-hole exchanges) along the principal lattice directions are always allowed (marked with probability $1$). Diagonal attempts are allowed with probability $\alpha$. The limit of $\alpha\rightarrow 0 (1)$ corresponds with NN (NNN) couplings. Once $\alpha$ is set at the beginning, the allowed exchanges are controlled by the Metropolis rate function Eq.~\eqref{eq:rate_alpha}.}
\end{figure}

Given a particle, its neighboring sites along the principal lattice directions has different probability to be reached than the sites along the diagonals (see Fig.~\ref{fig1_C}). The former are always accessible (probability 1), whereas the latter have probability $0\leq \alpha\leq 1$, parameter which is set at the beginning of the simulation (as well as $T$ and $E$). The algorithm proceed as follows\footnote{The algorithm as we use it is more complicated to obtain better (and faster) results. We do not present this algorithm due to its complexity. Instead, we show a simpler algorithm which would work like this.}:
\begin{enumerate}
\item
We randomly select a NNN pair $\mathbf{i},\mathbf{j}$ of sites with different occupancies, $\sigma_{\mathbf{i}}\neq \sigma_{\mathbf{j}}$. 
\item
As Fig.~\ref{fig1_C} illustrates, if the selected pair is along one of the principal lattice directions the exchange attempt is allowed (Go to next item). In other case, draw a random number $r_1$, and if $r_1\leq \alpha$ then the attempt is allowed (Go to next item), otherwise rejected (Go to item 1).
\item
Perform a standard MC trial move controlled by the Metropolis \textit{rate} function $w(z)=\min\left\lbrace 1,e^{-z} \right\rbrace$, i.e., controlled by the standard biased rate Eq.~\eqref{eq:rate_alpha}. To be specific, first compute $z$; if $z\leq 0$ the attempt (particle-hole exchange) is finally accepted; if, however, $x>0$, we draw a random number $r_2$ and perform the exchange only if $r_2\leq e^{-x}$.
\item
Go to item 1.
\end{enumerate}
Notice that when $\alpha\rightarrow 0$ only NN particle-hole exchanges are allowed, otherwise there is a non-zero probability for diagonal hops. The limit case of $\alpha=1$ corresponds with the NNN dynamics, i.e., with the NDLG model already introduced in Chapter~\ref{cap1}. Therefore, this model (henceforth referred to as the $\alpha$DLG) can capture both the DLG and NDLG phenomenology by varying the parameter $\alpha$ (the \textit{effective connectivity}). One may also think on $\alpha$ as a \textit{temperature} which controls the diagonal degrees of freedom. For $E=0$ this model reduces to the equilibrium lattice gas\footnote{Although, of course, the equilibrium critical temperature depends on $\alpha$, but not the critical properties. The $\alpha$DLG belongs to the Ising universality class for all $\alpha$.} (see Appendix~\ref{apendA}).

We carried out extensive MC simulations on the $\alpha$DLG by assuming half-filled lattices and periodic boundary conditions (toroidal conditions). Hence, nontrivial nonequilibrium steady state is set in asymptotically. As in the standard DLG, MC simulations reveal that this model undergoes a \textit{temperature-induced} second-order phase transition. This transition is analogous to the studied in Chapters~\ref{cap1} and \ref{cap2}. However, in this appendix we are rather interested in the stability of the interface under saturating fields. Specifically, we study the mechanisms which lead to an ordered or disordered low-temperature states in the large field limit, i.e., to a stable or unstable interface. To this end, we restrict ourselves to the $E\rightarrow \infty$ limit and to $T\rightarrow 0$. The latter limit is the most favorable case for ordering, which will enable us to identify the point in where the field effects outweigh the thermal effects. In addition, this limit is convenient for theoretical and computational purposes since $T$ is no longer a parameter. \\

For $\alpha=0$, we observe fully frozen striped configurations, which extends along the field direction. Here, a one-strip liquid-like (rich-particle) phase percolates along the field direction (the density of the vapor phase is zero). When increasing $\alpha$, i.e. increasing the weight of the diagonal dynamics, some particles are driven out the interface. Nevertheless, if $\alpha$ is not too large, the thermal effects still dominate. In such a case, a liquid--like (high-density) phase which is striped then coexists with its gas (low density phase). With a further increase in $\alpha$, the thermal effects are overweighted by the field effects and the interface becomes unstable. This results in fully disordered configurations (single gas-like phase). The transition between ordered and disordered states are rather sharp, which suggest a phase transition. This is confirmed. We found a \textit{topological} ($\alpha$-induced) extraordinary phase transition, which is found to be continuous (second order). In order to understand this second order phase transition, we introduce an order parameter. One quantity that has been used as a measure of order in the two-dimensional DLG and some related systems \cite{marro} is
\begin{equation}
m=\frac{1}{2\sqrt{\rho (1-\rho)}}\sqrt{\left| \langle M_{\parallel}\rangle-\langle M_{\perp}\rangle \right| }\,,
\label{eq:op}
\end{equation}
where $M_{\parallel}$ and $M_{\perp}$ are squared longitudinal and transverse \textit{magnetizations} given by,
\begin{equation}
M_{\parallel(\perp)}=\frac{1}{L_{\parallel}\times L_{\perp}\times L_{\parallel(\perp)}} \sum_{y(x)}\left[\sum_{x(y)}(1-2\sigma_{x,y}) \right]^2\,,
\label{eq:opmag}
\end{equation}
respectively, where $x$ and $y$ denote the principal lattice directions, and $\rho$ the systems density $\rho=N/(L_{\parallel}L_{\perp})$. The order parameter $m$ is a measure of the difference between the densities of the liquid (strip-like) and gas phases: one has $\langle M_{\parallel}\rangle=\langle M_{\perp}\rangle$ when configurations are fully disordered (as if $T=\infty$), which implies $m=0$, while $\langle M_{\parallel}\rangle\rightarrow 1$ and $\langle M_{\perp}\rangle\rightarrow 0$ in the limit of $T=0$ leading to $m\rightarrow 1$. We computed the order parameter $m$ for different values of $\alpha$. 
\begin{figure}
\centerline{\includegraphics[width=8cm,clip=]{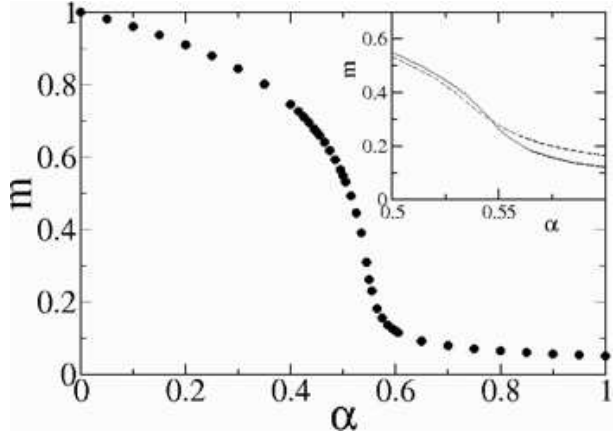}}
\caption[The $\alpha$-dependence of the order parameter $m$ in the $\alpha$DLG]{\label{fig2_C} The $\alpha$-dependence of the order parameter $m$ defined in Eq.~\eqref{eq:op} for half-filled lattices. The main graph shows the results from MC simulations for $L_{\parallel}\times L_{\perp}=256\times 256$. The inset shows a zooming of the transition region for system sizes $L_{\parallel}\times L_{\perp}=128\times 128$ (dashed line) and $L_{\parallel}\times L_{\perp}=256\times 256$ (solid line).}
\end{figure}
Figure~\ref{fig2_C} shows how $m$ varies between these two limits as a function of $\alpha$. As observed in Fig.~\ref{fig2_C}, the transition from a stable interface (high $m$ regime) into a unstable one (low $m$ regime) is rather sharp. As a matter of fact, due to the finite size effects some smoothness is introduced in the critical region (see the inset in Fig.~\ref{fig2_C}). It seems natural to try the power law behavior
\begin{equation}
m\approx A\left| \alpha_{c}-\alpha\right|^{\beta} \qquad \text{as} \qquad \alpha\rightarrow \alpha_{c}\: \text{(from below)}\,.
\label{eq:scaling_m}
\end{equation}
With this aim, we try to identify a value of $\alpha_{c}$ with the familiar Log-Log plots, which yields a linear region (see inset in Fig.~\ref{fig3_C}). The slope near $\alpha_c$ corresponds to $\beta$. Alternatively, one may plot $m$ raised to the power of $1/\beta$ versus $\alpha$ for different trials values of $\beta$, looking for also straight lines. This is also shown in Fig.~\ref{fig3_C}. The latter procedure has the advantage that no guess for $\alpha_c$ that might introduce further errors is involved. Both methods indicate that the data are consistent with Eq.~\eqref{eq:scaling_m}. The estimates obtained from this analysis are $\alpha_c=0.550\pm 0.04$ and $\beta=0.28\pm 0.02$.

\begin{figure}
\centerline{\includegraphics[width=8cm,clip=]{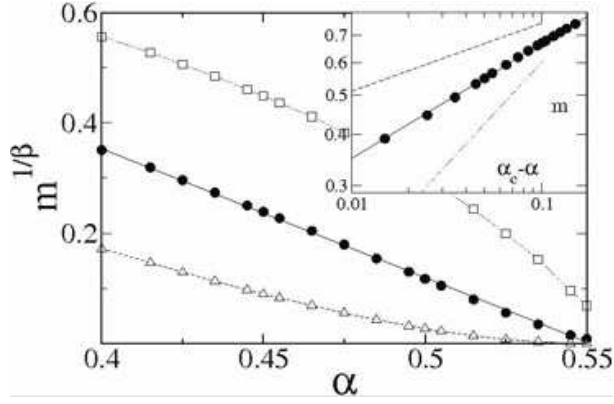}}
\caption[Order parameter critical exponent for the second order phase transition which occurs in the $\alpha$DLG]{\label{fig3_C} The main graph shows plots of $m^{1/\beta}$ for different values of $\beta$: $1/2$ (squares) and $1/6$ (triangles); this gives a the straight line only for $\beta\approx 0.28$ (circles). The inset is a Log-Log plot assuming $\alpha_{c}=0.550\pm 0.004$ which gives $\beta\approx 0.28$. Lines of slope $1/2$ (dotted line) and $1/6$ (dashed line) are also shown.}
\end{figure}

This connectivity-induced (nonequilibrium) phase transition characterizes the balance between thermal and field effects, and describes the stability of the interface in the $\alpha$DLG and related models. That is, for low enough connectivities the $\alpha$DLG is well ordered (the thermal effects dominate, and the interface is stable; this is indeed the case of the standard DLG), whereas for connectivities above the critical point the system exhibits disordered configurations (thermal effects are outweighed by field effects, and therefore the interace is unstable, as the NDLG). This second order (continuous) phase transition indicates that the temperature-field balance is not as trivial as the microscopic dynamic (by the Metropolis rate Eq.~\eqref{eq:rate_alpha}) may suggest, i.e., a mere linear superposition, which in particular would had led to a linear behavior in Fig.~\ref{fig2_C}. On the contrary, this is a clear expression of complexity originated in the cooperative behavior.

\chapter[Mesoscopic equation for the DLG with extended dynamics]{\label{apendB}Detailed derivation of a mesoscopic equation for the Driven Lattice Gas with next-nearest-neighbor Interactions}

In this appendix, we detail the derivation the (mesoscopic) Langevin-type equation for the \textit{Driven Lattice Gas} with next-nearest-neighbor interactions (NDLG) introduced in Chapter~\ref{cap1}. We employ and extend the method introduced in Ref.~\cite{paco} for the \textit{Driven Lattice Gas} with nearest-neighbor couplings (DLG).\\

The original microscopic model consists of a two-dimensional lattice, whose sites have occupation variables $\sigma_{i}=1$ or $0$, for site $i$. These variables evolve following a particle-hole exchange dynamics. Let us define at each point $\mathbf{r}\in \mathbb{Z}^2$ a density variable, $\phi_\mathbf{r}\in \mathbb{R}$, which is the averaged value of the occupation variables $\sigma_i$ in a region of volume $v$ around $\mathbf{r}$. The system evolves from a given configuration $\mathbf{\gamma}$ to another $\mathbf{\gamma}^{\prime}$ by choosing at random a particle at point $\mathbf{r}$ and exchanging it with next nearest neighbor in the $\mathbf{\mu}$ direction at time $t$, namely
\begin{equation}
\phi_{\mathbf{r}}^{\prime}= \phi_{\mathbf{r}}+v^{-1} \, \left( \delta_{\mathbf{r}}-\delta_{\mathbf{r}+\mathbf{\mu}}\right) \,.
\label{eq:conf1_apend}
\end{equation}
When $v$ is large enough, $\phi_{\mathbf{r}}$ is assumed to be a continuous function of $\mathbf{r}$, say $\phi(\mathbf{r},t)$. This represent the coarse grained excess particle density field. So that we have at time $t$, 
\begin{equation}
\mathbf{\gamma}^{\prime}=\left\lbrace \phi(\mathbf{r})+v^{-1} \, \nabla_{\mathbf{\mu}} \delta(\mathbf{r}^{\prime}-\mathbf{r}),\; \text{where} \; \phi \in \mathbf{\gamma}        \right\rbrace \,.
\label{eq:conf2_apend}
\end{equation}
We can also generalized this dynamics to consider exchanges of magnitude $\gamma/v$ with probability amplitude, the latter being an even function of $\eta$ \cite{garrido_eta}, e.g.,
\begin{equation}
f(\eta)=\left( \delta(\eta+1)+\delta(\eta-1)\right) /2\,.
\label{eq:even}
\end{equation}
Or in other words, Eq.~\eqref{eq:even} represents the probability distribution function of a $\eta$, amount of mass attempted to be displaced. The configurations $\eta$ evolves according to a stochastic hopping dynamics which conserves the number of particles, or equivalently the density. That is, configurations have a statistical weight $P(\mathbf{\gamma},t)$, which evolves accordingly to the following continuous \textit{master equation} \cite{vkampen,gardiner}:
\begin{eqnarray}
\partial_{t} P(\mathbf{\gamma},t)=\sum_{\mathbf{\mu}}\int d\eta f(\eta) \int d \mathbf{r} \, \left[ \Omega(\mathbf{\gamma} \rightarrow \mathbf{\gamma}^{\prime})P(\mathbf{\gamma},t)-\Omega(\mathbf{\gamma}^{\prime} \rightarrow \mathbf{\gamma})P(\mathbf{\gamma}^{\prime},t) \right] \,,
\label{eq:master_apen}
\end{eqnarray}
where $\Omega(\mathbf{\gamma} \rightarrow \mathbf{\gamma}^{\prime})$ stands for the transition probability per unit time (transition rate) from $\mathbf{\gamma} $ to $\mathbf{\gamma}^{\prime}$. As usual, the transition rates are given in terms of entropic and energetic contributions as
\begin{equation}
\Omega(\mathbf{\gamma} \rightarrow \mathbf{\gamma}^{\prime})=D(\Delta S(\eta))\cdot D(\Delta H(\eta)+H_{\varepsilon})\,.
\label{eq:ads_rate_apend}
\end{equation}
Here $D$ is a function satisfying the detailed balance constraint, i.e., $D(-x)=e^{x}D(x)$, which ensures that in the zero field limit ($\varepsilon=0$) the stationary distribution is the equilibrium one, $P_{st}\propto e^{-H}$. $\Delta H$ and $\Delta S$ are the increment of energy $H(\mathbf{\gamma}^{\prime})-H(\mathbf{\gamma})$ and entropy $S(\mathbf{\gamma}^{\prime})-S(\mathbf{\gamma})$, respectively. At a mesoscopic level, the equivalent to the microscopic Hamiltonian (Eq.~\eqref{eq:hamilt} in Chapter~\ref{cap1}) is the standard $\phi^{4}$ (Ginzburg--Landau) Hamiltonian. The structure of the free energy consist of two contributions \cite{amit}: entropic and energetic which are given by, respectively
\begin{eqnarray} 
S(\gamma)=v \int  dr \left( \frac{a}{2}\phi^{2} + \frac{g}{4!}\phi^{4}\right) \nonumber\\
H(\gamma)= v \int  dr \left( \frac{1}{2} \left( \nabla \phi \right)^{2}+\frac{\tau}{2}\right)\,,
\label{eq:ee_apend}
\end{eqnarray}
where $a$ and $g$ are entropic coefficients whereas the parameter $\tau$ comes from the energetic functional. Notice that, in Eq.~\eqref{eq:ads_rate_apend} the increment of energy from the drive $H_{\varepsilon}$ enters the dynamics though $D(\Delta H(\eta)+H_{\varepsilon})$, where $H_{\varepsilon}=\eta \, \mathbf{\mu} \cdot \mathbf{\varepsilon} \, (1-\phi^2)+\mathcal{O}(v^{-1})$. This means that the transition rates depend on the energy and entropy difference between configurations plus a term $H_{\varepsilon}$ whose dominant part in $v^{-1}$ is the natural choice to mirror the effects of the drive as far as it accounts for the local increment of energy due to the driving field. Notice also that the detailed dependence of the coarse--grained field $\mathbf{\varepsilon}$ on the microscopic field introduced in Eq.~\eqref{eq:metro} remain still as an open issue \cite{marro}. We also assumed the driving field acting along one of the principal lattice directions, say horizontal.\\ 

The next step is to get a Fokker-Planck equation by expanding the master equation Eq.~\eqref{eq:master_apen} in $v^{-1}$ up to $v^{2}$ order (Kramers--Moyal expansion \cite{vkampen}). To this end, we expand a functional $F$ around $\mathbf{\gamma}$ which read,
\begin{equation}
F(\mathbf{\gamma}^{\prime})=F(\mathbf{\gamma})+\sum_{k=1}^{\infty} \frac{-\eta^{k} v^{-k}}{k!} \left( \nabla_{\mathbf{\mu}}\frac{\delta}{\delta \phi} \right)^{k} F(\mathbf{\gamma}) \,,
\label{eq:expansion}
\end{equation}
where $F$ is a functional representing either $S$, $H$, or $P$. The operator $\frac{\delta}{\delta \phi}$ means the functional derivative of $F(\phi)$ with respect to $\phi$. To be specific, we also choose $H_{\varepsilon}$ as
\begin{equation}
H_{\varepsilon}(\mathbf{\gamma}^{\prime}\rightarrow \mathbf{\gamma}^{\prime})=\eta \lambda_{\mu}^{(\varepsilon)}+\sum_{k=1}^{\infty} \frac{(-1)^{k}\eta^{k+1}v^{-k}}{(k+1)!}\left( \nabla_{\mathbf{\mu}}\frac{\delta}{\delta \phi}\right)^{k} \lambda_{\mu}^{(\varepsilon)}\,,
\label{eq:he}
\end{equation}
with $\lambda_{\mu}^{(\varepsilon)}=\mathbf{\mu}\cdot \mathbf{\varepsilon} \left( 1-\phi(\mathbf{r})^{2}\right) $. These expansions yield the following Fokker-Planck equation \cite{paco}.
\begin{eqnarray}
\partial_{t} P_t=\sum_{\mathbf{\mu}}\int d \mathbf{r} \, \left( \nabla_{\mathbf{\mu}}\frac{\delta}{\delta \phi}\right) \times \nonumber\\ 
\left[ v^{-1}p(\lambda_{\mu}^{S},\lambda_{\mu}^{H}+\lambda_{\mu}^{\varepsilon})P_t+ \frac{1}{2v^2}q(\lambda_{\mu}^{S},\lambda_{\mu}^{H}+\lambda_{\mu}^{\varepsilon}) \left(  \nabla_{\mathbf{\mu}}\frac{\delta}{\delta \phi} \right) P_t \right]  \,,
\label{eq:fokker}
\end{eqnarray}
where
\begin{eqnarray}
P_t\equiv P(\mathbf{\gamma},t), \nonumber\\
\lambda_{\mu}^{X}\equiv-\left( \nabla_{\mu} \frac{\delta}{\delta \phi} \right) X, \nonumber\\
p(z_{1},z_{2})\equiv\int d\eta\; \eta \,f(\eta) \,D(\eta z_{1})\,D(\eta z_{2}), \nonumber\\
q(z_{1},z_{2})\equiv\int d\eta\; \eta^{2}\, f(\eta)\, D(\eta z_{1})\,D(\eta z_{2}).
\label{ea:proto-fL_apend}
\end{eqnarray}
After using standard techniques in the theory of stochastic processes \cite{vkampen} we derive from the Fokker--Planck equation its stochastically equivalent Langevin equation using the Ito prescription, which reads
\begin{equation}
\partial_t \phi(\mathbf{r},t)= \sum_{\mu} \nabla_{\mu} \left[ p(\lambda_{\mu}^{S},\lambda_{\mu}^{H}+\lambda_{\mu}^{\varepsilon})+q(\lambda_{\mu}^{S},\lambda_{\mu}^{H}+\lambda_{\mu}^{\varepsilon})^{1/2}\,\xi_{\mu}(r,t)  \right].
\label{eq:proto-L_apend}
\end{equation}
The time has been rescaled by $v^{-1}$ and $\xi_{\mu}(\mathbf{r},t)$ is a delta-correlated Gaussian white noise, i.e., $\langle \xi_{\mu}(\mathbf{r},t) \rangle_t=0$ and $\langle \xi_{\mu}(\mathbf{r},t)\,\xi_{\mu^{\prime}}(\mathbf{r}^{\prime},t^{\prime}) \rangle=\delta_{\mu \mu^{\prime}} \delta(t-t^{\prime}) \delta(\mathbf{r}-\mathbf{r}^{\prime})$.

We focus on the critical region where large fluctuations on all length scales dominate. Further simplification in Eq.~\eqref{eq:proto-L_apend} is possible in this regime by dropping the irrelevant terms in the renormalization group sense. Following the standard field theoretic methods let us introduce an external momentum scale $\tau$ and make the following anisotropic scale transformations
\begin{equation}
\begin{split}
t\rightarrow \tau^{-z} t\,,\\
r_{\perp}\rightarrow \tau^{-1}r_{\perp}\,,\\
r_{\parallel}\rightarrow \tau^{-s}r_{\parallel}\,,\\
\phi \rightarrow \tau^{\delta}\phi\,.
\end{split}
\label{eq:adsscale_apend}
\end{equation}
Next, we expand the Langevin equation in powers of $\tau$ around $\tau=0$, keaping only the leading terms. For next-nearest-neighbor interactions, the sum $\sum_{\mu}$ involves a sum for each possible direction, namely $\sum_{\parallel}\sum_{\perp}\sum_{\diagup}\sum_{\diagdown}$ which are respectively the parallel, transverse, and the two diagonals directions. After some cumbersome algebra, one can split the terms depending on both diagonal ($\diagup$ and $\diagdown$) into the two parallel and transversal to the driving field $\varepsilon$ components. The time scale, the transverse noise, and the transverse spatial interaction are forced to remain invariant under transformation. With this, one gets $z=4$ and $\delta=(s+d-3)/2$. Different scenarios are now possible depending on the value of $s$. Demanding that the most relevant terms in the parallel and transverse to the field scale in the same way, as in the standard analysis of the critical behavior of the NDLG (see subsection~\ref{cap1:sims}), would lead to $s=2$. According to this, the Langevin equation for finite (coarse--grained) driving field contains a large bunch of $\varepsilon$-dependent terms. Similarly to what occurs in Ref.~\cite{paco}, just taking a large enough value of the driving field most of those terms vanish. By doing this, one is led to the following Langevin equation
\begin{equation}
{\partial}_t \phi(\mathbf{r},t)=-\nabla_{\bot}^4 \phi+(\tau+\frac{3}{2}a)\nabla_{\bot}^2 \phi + \frac{g}{4}\nabla_{\bot}^2\phi^3 +a \nabla_{\Vert}^2 \phi+\xi(\mathbf{r},t)
\label{eq:adsNNN_apend}
\end{equation}
The resulting Langevin equation is analogous to the one for nearest-neighbor interactions (see Eq.~\eqref{eq:adsNN} in Chapter~\ref{cap1}), although Eq.~\eqref{eq:adsNNN_apend} contains additional entropic, not energetic, contributions due to the presence of larger number of bonds than in the NN case.

\chapter[Hydrodynamics for three-dimensional granular gases]{\label{apendD}Hydrodynamic description for a three-dimensional granular gas: Clustering and symmetry breaking}

\begin{figure}
\centerline{\includegraphics[width=9cm,clip=]{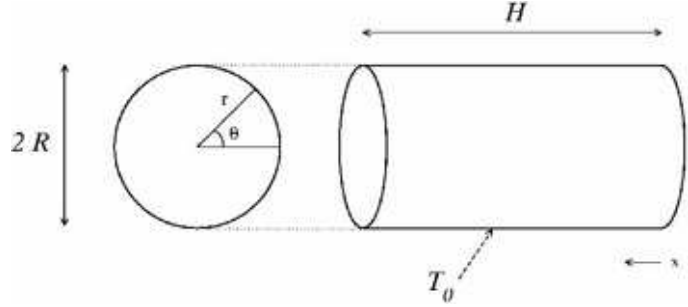}}
\caption[Perspective sketch of the three-dimensional experimental \textit{set-up}]{\label{fig1_D} A perspective sketch of a cylinder of volume $\pi H R^{2}$. The curved surface is thermalized to temperature $T_{0}$ whereas the remaining two surfaces are considered elastic. Definition of the cylindrical coordinates, which we will employ along this appendix, are also indicated.}
\end{figure}

A straightforward extension to three dimensions of the circular system studied in Chapter~\ref{cap4} is a \textit{cylinder} in which the curved wall is maintained at a constant temperature and the other two are elastic (see Fig.~\ref{fig1_D}). Particles of diameter $d$ move inside the cylinder underging inelastic collisions with a constant restitution coefficient $\mu$. In this appendix we study the clustering and symmetry-breaking instabilities employing granular hydrodynamics in such a geometry. To this end, we generalize the Grossman \textit{et al.} \cite{grossman} constitutive relations to three-dimensional situations.\\

\textit{Constitutive Relations}. Before going any further, we derive heuristic constitutive relations for inelastic gases in three dimensions by employing free volume arguments in the vicinity of the close packing, suggesting an interpolation between the close-packing limit and the usual dilute-limit relations. Here, we follow the strategy taken by Grossman \textit{et al.} \cite{grossman} in two dimensions. 

Let us consider our three-dimensional system, in which there is a temperature gradient in the radial direction because of the thermalized wall (set at temperature $T_{0}$). As a result, there will be an energy flux along this direction. Consequently ---assuming both azimuthal and longitudinal symmetry, i.e, the state do not depend neither $\theta$ nor $x$---, the energy balance equation, which is derived from the hydrostatic version of Eqs.~\eqref{eq:hydrodyn}, reduces to 
\begin{equation}
\frac{1}{r}\frac{d}{dr}\left( r \kappa \frac{dT}{dr}\right) =I\,,
\label{eq:apend_flux}
\end{equation}
where $I$ is the sink term (mean energy lost per unit volume per unit time) and $\kappa$ is the coefficient of thermal diffusivity. Constitutive relations for $I$ and $\kappa$ are estimated heuristically following the approach devised in Ref.~\cite{grossman}, yielding the same expressions as in Eq.~\eqref{state}. These are
\begin{equation}
\begin{split}
\kappa=\frac{\sigma\,n\,(\alpha l + d)^2 T^{1/2}}{l} \:,\: \text{and}\\
I= \frac{\sigma}{\gamma l}\, (1-\mu^2) \,n \, T^{3/2}\,,
\end{split}
\label{eq:state_apend}
\end{equation}
where $\alpha$ and $\gamma$ are numerical factors of the order unity. The effect of the system dimensionality enters only in $\sigma$ and $\gamma$, which will differ from the two-dimensional case. Concerning the equation of state in the low density limit, as is well-known, the ideal gas law holds, $P=nT$. Contrary, in the high density limit, the mean free path $l$ is much less than the particle diameter $d$, that is, $l\ll d$. Thus, one finds
\begin{equation}
\frac{n}{n_{c}}=\frac{d^{3}}{(d+l)^{3}}\approx 1-\frac{3l}{d}
\label{eq:apend_hd}
\end{equation}
where $n_{c}=\sqrt{2}/d^{3}$ is the close packing value in three-dimensions \cite{sloane}. In this limit \cite{grossman} the entropy per particle is $S\sim ln(l^{3})+g(T)$, with $g$ an arbitrary function of $T$. Therefore, employing Eq.~\eqref{eq:apend_hd} we obtain the pressure in the limit $n\rightarrow n_{c}$,
\begin{equation}
P=\frac{3n^{2}T}{n_{c}-n}
\label{eq:hdeos_apend}
\end{equation}
We therefore propose the following interpolation formula for the pressure
\begin{equation}
P=nT\frac{n_{c}+2n}{n_{c}-n}\,.
\label{eq:eos_apend}
\end{equation}
The mean free path can be also expressed in terms of the density and temperature. In the dilute limit one has (in three dimensions) $l=1/\left( \pi \sqrt{2}nd^{2}\right) $, while from Eq.~\eqref{eq:apend_hd} one has in the high density limit
\begin{equation}
l\approx \frac{n_{c}-n}{3n_{c}}d\,.
\label{eq:hdl_apend}
\end{equation}
Again, by using these limits to interpolate a global expression for the mean free path, we find
\begin{equation}
l= \frac{1}{\pi \sqrt{2}nd^{2}} \frac{n_{c}-n}{n_{c}-an}\,,
\label{eq:meanfreepath_apend}
\end{equation}
where $a=1-\frac{3}{2\pi}$.\\

\begin{figure}
\centerline{\includegraphics[width=8cm,clip=]{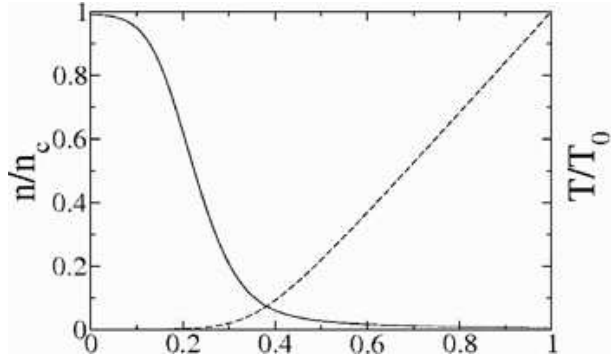}}
\caption[Scaled temperature and density profiles as predicted by granular hydrostatic theory in three dimensions]{\label{fig2_D} Scaled temperature (dashed line) and density (solid line) profiles predicted by the hydrostatic theory for $\eta=10^{4}$ and $f=0.036$.}
\end{figure}

\textit{The Density Equation}. Substituting Eqs.~\eqref{eq:state_apend} into Eq.~\eqref{eq:apend_flux}, and eliminating the temperature dependence with Eqs.~\eqref{eq:eos_apend} and \eqref{eq:meanfreepath_apend} we arrive to a second order differential equation for $n$:
\begin{equation}
\frac{1}{r}\frac{d}{dr}\left[ r \mathcal{F}(z)\frac{dz}{dr}\right] =\eta \mathcal{Q}(z)\,.
\label{eq:density_eq_apend}
\end{equation}
This is \textit{the density equation}. For convenience, the radial coordinate has been rescaled by $R$ and a normalized inverse density $z(r)=n_{c}/n$ is introduced. The functions $\mathcal{F}$ and $\mathcal{Q}$ read
\begin{equation}
  \begin{split}
\mathcal{F}(z)=\frac{(z^2+4z-2)\left[\alpha z(z-1)+2\pi(z-a)\right]^2}
{(z-a)(z-1)^{1/2}z^{3/2}(z+2)^{5/2}}\\
\mathcal{Q}(z)=\frac{(z-a) (z-1)^{1/2}}{(z+2)^{3/2} z^{1/2}}\,,
  \end{split}
\label{eq:FQ_apend}
\end{equation}
and $\eta=\frac{4\pi^{2}}{\gamma}(1-\mu^{2})\left( \frac{R}{d}\right)^{2}$. The fixed total number of particles $N$ yields a normalization condition, which read
\begin{equation}
\int_{0}^{1} \, z^{-1}(r)\,r \,dr = f/2\,,
\label{eq:normalization_apend}
\end{equation}
where $f=\frac{N}{\pi\sqrt{2}\Delta}\left( \frac{d}{R}\right)^{3} $ is the average grain area fraction, and $\Delta=H/R$ the system aspect ratio. Equations~\eqref{eq:density_eq_apend} and \eqref{eq:normalization_apend}, together with the boundary conditions $dz(0)/dr= 0$ and $z(1)=\textit{const}$ form a complete set. The hydrostatic problem is completely determined by three scaled parameters $\eta$, $f$, and $\Delta$. Notice that the definitions of $\mathcal{F}$, $\mathcal{Q}$, $\eta$, and $f$ differ from the two-dimensional case (see Eqs.~\eqref{Q}--\eqref{normalization}). Figure~\ref{fig2_D} shows the numerical solution of Eq.~\eqref{eq:density_eq_apend}. This is actually a one-dimensional solution, which corresponds with the \textit{azimuthally- and longitudinally}-symmetric state. Once the density $z(r)=n_{c}/n(r)$ is obtained, the temperature profile can be determined from the equation of state (Eq.~\eqref{eq:eos_apend}).\\

\begin{figure}
\centerline{\includegraphics[width=8cm,clip=]{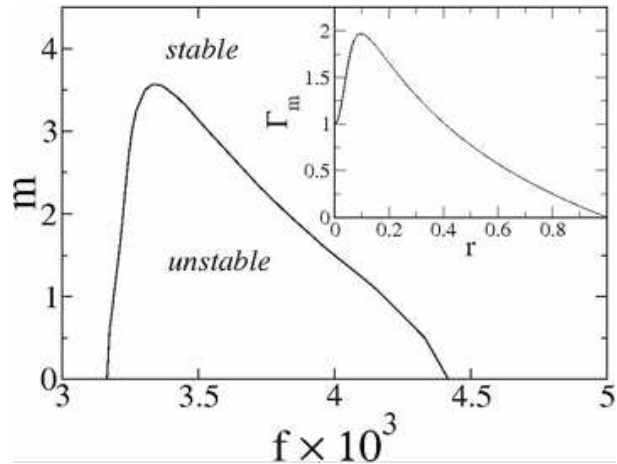}}
\caption[Marginal stability curve $m=m(f)$ (main graph) and eigenfunctions $\Gamma_{m}(r)$ (inset)]{\label{fig3_D} Main graph: Marginal stability curve $m=m(f)$ for $\eta=10^{5}$. The symmetry-breaking instability develops in the parameter region above this curve. Inset: Eigenfunction $\Gamma_{m}(r)$ corresponding to the eigenvalue $m=0.585$ for $\eta=10^{5}$ and $f=2.16\cdot 10^{-3}$.}
\end{figure}

\textit{Marginal Stability Analysis}. In general, Eq.~\eqref{eq:density_eq_apend} provides only one of the possible solutions ---actually, a one-dimensional solution---. However, as we will demonstrate, this system exhibits symmetry-breaking instability\footnote{This is in contrast with its two-dimensional counterpart, previously studied in Chapter~\ref{cap4}.}. That is, there are truly two-dimensional solutions in which the longitudinal symmetry along $x$ is broken. These solutions can be found by linearizing Eq.~\eqref{eq:density_eq_apend} around the one-dimensional solution given by Eqs.~\eqref{eq:density_eq_apend}--\eqref{eq:normalization_apend} ---a similar analysis was performed in Chapters~\ref{cap3} and \ref{cap4}---. In general, one can rewrite the energy balance equation in terms of the three coordinates $r$, $\theta$, and $x$:
\begin{equation}
\nabla \cdot \left( \mathcal{F}(r) \nabla z\right) =\eta \mathcal{Q}(z)\,,
\label{eq:general_apend}
\end{equation}
with $\mathcal{F}$ and $\mathcal{Q}$ given by Eq.~\eqref{eq:FQ_apend}. Substituting $z(r,\theta,x)\sim z(r)+\varepsilon \omega(r,\theta,x)$, assuming $\omega=\Xi(r)\Pi(\theta)\Phi(x)$, and linearizing Eq.~\eqref{eq:general_apend} with respect to the small correction $\varepsilon$, one obtains:
\begin{equation}
\begin{split}
\Phi(x)=A\sin(mx)+B\cos(mx)\,,\; x\in[0,\Delta]\\
\Pi(\theta)=C\sin(k\theta)+D\cos(k\theta)\,,\; \theta\in [0,2\pi)\\
\Gamma^{\prime\prime}+\frac{1}{r}\Gamma^{\prime}-\left( m^{2}+\frac{k^{2}}{r}+\frac{\eta \mathcal{Q}^{\prime}}{\mathcal{F}}\right) \Gamma=0\,,\; r\in[0,1]\,,
\end{split}
\label{eq:relations_apend}
\end{equation}
where $A$,$B$ $C$, and $D$ are constants, and $m$ and $k$ are, respectively, the longitudinal and azimuthal wave number determined by the boundary conditions. Here, $\Gamma(r)\equiv \Xi(r) \mathcal{F}$, $\mathcal{Q}^{\prime}$ denotes the $z$ derivative of $\mathcal{Q}$, and functions $\mathcal{F}$ and $\mathcal{Q}$ are evaluated at $z=z(r)$. For fixed values of $\eta$ and $f$, Eq.~\eqref{eq:relations_apend} and the boundary conditions
\begin{equation}
\Phi^{\prime}(0)=\Phi^{\prime}(\Delta)=0
\label{eq:bound_apend}
\end{equation}
represent a linear eigenvalue problem for $k$ and $m$. We focus here only in the longitudinal wave number $m$, so that $m=j\pi/\Delta$, where $j\in \left\lbrace 0, \mathbb{N}\right\rbrace $ because of the boundary conditions Eq.~\eqref{eq:bound_apend}. That is, we solve numerically the eigenvalue problem for $k=0$. If there are nontrivial solutions for the lowest mode $k=0$ then there will be solutions for higher modes. Moreover, in Chapter~\ref{cap4} we proved the stability of the azimuthally-symmetric state ---against linear perturbations---, hence one expects that the azimuthal wave number $k$ does not play a relevant role here. The symmetry breaking instability along the longitudinal (not azimuthal) direction is confirmed. Figure~\ref{fig3_D} shows the marginal stability curve $m=m(f)$ found numerically. For a fixed value of $\eta$ the azimuthally- and longitudinally-symmetric state is \textit{unstable} (\textit{stable}) for any $m$ below (above) the marginal stability curve. Interestingly, the symmetric state remains stable for any $m$ beyond a finite interval of $f\in(f_{\text{min}},f_{\text{max}})$. Or in other words, for each aspect ratio $\Delta>\Delta_{\text{crit}}=\pi j/m$, where $j=0,1,2,\ldots$, the symmetric state loses stability with respect the longitudinal mode $j$ \cite{livne1}.

Therefore, granular hydrodynamics, employing our heuristic constitutive relations ---an extension of the Grossman \textit{et al.} approach to three-dimensional cases---, may be a good candidate tfor describing the clustering and symmetry-breaking instabilities which may occur in inelastic hard-sphere systems. 

\backmatter
\part{\label{part:Concl}Conclusions}
\chapter{\label{capConclusion}Conclusions and future work}
\thispagestyle{empty}
Unfortunately, despite substantial research, a rigorous statistical-mechan-ical description of nonequilibrium instabilities remains elusive. The work presented here delves into the nonequilibrium realm, seeking a better understanding of the basic features of nonequilibrium phase transitions. Indeed, the study of \textit{driven diffusive fluids} and \textit{driven granular gases}, besides its technological importance, is of general interest because it will contribute to the rationalization of nonequilibrium phenomena. The emphasis in our research is on the study ---in most cases at different levels of description, namely, macroscopic, mesoscopic, and microscopic--- of the instabilities which occur in those systems. This task was approached with both state-of-the-art and novel theoretical as well as computational techniques in statistical physics.\\

This thesis was structured in four parts. The first Part, comprised of Chapters~\ref{intro:DDF}, \ref{cap1}, and \ref{cap2}, was devoted to the early--time kinetic and steady-state properties of \textit{driven diffusive fluids}; Part~\ref{part:Granos} (Chapters ~\ref{intro:Granos}, \ref{cap3}, and \ref{cap4}) dealt with both the atomistic and macroscopic description of two-dimensional driven \textit{inelastic} gases; Part~\ref{part:apend} provided the appendices (Appendices ~\ref{apendA}, \ref{apendC}, \ref{apendB}, and \ref{apendD}). Finally, Part~\ref{part:Concl} summarizes the main results and the original contributions presented in this work. We also offer an outlook for principal issues to be investigated in future work.\\

\textit{Chapter~\ref{cap0}.} In the first chapter, we reviewed the general features, phenomenology, and open issues of the class of systems which we are dealing with in the text from a mechanical-statistical perspective. Likewise, we discuss the subject of this thesis as well as the main rationale behind it.\\

\textit{Chapter~\ref{intro:DDF}.} In this chapter, we presented the model ---i.e., the \textit{driven lattice gas} (DLG)--- which was the prototype for \textit{driven diffusive fluids}, and we summarized some of its already known unusual properties and controversies. The relevant background for Chapters~\ref{cap1} and \ref{cap2} is also provided. This comprises state-of-the-art simulation schemes (lattice and off-lattice Metropolis \textit{Monte Carlo}) as well as field--theoretical approaches (statistical field theories).\\

\textit{Chapter~\ref{cap1}.} We have described Monte Carlo simulations and field theoretical calculations that aim at illustrating how slight modifications of dynamics at the microscopic level may influence, even quantitatively, both the resulting nonequilibrium steady state and the kinetic properties. With this aim, we took as a reference the DLG. We introduced related lattice and off-lattice microscopic models in which particles, as in the DLG, interact via a local anisotropic rule. The rule induces preferential hopping along one direction, so that a net current sets in if allowed by boundary conditions. In particular, we have discussed on the similarities and differences between the DLG and its continuous counterpart, namely, a Lennard--Jones analogue in which the particles' coordinates vary continuously. A comparison between the two models allowed us to discuss some exceptional, hardly realistic features of the original discrete system ---which has been considered a prototype for nonequilibrium anisotropic phase transitions (although we claim here is not justified). We outline the main conclusions as follows:
\begin{itemize}
\item
We found that the continuous, off-lattice model closely resembles the DLG in that both depict a particle current, a highly anisotropic liquid-vapor interface which extends along the field direction, and the corresponding second order phase transition. 
\item
However, they differ in some essential features. Contrary to the DLG, its off-lattice counterpart shows a transition temperature which decreases with the field strength. Moreover, unlike for the DLG, there is no phase transition for a large enough field. Concerning the early process of kinetic ordering, early--time anisotropies in the off--lattice case point along the field, contrary to the ones observed in the discrete DLG.
\item
These observations suggest that the DLG behavior is highly conditioned by the lattice geometry, which acts more efficiently in the DLG as an ordering agent than the field itself. Therefore, the striking, counterintuitive feature of the DLG that a strong field raises the critical temperature above the equilibrium one is a \textit{geometrical} effect ---induced by the lattice topology--- rather than \textit{dynamic}.
\item
In particular, it ensues that, due to its uniqueness, the DLG does not have a simple off-lattice analog. 
\item
In order to deepen in this situation, we studied the DLG with an infinite drive extending hops and interaction to next--nearest--neighbors (this model was named NDLG). We confirmed that the NDLG behaves closer to the off-lattice case: the critical temperature decreases with the field and early--time anisotropies point along the field.
\item
However, we found that allowing jumping along intermediate directions leaves invariant both spatial correlations and criticality. The former was monitorized by the two-point correlation function and the equal-time structure factor, whereas the latter was determined by standard finite size scaling, and corroborated by the \textit{anisotropic driven system} (ADS) (mesoscopic) approach.
\item
Therefore, allowing jumping along intermediate directions modifies essentially the phase diagram ---which is not determined by bare symmetry arguments--- but not features (such as generic power--laws in spatial correlations) that seem intrinsic of the nonequilibrium nature of the phenomenon. 
\end{itemize}
These conclusions are properly complemented by the results in the Chapter~\ref{cap2} on criticality in a related off-lattice model. Moreover, we propose that the fact that particles are constrained to travel along discrete lattice directions may condition not only the properties of the DLG, but also the properties of many other lattice models. Further research, under way at present, confirms this on the off-lattice versions of the \textit{Time Asymmetric Exclusion Process}\cite{liggett,antal} (TASEP), and of the DLG with nearest neighbor \textit{repulsion} \cite{szabo,dickman}. Further issues include the characterization of the stability of the interface under strong fields (see Appendix~\ref{apendC}), and the study of the early--time kinetic of phase separation in the NDLG from the field--theoretical standpoint, by using the successful ADS approach. \\
%





\textit{Chapter~\ref{cap2}.} Since the DLG was shown in Chapter~\ref{cap1} to be unrealistic in some essential sense, we introduced here a novel, \textit{realistic} off--lattice driven fluid, which is a candidate to portray some of the anisotropic behavior in nature. This is a nonequilibrium Lennard-Jones fluid with an external constant driving field, settled on a thermal bath. We studied short--time kinetic and steady--state properties of its non--equilibrium phases, namely, solid (mono- and polycrystalline), liquid and gas anisotropic phases. Specifically, we described the early--time segregation process as monitored by the excess energy, which measures the droplets surface; structural properties of the steady state, namely, the radial and azimuthal distribution functions, and the degree of anisotropy; transport properties; and also an accurate estimate of the liquid--vapor coexistence curve and the associated critical indeces. The principal conclusions are:
\begin{itemize}
\item
This model seems to contain the necessary essential physics to be useful as a prototypical model for anisotropic behavior in nature.
\item
This case is more convenient for computational purposes, than others such as, for instance, standard molecular--dynamics realizations of driven fluid systems. 
\item
This is the natural extension to nonequilibrium anisotropic cases of the familiar Lennard-Jones fluid, which has played an important role in analysing equilibrium fluids. Indeed, our model reduces to the (equilibrium) Lennard-Jones fluid for zero field. Otherwise, it exhibits a net current, and striped structures below a critical point.
\item
Concerning kinetics properties, in spite of the anisotropy of the late--time \textquotedblleft spinodal decomposition\textquotedblright\ process, earlier nucleation seems to proceed by \textit{Smoluchowski coagulation} and \textit{Ostwald ripening}, which are known to account for nucleation in equilibrium, isotropic lattice systems and actual fluids. 
\item
Unexpectedly, we have also found that the model critical behavior is consistent with the Ising, equilibrium one but not with the one for the DLG. The main reason for this and other disagreements discussed in this chapter might be the particle--hole symmetry violation in the driven Lennard--Jones fluid.
\item
In addition, regarding the modeling of complex systems, one may infer that spatial discretization may change significantly not only morphological and early--time kinetics properties, but also critical properties. This is in stark contrast with the concept of universality in equilibrium systems, where critical properties are independent of dynamic details.
\item
Finally, we believe this model will motivate new experiments on anisotropic spinodal decomposition besides theoretical work. In fact, our observations on the early-time separation process and structural properties are easily accessible by micro calorimetric and spectroscopy experiments, for instance.
\end{itemize}
Many other issues remain. Further research, currently in progress, involves detailed studies of its interfacial properties, the role of the finite size effects, and sheared versions of this driven Lennard--Jones fluid. Other questions which we would like to raise for future work concern how the particle-hole symmetry and the shape of the interaction potential influence the critical properties. A detailed analysis of the late--time segregation process, which has already been studied both for equilibrium \cite{marro2,bray} and non--equilibrium cases, including the DLG \cite{hurtado,levine}, will be the subject of future investigation. Also useful is the detailed study of the structure and morphological features of the (monocrystalline) solid and glass--like (polycrystalline solid) phases. The fact that our off-lattice model and a related driven-diffusive model \cite{szabo} exhibit a critical behavior consistent with Ising is rather remarkable, and merits further investigations.\\ 

\textit{Chapter~\ref{intro:Granos}.} In this chapter, we have provided the framework relevant to Chapters~\ref{cap3} and \ref{cap4}. We discussed some unresolved puzzles and recent developments in granular fluids, in particular, on the applicability of granular hydrodynamics to granular flows. We also presented the basic ingredients of the studied models and the computer simulation methods (event-driven molecular dynamics).\\

\textit{Chapter~\ref{cap3}.} We have described hydrodynamic derivations as well as event-driven molecular dynamics simulations in a monodisperse granular gas confined in an annulus, the inner circle of
which represents a ``thermal wall". The quasi-elastic limit was considered, and the granular hydrodynamics with Enskog-type transport coefficients \cite{jenkins} was employed. We performed a comprehensive study of the clustering, symmetry breaking and phase separation instabilities, which enable us to establish a detailed phase diagram. To test and complement our theoretical predictions, we performed event-driven molecular dynamics simulations of this system. This led us to discuss the physics of the instabilities which occur in this model. Here we itemize the main conclusions:
\begin{itemize}
\item
The zero-flow regime was completely determined by three scaled parameters: the grain area fraction, the inelastic heat loss parameter, and the aspect ratio of the annulus. 
\item
In our event-driven molecular dynamics simulations, azimuthally symmetric steady states were observed far from the driving wall. In spite of the small number of particles employed, these \textit{annular states} were accurately described by the numerical solutions of our granular hydrostatic equations. 
\item
A marginal stability analysis yielded a region of the (three-dimensional) parameter space where the annular state ---the basic, azimuthally symmetric steady state of the system---  is unstable with respect to small perturbations which break the azimuthal symmetry. We computed the marginal stability curves and compared them to the borders of the spinodal (negative compressibility) interval of the system. 
\item
We found that, the physical mechanism of the phase separation \textit{instability} is the negative compressibility of the granular gas in the azimuthal direction, caused by the inelastic energy loss. Mathematically, phase separation manifests itself in the existence of \textit{additional} solutions to the \textit{density equation} in some region of the parameter space.
\item
Our event-driven simulations of this system also showed phase separation, but it is masked by large spatio-temporal
fluctuations. By measuring the probability distribution of the amplitude of the
fundamental Fourier mode of the azimuthal spectrum of the particle density we
were able to clearly identify the transition to phase separated states in the simulations. 
\item
We found that the instability region of the parameter space,
predicted from hydrostatics, is located within the phase separation region
observed in our simulations. This implies the presence of a binodal 
(coexistence) region, where the annular state is \textit{metastable}. 
\item
Clustering, symmetry breaking, and phase separation instabilities should be observable in actual experiments. In fact, the model we proposed is experimentally accessible, and, therefore, if brought to the attention of the appropriate community, might stimulate experimental work on these instabilities.
\end{itemize}
Of course, it would be interesting to test the predictions of our theory/simulations in experiment. By focusing on the annular geometry, we hope to motivate experimental studies of these granular instabilities which may be
advantageous in this geometry. The annular setting avoids lateral side walls
(with an unnecessary/unaccounted for energy loss of the particles). A possible experiment can employ metallic spheres rolling on a slightly concave smooth surface and driven by a rapidly vibrating (slightly eccentric and possibly rough) interior circle. While particle rotation and rolling friction may become important, we expect that the main predictions of the theory will persist. It would be also interesting to determine the nature (sub- or supercritical?) of the solutions which bifurcate from the azimuthally symmetric state. Further studies may exploit the similarities between the phenomenology observed in our model system and the one in planetary rings, in which clustering, spontaneous symmetry breaking, oscillations, and many other instabilities occur.\\

\textit{Chapter~\ref{cap4}.} In this chapter we addressed granular hydrodynamics and fluctuations in a simple two--dimensional granular system under conditions when first-principle hydrodynamic descriptions break down because of large density, \textit{not} large inelasticity. We put it into a test ---using event-driven molecular dynamics simulations--- in an extreme case when granular clusters with the maximum density close to the hexagonal close packing form (we defined them as \textit{macro--particles}). The model system we considered here, comprises inelastically colliding hard disks enclosed by circular boundaries, driven by a thermal wall at zero gravity. The main conclusions are summarized as follows:
\begin{itemize}
\item
Molecular dynamics simulations showed a sharp--edged close--packed cluster (macroparticle) of an almost circular shape, weakly fluctuating in space and isolated from the thermal wall by a low-density gas. This is due to collisional cooling. The macroparticle formation requires a certain number of particles, and, as the particle density is increased, the transition from a (quasi-) homogeneous state to a macroparticle is rather sharp.  
\item
We were able to solve numerically a set of granular hydrostatic equations which employ the constitutive relations by Grossman \textit{et al.} \cite{grossman}. We found that the density profiles, in a wide range of parameters, agrees almost perfectly ---including the close-packed part--- with the azimuthally symmetric solution of granular hydrostatic equations. We also showed that, for the same setting, first-principle Enskog-type relations \cite{jenkins} perform poorly.
\item
This agrees with previous results in rectangular geometry, with and without gravity, and extends the validity of the Grossman \textit{et al.} relations for a slightly lower restitution coefficient, and for the circular geometry.
\item
A marginal stability analysis have shown ---independently of the constitutive relations employed--- that there are no steady--state solutions with broken azimuthal symmetry which would bifurcate from the azimuthally symmetric solution. This agrees with the persistence of circular shapes for the macroparticle as observed in our simulations. 
\item
We addressed, for the first time, fluctuations of the macroparticles, by measuring the radial probability distribution function of the center of mass of the system. These fluctuations turn out to be Gaussian.
\item
The observed Gaussian fluctuations of the center of mass suggest an effective Langevin description in terms of a macroparticle in a confining potential of hydrodynamic nature, driven by discrete particle noise ---i.e., the cluster performs a Brownian motion inside the system. The associated Langevin equation is provided.
\item
In stark contrast with equilibrium systems, in which the relative magnitude of fluctuations decreases with increasing the number of particles\footnote{For an ideal gas in equilibrium: $\sigma \sim {\cal O}(N^{-1/2})$, where $\sigma$ is the relative magnitude of fluctuations and $N$ is the number of particles.}, we found that the fluctuations persist as the number of particle in the system is increased. 
\end{itemize}
There are still many open issues. A question that should be addressed is whether are there additional instabilities in this model system. For instance, an important insight, in analogy with the KTHNY theory \cite{melting} of two-dimensional melting (for fluids at thermal equilibrium), might be given by the positional and orientational ordering of the macroparticles at different number densities. This can be achieved by measuring their respective correlation functions. In addition, both a general solution of the time-dependent hydrodynamic equations and a non-linear stability analysis will contribute to the quest. Other future issue may involve the detailed study of instabilities in related three-dimensional systems. An outline can be found in Appendix~\ref{apendD}. It should be also worthy to study how polydisperse granulates behave in this system.\\

Despite that the appendices are commonly pushed into the background, we consider that our Appendices~\ref{apendC} and \ref{apendD} to merit their own conclusions.\\

\textit{Appendix~\ref{apendC}.} In this appendix, we introduced a driven-diffusive lattice model (the $\alpha$DLG) intended to aid in understanding the balance between thermal and field effects which occur in these models, such as the DLG and related models (specifically the models studied in Chapter~\ref{cap1}). This balance can be understood in terms of stability of the interface between the condensed and vapor phases (which are elongated along the field direction). Thermal effects favor the stability (the order) of the interface, whereas the driving effects would drive particles out of the interface, favoring disorder. We studied this \textit{dynamic} balance in the extreme case of the saturating field, zero-temperature limit. The main results are:
\begin{itemize}
\item
Our model undergoes a second-order phase transition employing as an order parameter the measure of the \textit{ordering on the lattice} as a function of the \textit{effective connectivity}. That is, for low enough connectivities the system is well ordered (the thermal effects dominate, and the interface is stable), whereas for connectivities above the critical point $\alpha_c$ the system exhibits disordered configurations (thermal effects are outweighed by the field effects, and therefore the interace is unstable).
\item
We computed the critical indexes associated with the transition.
\item
This second order (continuous) phase transition indicates that the tempera-ture-field balance is not as trivial as the microscopic dynamic may suggest, i.e., a mere linear superposition \cite{metropolis}. On the contrary, this is a clear expression of complexity originated in the cooperative behavior.
\end{itemize}

\textit{Appendix~\ref{apendD}.} 
The last appendix addressed clustering and symmetry-breaking instabilities in a monodisperse gas of inelastic hard spheres confined in a cylinder, the curved wall of which represents a ``thermal wall". We employed a set of granular hydrodynamic equations with heuristic constitutive relations, which we derived from free-volume arguments. This model is a simple extension for three-dimensional cases of the circular system studied in Chapter~\ref{cap4}. 
\begin{itemize}
\item
Following the Grossman \textit{et al.} \cite{grossman} strategy, we proposed phenomenological expressions for the equation of state and the transport coefficients for three-dimensional cases. We employed free-volume arguments in the vicinity of the close packing, suggesting an interpolation between the hexagonal-packing limit and the well-known low-density relations.
\item
The stationary solutions of the granular hydrodynamic equations predict symmetric clustered states far from the thermal wall. Some of the solutions bifurcate from the (steady-state) symmetric solution, i.e., hydrodynamic also predicts the existence of broken-symmetry states. The presence of symmetry-breaking instability in this system is in contrast with our findings in its two-dimensional counterpart (see Chapter~\ref{cap4}) in which no symmetry-breaking instability was found.
\item
Granular hydrodynamics, employing our heuristic constitutive relations, may be a good candidate for describing the clustering and symmetry-breaking instabilities which may occur in inelastic hard-sphere systems.
\end{itemize}

\chapter{Published work}
\thispagestyle{empty}
\begin{itemize}
\item
Manuel D\'iez-Minguito, Pedro L. Garrido, and Joaqu\'in Marro, \emph{Lennard-Jones and lattice models of driven fluids}, Phys. Rev. E \textbf{72}, 026103 (2005).
\item
Joaqu\'in Marro, Pedro L. Garrido, and Manuel D{\'i}ez-Minguito, \emph{Nonequili-brium anisotropic phases, nucleation and critical behavior in a driven Lennard-Jones fluid}, Phys. Rev. B \textbf{73}, 184115 (2006).
\item
Manuel D\'iez-Minguito, Pedro L. Garrido, and Joaqu\'in Marro, \emph{Lattice versus Lennard-Jones models with a net particle flow} in \textit{Traffic and Granular Flows '05}, Edited by A. Schadschneider, T. P\"oschel, R. K\"uhne, M. Schreckenberg, D. E. Wolf, (Springer, Berlin, 2007).
\item
Baruch Meerson, Manuel D\'iez-Minguito, Thomas Schwager, and Thorsten P\"{o}schel, \emph{Close-packed granular clusters: hydrostatics and persistent Gaussian fluctuations}, Granular Matter, Granular Matter \textbf{10}(1), 21-27 (2007).
\item
Manuel D\'iez-Minguito, J. Marro, and P. L. Garrido, \emph{On the similarities and differences between lattice and off-lattice models of driven fluids} in \textit{Workshop on Complex Systems: New Trends and Expectations}, Euro. Phys. J.-Special Topics \textbf{143}, 269 (2007).
\item
Manuel D\'iez-Minguito and Baruch Meerson, \emph{Phase separation of a driven granular gas in annular geometry}, Phys. Rev. E \textbf{75}, 011304 (2007).
\end{itemize}

\vskip 0.3cm
{ABSTRACTS}

\begin{itemize}
\item
Manuel D\'iez-Minguito, Pedro L. Garrido, Joaqu\'in Marro, and Francisco de los Santos, \emph{Driven two-dimensional Lennard-Jones fluid} in \textit{Modeling Cooperative Behavior in the Social Sciences}, P. L. Garrido, J. Marro, M. A. Mu\~noz (eds.), AIP Conference Proceedings \textbf{779}, 199 (2005).
\item
Manuel D\'iez-Minguito, Pedro L. Garrido, Joaqu\'in Marro, and Francisco de los Santos, \emph{Simple computational Lennard-Jones fluid driven out-of-equilibrium}, in \textit{XIII Congreso de F\'isica Estad\'istica FISES'05}, Abstract Book (2005).
\item
Manuel D\'iez-Minguito and Baruch Meerson, \emph{Phase separation of a driven granular gas in annular geometry}, in \textit{XIV Congreso de F\'isica Estad\'istica FISES'06}, Abstract Book (2006).
\end{itemize}

\vskip 0.3cm

\vspace*{1cm}
\begin{center}
The author acknowledge financial support from the Spanish Ministry for Science (MEC) and FEDER (project FIS2005-00791).
\end{center}

\chapter[Resumen en Espa\~nol]{Resumen en Espa\~nol\label{resumen}}
\thispagestyle{empty}
La naturaleza presenta una estructura jer\'arquica en niveles, con escalas temporales, espaciales y de energ\'ia que se extienden desde el mundo \textit{microsc\'opico} al \textit{macrosc\'opico}. Aunque pueda sorprendernos, a menudo es posible tratarlos de forma independiente, e.g., en el caso de gases moleculares. El nivel macrosc\'opico, que no es sino el mundo percibido directamente por nuestros sentidos, se describe hidrodin\'amicamente. Esto es, es descrito por funciones \textit{continuas} (o continuas a trozos) de sus coordenadas espaciales \textbf{r} y del tiempo $t$ (\textit{campos hidrodin\'amicos}). Por consiguiente, las grandes y exitosas disciplinas de la f\'isica macrosc\'opica, tales como la mec\'anica de fluidos, elasticidad, ac\'ustica, electromagnetismo, son teor\'ias de campos. Estos campos est\'an determinados por ecuaciones integro-diferenciales que involucran funciones, en principio desconocidas, de \textbf{r} y $t$. En cambio, el nivel microsc\'opico describe colecciones de un gran n\'umero de constituyentes, t\'ipicamente \'atomos, mol\'eculas o entidades m\'as complejas, los cuales interact\'uan los unos con los otros de acuerdo con ciertas fuerzas electro-mec\'anicas. Aunque generalmente, la evoluci\'on temporal de cada individuo est\'a dada por las leyes de la mec\'anica cu\'antica, en muchas situaciones la aproximaci\'on cl\'asica resulta ser excelente. La diferencia entre niveles macrosc\'opico y microsc\'opico es esencialmente relativa: el concepto clave es el n\'umero de cuerpos, no su tama\~no\footnote{Por ejemplo, una galaxia es un objeto (macrosc\'opico) compuesto por un gran n\'umero de estrellas (microsc\'opicas); del mismo modo, una comunidad de seres vivos es considerada como un conjunto de individuos; un gas como conjunto de mol\'eculas, etc.}.

Si hay una clara separaci\'on entre microsc\'opico y macrosc\'opico, a\'un es posible introducir una nueva escala intermedia: el nivel de descripci\'on \textit{mesosc\'opico}. \'Esta es una representaci\'on \textit{gruesa}\footnote{O ``coarse--grain" (grano grueso) seg\'un fuentes bibliogr\'aficas en lengua inglesa.} que tiene por objeto dar cuenta de la f\'isica de los modelos microsc\'opicos a escalas temporales y espaciales m\'as altas, y que, matem\'aticamente, est\'a caracterizada por ecuaciones diferenciales en derivadas parciales estoc\'asticas \cite{vkampen,gardiner}. Para ser exactos, la descripci\'on mesosc\'opica de un sistema dado, cuya din\'amica es, en promedio, gobernada por las leyes de evoluci\'on temporal macrosc\'opica, es el resultado del recuento y acumulaci\'on de fluctuaciones microsc\'opicas. Este enfoque nos facilita el estudio de las propiedades de variaci\'on lenta de los sistemas de muchos cuerpos y, por otra parte, conforma el fundamento te\'orico de los estudios m\'as modernos de fen\'omenos cr\'iticos. Es com\'un encontrar en la literatura trabajos que incluyen en este nivel intermedio descripciones de teor\'ia cin\'etica de los gases, e.g., la ecuaci\'on de Boltzmann. Por ello, este nivel es a veces conocido como \textit{cin\'etico}\footnote{Aunque ecuaciones diferenciales estoc\'asticas y ecuaciones cin\'eticas puedan ser clasificadas en un mismo nivel, involucran diferentes escalas (mesosc\'opicas) tanto espaciales como temporales.}.

Estos tres niveles de descripci\'on\footnote{Nos restringiremos a los niveles aqu\'i descritos, aunque existen escalas hidrodin\'amicas a\'un m\'as altas, e.g., el as\'i llamado \textit{tren de v\'ortices} de Karman.} son, en principio, enfoques equivalentes de la misma realidad f\'isica. No obstante, las leyes y simetr\'ias asociadas a cada nivel son tan diferentes entre s\'i que desde muy pronto surgi\'o en la comunidad cient\'ifica la necesidad de entender c\'omo se relacionan, o, dicho de otro modo, c\'omo las propiedades hidrodin\'amicas emergen en t\'erminos de la din\'amica microsc\'opica de los constituyentes subyacentes. As\'i se desarroll\'o la \textit{Mec\'anica Estad\'istica}, disciplina que proporciona esas relaciones. En efecto, la mec\'anica estad\'istica establece como su propio objetivo el determinar el comportamiento macrosc\'opico de la materia originada en el comportamiento colectivo de entidades (microsc\'opicas) individuales. Algunos de los fen\'omenos que observamos a alto nivel son consecuencia de simples efectos sinerg\'eticos de las acciones de los constituyentes, por ejemplo, la presi\'on ejercida por un gas molecular en las paredes del recipiente que lo contiene o miles de luci\'ernagas centelleando al un\'isono, mientras que otros son ejemplos paradigm\'aticos de comportamiento colectivo emergente, e.g., el cambio de r\'egimen laminar a flujo turbulento en fluidos o el mill\'on de \'atomos que forman el programa de la vida: el \'acido desoxirribunocleico o ADN. En estos \'ultimos casos, el comportamiento de los constituyentes llega a ser singular y muy diferente del que tendr\'ia en ausencia del resto, dando lugar a un comportamiento sin hom\'ologo directo en las propiedades o din\'amica de los constituyentes.

El mayor logro de la mec\'anica estad\'istica es la \textit{Teor\'ia de Colectividades} \cite{libros_eq}. \'Esta establece formalmente el nexo de uni\'on entre las propiedades macrosc\'opicas de sistemas \textit{en equilibrio} y las leyes que gobiernan las interacciones a nivel microsc\'opico entre part\'iculas individuales. Se dice que un sistema dado se encuentra \textit{en equilibrio} cuando est\'a aislado, no presenta hist\'eresis y se haya en un estado estacionario en el que todas sus propiedades macrosc\'opicas permanecen fijas \cite{biel}. En tal caso, las propiedades macrosc\'opicas son expresadas en t\'erminos de variables \textit{termodin\'amicas}. En efecto, la termodin\'amica es, en este sentido, una descripci\'on hidrodin\'amica que consiste en leyes y relaciones entre magnitudes termodin\'amicas, y cuya base te\'orica microsc\'opica es la mec\'anica estad\'istica. Desde el punto de vista matem\'atico, la teor\'ia de colectividades se ha axiomatizado por completo. Esto nos proporciona, al menos en principio, expresiones anal\'iticas para estas relaciones termodin\'amicas, permitiendo obtener toda la informaci\'on macrosc\'opica relevante del sistema en cuesti\'on\footnote{Una vez especificado el Hamiltoniano microsc\'opico del sistema (llam\'emosle $\mathcal{H}$), la distribuci\'on ``can\'onica" de probabilidad de estados estacionaria $\mathcal{P}$ est\'a dada en t\'erminos del factor de Boltzmann $\mathcal{P}=e^{-\mathcal{H}/k_{B}T}/Z$, donde $Z$ es la funci\'on de partici\'on, $k_{B}$ es la constante de Boltzmann y $T$ la temperatura. A partir de los adecuados promedios sobre \'esta, los observables estacionarios pueden ser calculados. El resto de dificultades son ``meramente" t\'ecnicas.}, por ejemplo, la energ\'ia libre, la ecuaci\'on de estado, la radiancia espectral, etc.

\begin{figure}
\centerline{\includegraphics[width=5.0cm]{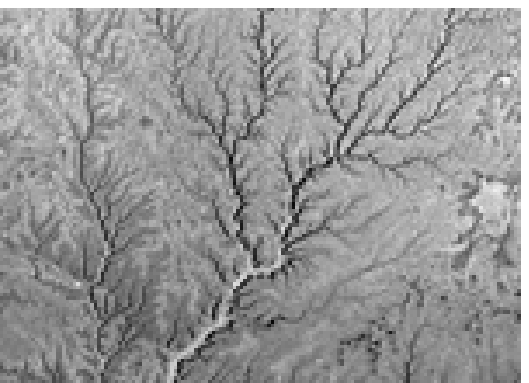}
\includegraphics[width=5.0cm]{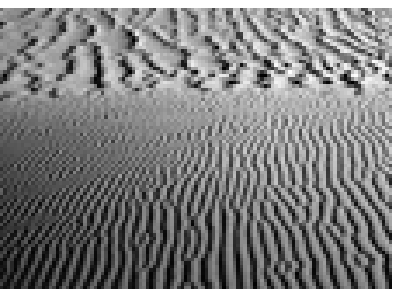}}
\centerline{\includegraphics[width=5.0cm]{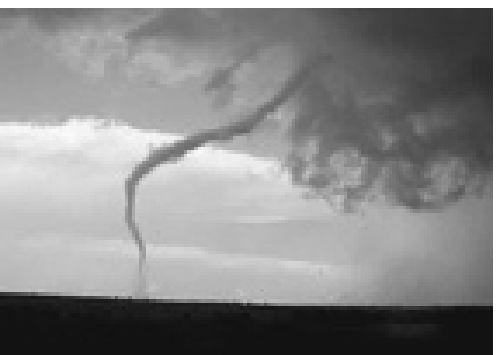}
\includegraphics[width=5.0cm]{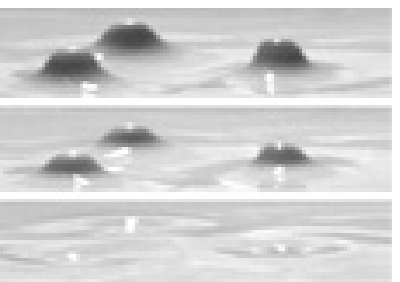}}
\caption[Complejidad del comportamiento macrosc\'opico]{La complejidad en la naturaleza se manifiesta a todas las escalas. Algunos ejemplos llamativos son los siguientes. Arriba izquierda: geometr\'ias fractales en cuencas fluviales en El Yemen (fuente NASA). Arriba derecha: modelado e\'olico franjiforme en las \textit{Great Sand Dunes National Monument}, EE.UU. (fotografiado por Bob Bauer). Abajo izquierda: tornado como un flujo geof\'isico en rotaci\'on. Abajo derecha: \textit{oscilones} en suspensiones coloidales agitadas verticalmente \cite{clay}.}
\label{ejemplos}
\end{figure}

Por contra (!`y afortunadamente!), en la naturaleza los sistemas en equilibrio son la excepci\'on (incluso una idealizaci\'on) m\'as que la regla: fen\'omenos \textit{fuera del equilibrio} son, con creces, mucho m\'as abundantes. Galaxias, seres humanos, reacciones qu\'imicas, flujos geof\'isicos, portadores de carga en dispositivos semiconductores, mercados financieros, tr\'afico en autopistas, por citar unos pocos, son sistemas de muchos cuerpos en condiciones de no-equilibrio\footnote{Fuera del equilibrio se encuentra tambi\'en cualquier sistema que se encuentre relajando hacia un estado de equilibrio. No obstante, en la mayor\'ia de estas situaciones su din\'amica puede comprenderse adecuadamente a partir de conceptos de equilibrio.}. Los sistemas fuera del equilibrio se caracterizan por no estar cerrados, es decir, por intercambiar energ\'ia, part\'iculas y/o informaci\'on con su entorno. En general, el estado de un sistema fuera del equilibrio no es determinado \'unicamente por ligaduras externas, sino que tambi\'en depende de su historia (presentan hist\'eresis). Esto da lugar a la enorme complejidad del mundo real, la cual surge a todos los niveles de descripci\'on \cite{kadanoff_complex,cross,gollub,barabasi,bak} y se manifiesta en formaci\'on de patrones, fractales, sincronizaci\'on, caos, criticalidad auto-organizada, etc. Toda esta sorprendente y complicada fenomenolog\'ia es com\'unmente asociada a \textit{inestabilidades}\footnote{Las inestabilidades presentes en sistemas lejos del equilibrio presentan similitudes a nivel morfol\'ogico con las observadas en el equilibrio, descritas \'estas de forma mecano-estad\'istica. Por tanto, suelen referirse como \textit{cambios de fase del no-equilibrio}, aunque los cambios de fase fuera del equilibrio, sin las restricciones de \'este, son mucho m\'as ricos desde un punto de vista fenomenol\'ogico.}, las cuales se describen como cambios de fase \cite{cross,marro,haken,gar}, bifurcaciones, sinergia, etc. con el prop\'osito de conectar lo microsc\'opico con las estructuras coherentes observadas a alto nivel. 

Sin embargo, la mayor\'ia de los estudios de sistemas fuera del equilibrio realizados hasta la fecha adoptan un punto de vista eminentemente fenomenol\'ogico (macrosc\'opico). En general, poco se sabe del porqu\'e de tales estructuras o de c\'omo esta complejidad emerge desde interacciones colectivas a nivel microsc\'opico. Consecuentemente, no existe una teor\'ia apropiada y el desarrollo de s\'olidos fundamentos matem\'aticos est\'a s\'olo en sus primeros pasos comparado con el del equilibrio, donde la teor\'ia de colectividades funciona con \'exito.

Una diferencia fundamental entre mec\'anica estad\'istica del equilibrio y del no-equilibrio es que, mientras en la primera la distribuci\'on de probabilidad de estados, de la cual se deriva toda la informaci\'on macrosc\'opica \'util, es conocida, fuera del equilibrio es, a priori, desconocida. Uno debe encontrar una distribuci\'on de estados, en este caso dependiente del tiempo, la cual verifica una ecuaci\'on general de evoluci\'on, a saber, la \textit{ecuaci\'on maestra} \cite{vkampen}. Los casos resolubles se limitan a unos pocos, y lo son s\'olo de forma aproximada. \'Unicamente es posible ofrecer un enfoque unificado para sistemas que se encuentran \textit{no demasiado lejos}\footnote{Como es de suponer, una definici\'on precisa de ``no demasiado lejos" es muy dif\'icil de dar y depende en gran medida del sistema estudiado y debe darse \textit{ad hoc}.} del equilibrio \cite{mazur}, donde el sistema puede ser tratado perturbativamente alrededor del estado del equilibrio empleando las t\'ecnicas cl\'asicas de las \textit{teor\'ias de respuesta lineal}. En cualquier caso, nuestra atenci\'on se centra precisamente en sistemas lejos del equilibrio, donde tales enfoques dejan de tener utilidad.\\

\textbf{\textit{En este contexto, el objeto principal de esta tesis reside en el estudio de inestabilidades ---a diferentes niveles de descripci\'on--- en sistemas} forzados\textit{, i.e., mantenidos lejos del equilibrio por un agente externo. Nos centraremos en dos importantes clases de sistemas, a saber,} fluidos difusivos con arrastre  \textit{y} gases granulares vibro-fluidificados.}\\

La primera familia (fluidos difusivos con arrastre) incluye\footnote{Aqu\'i seguimos la definici\'on dada por Schmittmann \& Zia en Ref.~\cite{zia}.} sistemas que se encuentran acoplados a dos fuentes de energ\'ia constante, de tal modo que se establece un flujo estacionario de energ\'ia a trav\'es de aqu\'ellos. Esta definici\'on es ciertamente bastante amplia e incluye sistemas de muy diversa \'indole. Aqu\'i nos restringiremos a sistemas que presentan una corriente de part\'iculas (cuyo n\'umero es una magnitud conservada) a su trav\'es, anisotrop\'ias espaciales asociadas a la acci\'on de un campo externo y en los cuales finalmente se alcanza un estado estacionario de no-equilibrio. Un ejemplo sencillo y protot\'ipico podr\'ia ser el de una impedancia en un circuito el\'ectrico disipando a la atm\'osfera la energ\'ia suministrada por una bater\'ia. Pero incluso en \'esta reducida clase de sistemas, la distribuci\'on estacionaria de probabilidad es desconocida\footnote{Es necesario hacer notar que en la literatura estos sistemas son referidos como \textit{sistemas} difusivos con arrastre en vez de \textit{fluidos} difusivos con arrastre, denominaci\'on que adoptamos en esta tesis. Como explicaremos con cierto detalle en el Cap\'itulo~\ref{cap1}, reservamos aquella denominaci\'on para una descripci\'on mesosc\'opica de campo medio.}.

La motivaci\'on fundamental subyacente tras estudio de fluidos difusivos con arrastre es la necesidad de desentra\~nar los ingredientes b\'asicos del comportamiento observado en sistemas lejos del equilibrio. Por ello, es justificado, pertinente y necesario el estudio de modelos simplificados que capturen lo esencial del comportamiento microsc\'opico que da lugar a las complicadas estructuras macrosc\'opicas. Estos modelos son a menudo caricaturas de los sistemas reales en los cuales la din\'amica de los constituyentes est\'a dada por simples reglas estoc\'asticas. Entre ellos, los \textit{modelos reticulares} \cite{marro,zia,liggett,privman,droz,hinrichsen,odor} han jugado un papel preponderante debido primordialmente al hecho de que, a veces, aportan resultados anal\'iticos exactos y permiten aislar las caracter\'isticas esenciales de un cierto sistema. Adem\'as, los modelos reticulares nos ayudan a ganar la intuici\'on necesaria para desarrollar teor\'ias y, desde un punto de vista experimental, su implementaci\'on en computadores resulta ser muy eficiente. Todo esto ha contribuido al desarrollo y aparici\'on de numerosas y novedosas t\'ecnicas, incluyendo la \textit{teor\'ia estad\'istica de campos} lejos del equilibrio. En concreto, modelos reticulares han modelado con \'exito algunos aspectos esenciales de organismos con estructura social \cite{ferreira,treves,migrating}, formaci\'on de redes fluviales \cite{river_networks}, propagaci\'on de epidemias \cite{epidemics}, vidrios, circuitos el\'ectricos, enzimas \cite{enzyme}, tr\'afico \cite{traffic,helbing,lanes}, coloides y espumas, mercados financieros \cite{econophys}, etc. 

Los cambios de fase (de equilibrio) y los fen\'omenos cr\'iticos son actualmente bien entendidos, en muchos aspectos, gracias a la aplicaci\'on de los m\'etodos del \textit{grupo de renormalizaci\'on} \cite{amit,renormalization} a modelos reticulares. El resultado primordial derivado del estudio de fen\'omenos cr\'iticos (en equilibrio, insistimos) es el concepto de \textit{universalidad}: en las cercan\'ias de los puntos cr\'iticos el comportamiento de sistemas dispares est\'a determinado \'unicamente por ciertas caracter\'isticas comunes b\'asicas ---dimensionalidad, el rango de las interacciones, simetr\'ias, etc.--- y no parece depender en gran medida del sistema concreto \cite{amit,ma,binney}.

La pregunta de si este concepto es tambi\'en aplicable a cambios de fase lejos del equilibrio est\'a todav\'ia lejos de ser respondida. Por ello, se espera que, al igual que para sistemas en equilibrio, los modelos reticulares jueguen tambi\'en un importante papel aqu\'i. Trataremos con detalle estos aspectos en el Cap\'itulo~\ref{cap1}, en el que estudiaremos un modelo reticular difusivo con arrastre, paradigma de cambios de fase anisotr\'opicos lejos del equilibrio. Sin embargo, es f\'acil imaginar que los modelos reticulares, comparados directamente con sistemas reales, presentan la desventaja de que son demasiado ``crudos" o simplificados. Esto debe entenderse en el sentido de que los modelos reticulares no capturan todos los detalles, especialmente los morfol\'ogicos, del diagrama de fases. Discutiremos \'este y otros problemas en el Cap\'itulo~\ref{cap2}, donde presentaremos un nuevo modelo m\'as \textit{realista} para simulaci\'on de fluidos anisotr\'opicos.\\

La segunda clase de sistemas considerados en esta tesis versa acerca de \textit{gases granulares agitados mec\'anicamente}. Un medio granular \cite{jaeger} es un conglomerado de muchas part\'iculas de tama\~no macrosc\'opico, y por lo tanto, susceptible de ser modelados cl\'asicamente. Hay ejemplos por doquier y se presentan a muy diferentes escalas: desde materiales polvoreados hasta nubes de polvo intergal\'actico. Dentro de ese intervalo podemos encontrar arena, hormig\'on, cereales, flujos volc\'anicos, los anillos de Saturno y otros muchos ejemplos. Su importancia cient\'ifica y tecnol\'ogica es indudable, con numerosas aplicaciones farmac\'euticas, en la construcci\'on e ingenier\'ia civil, qu\'imicas, agrarias y alimentarias, y, a mayor escala, en procesos geol\'ogicos y astrof\'isicos. Adem\'as, su estudio est\'a correlacionado con otras clases de sistemas como coloides, espumas y vidrios. As\'i pues la fenomenolog\'ia mostrada por estos sistemas es ampl\'isima, yendo desde unas pocas micras a miles de kil\'ometros: formaci\'on de patrones en granulados fluidificados \cite{melo,umbanhowar,olafsen}, agregaci\'on \cite{macnamara,goldhirsch2}, flujos y bloqueos en silos y tolvas \cite{hoppers}, avalanchas y deslizamientos \cite{avalanches}, mezcla y segregaci\'on \cite{segrega}, convecci\'on \cite{convect1,convect2}, erupciones \cite{eruptions,thorsten3}, por citar unos pocos. Por todo ello, su importancia est\'a lejos de toda duda y la adecuada comprensi\'on de sus propiedades no es s\'olo una necesidad industrial urgente, sino que tambi\'en supone un reto importante para la f\'isica.

Puesto que los granos individuales tienen dimensiones macrosc\'opicas, la fricci\'on, deformaci\'on, ruptura y las p\'erdidas energ\'eticas por colisi\'on producen disipaci\'on de energ\'ia, esto es, la energ\'ia cin\'etica es transferida continuamente al medio en forma de calor. Asimismo, la escala de energ\'ia t\'ipica para, por ejemplo, un grano de arroz de, digamos, longitud $l\approx 5mm$ y masa $m$ es su energ\'ia potencial $mgl$, donde $g$ es la aceleraci\'on gravitatoria. Es para tama\~nos de grano mucho m\'as peque\~nos cuando otros tipos de interacci\'on ganan en importancia\footnote{En granos con $d\approx 0.1$ aparecen cargas electrost\'aticas superficiales, o para $d\approx 0.01$ los efectos magn\'eticos y de adhesi\'on superficial comienzan a ser relevantes.}. De ah\'i que en los medios granulares, y a diferencia de gases moleculares, la temperatura $T$ no juega ning\'un papel, es decir, $k_BT\ll mgl$. Los efectos din\'amicos pesan m\'as que cualquier tipo de consideraciones entr\'opicas \cite{jaeger} y, consecuentemente, argumentos mecano-estad\'isticos o termodin\'amicos resultan de poca utilidad. Adem\'as, dado que cualquier granulado est\'a formado por un n\'umero finito de constituyentes\footnote{A diferencia de gases moleculares, $N\ll N_A$ donde $N$ es el n\'umero de part\'iculas y $N_A$ el n\'umero de Avogadro. No obstante, si $N$ fuera lo suficientemente grande (a determinar \textit{ad hoc}) a\'un ser\'ia posible emplear un tratamiento estad\'istico.}, frecuentemente el efecto de las fluctuaciones en observables macrosc\'opicos es enorme y, por tanto, su estad\'istica es dominada por los, as\'i llamados, \textit{eventos raros}. Otros efectos tambi\'en importantes est\'an relacionados con el fluido intersticial, como aire o agua, aunque en muchas situaciones y, en buena aproximaci\'on, puede ignorarse considerando \'unicamente las interacciones de grano con grano. En esta tesis (concretamente en la Parte~\ref{part:Granos}) nos restringiremos al estudio de sistemas granulares \textit{secos}, donde la interacci\'on entre part\'iculas predomina sobre la interacci\'on con el posible fluido intersticial.

Sin embargo, y a pesar de la din\'amica cl\'asica que los caracteriza, los medios granulares se comportan de forma poco convencional: sus fases ---s\'olida, l\'iquida y gaseosa--- ponen de manifiesto una compleja fenomenolog\'ia de car\'acter colectivo que los distinguen fuertemente de sus hom\'ologos moleculares. El resultado es una paup\'errima comprensi\'on de sus propiedades din\'amicas y estad\'isticas. De hecho, desde un punto de vista te\'orico, s\'olo existen predicciones de car\'acter muy general y muchas cuestiones permanecen sin responder. Los s\'olidos granulares son altamente hister\'eticos, muestran indeterminaci\'on de tensiones est\'aticas, y el entramado de granos da lugar a cadenas de fuerzas, tensiones y esfuerzos internos, bloqueos, estr\'es conducido, dilataci\'on o compresi\'on por cizalladura. Los l\'iquidos son viscosos, presentan avalanchas, segregaci\'on por tama\~nos, bandas de cizalladura, formaci\'on de patrones. Los gases, los cuales s\'olo perduran con un continuo aporte de energ\'ia (por ejemplo, agitaci\'on mec\'anica\footnote{Se habla entonces de gases fluidificados o vibrofluidificados}), son altamente compresibles y muestran agregados inhomog\'eneos, colapso inel\'astico en modelos computacionales, distribuciones de velocidad no maxwellianas, falta de separaci\'on entre escalas macrosc\'opica y microsc\'opica.

Toda esta fenomenolog\'ia hace que resulte muy dif\'icil enmarcarlos dentro de la clasificaci\'on tradicional de estados de agregaci\'on de la materia, dificultando considerablemente su estudio. En Cap\'itulos~\ref{cap3} y \ref{cap4} abordamos algunos problemas no resueltos y controversias, a saber: cu\'ales son las propiedades estad\'isticas de los medios granulares, c\'omo y bajo qu\'e condiciones cambian de fase y cu\'ales son los modelos continuos \'optimos y cu\'ando aplicarlos. Especialmente nos centramos en la descripci\'on hidrodin\'amica de las inestabilidades de agregaci\'on heterog\'enea y ruptura espont\'anea de simetr\'ia en gases granulares fluidificados. 

\subsubsection{Perspectiva\label{resumen:over}}
A lo largo de esta tesis abordamos el estudio de estas dos familias de sistemas. Nuestro trabajo se divide en cuatro partes. En el Cap\'itulo~\ref{cap0} se presenta una introducci\'on general que describe el objeto principal de esta tesis. Los Cap\'itulos~\ref{intro:DDF},~\ref{cap1} y \ref{cap2} conforman la Parte~\ref{part:DDF}, que est\'a dedicada a \textit{fluidos difusivos con arrastre}. La Parte~\ref{part:Granos} trata de \textit{gases granulares} y re\'une los Cap\'itulos~\ref{intro:Granos},~\ref{cap3} y \ref{cap4}. La Parte~\ref{part:apend} est\'a formada por los ap\'endices~\ref{apendA}, \ref{apendC}, \ref{apendB} y \ref{apendD}. Finalmente la Parte~\ref{part:Concl} cierra con un resumen y las principales conclusiones.\\

La primera parte comienza en el \textbf{Cap\'itulo~\ref{intro:DDF}} presentando el Gas Reticular con Arrastre, conocido como \textit{Driven Lattice Gas} \cite{marro,zia,kls} ---modelo protot\'ipico para el estudio de cambios de fase anisotr\'opicos lejos del equilibrio. Las part\'iculas que lo forman se mueven en un ret\'iculo e interaccionan de acuerdo con ciertas reglas locales de car\'acter anisotr\'opico. \'Estas inducen un flujo de part\'iculas en una direcci\'on de tal modo que, si las condiciones de contorno as\'i lo permiten, se establece una corriente. Este modelo es considerado una simplificaci\'on de ciertos problemas de tr\'afico y fluidos. Presentamos algunas de sus ya conocidas propiedades y tambi\'en controversias, as\'i como los fundamentos de los m\'etodos te\'oricos y computacionales empleados en los Cap\'itulos~\ref{cap1} y \ref{cap2}.\\

Una de las preguntas que cabe plantearse sobre cualquier modelo f\'isico es si el comportamiento que manifiesta es universal. En el \textbf{Cap\'itulo~\ref{cap1}} abordamos esta importante cuesti\'on estudiando la segregaci\'on anisotr\'opica de fases en el modelo protot\'ipico introducido en el Cap\'itulo~\ref{intro:DDF}. El \'enfasis de nuestro estudio est\'a puesto en la influencia de los detalles de la din\'amica microsc\'opica en el estado estacionario del no-equilibrio resultante, aunque tambi\'en se presta cierta atenci\'on a aspectos cin\'eticos. En especial trataremos, tanto desde un punto de vista computacional como te\'orico, las similitudes y diferencias entre el \textit{driven lattice gas} y otros modelos reticulares y no reticulares relacionados con aqu\'el. La comparativa nos permite discutir ciertas propiedades excepcionales y poco realistas del modelo reticular original. Adem\'as, ponemos a prueba la validez de dos teor\'ias mesosc\'opicas actuales que pretenden dar cuenta del comportamiento cr\'itico de esta clase de sistemas.\\

En el \textbf{Cap\'itulo~\ref{cap2}} introducimos un novedoso modelo computacional de fluido no reticular y de no-equilibrio dise\~nado para el estudio de fen\'omenos anisotr\'opicos. Su din\'amica e interacciones son m\'as realistas que en, por ejemplo, el \textit{driven lattice gas}. Se trata de un modelo muy accesible desde el punto de vista computacional, que exhibe una corriente neta de part\'iculas y estructuras franjiformes a bajas temperaturas orientadas seg\'un la direcci\'on de un campo externo, asemej\'andose a muchas situaciones reales. Describiremos tanto las propiedades cin\'eticas a tiempos cortos (nucleaci\'on) como las estacionarias del modelo; en especial nos centraremos en su diagrama de fases: identificamos y caracterizamos sus fases s\'olida (mono- y policristalina), l\'iquida y gaseosa, todas ellas altamente anisotr\'opicas. El c\'alculo preciso de sus par\'ametros cr\'iticos nos lleva a discutir el papel de las simetr\'ias en el modelado de fluidos fuera del equilibrio con teor\'ias de campos y simulaciones. Adem\'as se esbozan algunas comparaciones con experimentos reales.\\

El \textbf{Cap\'itulo~\ref{intro:Granos}} sirve de introducci\'on a la Parte~\ref{part:Granos} de esta memoria; parte que es dedicada a los gases granulares agitados mec\'anicamente o vibrofluidificados. Perfilamos el estado del arte con respecto a la investigaci\'on en gases granulares. Concretamente discutiremos la aplicabilidad de las teor\'ias hidrodin\'amicas al estudio de flujos granulares. Tambi\'en proporcionamos brevemente los ingredientes b\'asicos de los modelos estudiados en los cap\'itulos siguientes y los m\'etodos de simulaci\'on.\\

En el \textbf{Cap\'itulo~\ref{cap3}} empleamos una descripci\'on hidrodin\'amica granular para estudiar la fenomenolog\'ia observada en un gas inel\'astico confinado en un recipiente anular sin gravedad. Nuestro principal prop\'osito es caracterizar apropiadamente sus estados estacionarios (de no-equilibrio) y establecer de forma precisa un diagrama de fases mediante la descripci\'on hidrodin\'amica y la computacional. Empleamos relaciones constitutivas de tipo Chapman-Enskog, derivadas \'estas de primeros principios. Prestamos especial atenci\'on a las inestabilidades observadas, a saber, separaci\'on de fases, agregaci\'on heterog\'enea de part\'iculas y ruptura de simetr\'ia. \'Esta \'ultima constituye una prueba precisa para modelos de flujos granulares y fomenta el estudio de formaci\'on de patrones lejos del equilibrio. Con este estudio pretendemos motivar experimentos en inestabilidades de separaci\'on de fases en gases granulares, que podr\'ian ser f\'aciles de implementar en esta geometr\'ia. Hasta ahora este tipo de estudios ha empleado geometr\'ias planas en vez de radiales.\\

El objeto del \textbf{Cap\'itulo~\ref{cap4}} es el estudio de hidrodin\'amica granular y fluctuaciones en un gas granular bidimensional a densidades elevadas, condici\'on \'esta que impide la aplicabilidad de descripciones hidrodin\'amicas derivadas de primeros principios. Estudiamos un modelo de discos r\'igidos cuasiel\'asticos interactuando en un recipiente circular y ganando energ\'ia del contorno. Mediante enfriamiento por colisi\'on entre part\'iculas se forma en la regi\'on central un granulado de alta densidad, el cual se comporta como una \textit{macro-part\'icula}. Algunas caracter\'isticas de esta macropart\'icula son bien descritas por la soluci\'on num\'erica de un conjunto de ecuaciones hidrost\'aticas granulares, las cuales hacen uso de relaciones constitutivas m\'as fenomenolog\'icas. Las predicciones hidrost\'aticas se comparan de forma excelente con simulaciones de \textit{din\'amica molecular por eventos}. Adem\'as desarrollamos una descripci\'on mesosc\'opica (tipo Langevin) para la macropart\'icula, confinada por un potencial arm\'onico de naturaleza hidrodin\'amica y forzada por un ruido blanco de n\'umero finito.\\

La Parte~\ref{part:apend} est\'a compuesta de cuatro ap\'endices: en el \textbf{Ap\'endice~\ref{apendA}} se compendian las propiedades m\'as relevantes del \textit{Gas Reticular} (o \textit{Lattice Gas}\footnote{Modelo de Ising conservado.}), el cual es el hom\'ologo de equilibrio del \textit{driven lattice gas} tratado en el Cap\'itulo~\ref{cap1}; en el \textbf{Ap\'endice~\ref{apendC}} introducimos un modelo cuyo prop\'osito es caracterizar la estabilidad de la interfase en modelos reticulares con arrastre; \textbf{Ap\'endice~\ref{apendB}} detalla las derivaciones de teor\'ia estad\'istica de campos desarrolladas en el Cap\'itulo~\ref{cap1} concernientes a ecuaciones de tipo Langevin; y finalmente en el \textbf{Ap\'endice~\ref{apendD}} proponemos un conjunto de relaciones constitutivas para la descripci\'on hidrodin\'amica de gases granulares tridimensionales.\\

Finalmente, tras los ap\'endices, se incluye un amplio resumen (que adjuntamos a continuaci\'on) de las principales conclusiones, resaltando las contribuciones originales de \'esta tesis y se\~nalando las l\'ineas maestras a seguir en trabajos futuros.\\

\subsubsection{Conclusiones\label{resumen:conclus}}
Desafortunadamente, y a pesar de los muchos esfuerzos cient\'ificos realizados, la descripci\'on mecano-estad\'istica rigurosa de inestabilidades lejos del equilibrio sigue siendo esquiva. Los trabajos presentados en esta tesis suponen un nuevo paso adelante para entender los cambios de fase lejos del equilibrio. En efecto, el estudio de \textit{fluidos difusivos con arrastre} y \textit{gases granulares vibrofluidizados}, aparte de su importancia tecnol\'ogica, ahonda en este aspecto y contribuye a la racionalizaci\'on de los fen\'omenos del no-equilibrio. Se ha hecho especial \'enfasis en el estudio y caracterizaci\'on de las inestabilidades ---en la mayor\'ia de los casos a distintos niveles de descripci\'on, a saber, macrosc\'opico, mesosc\'opico y microsc\'opico--- que tienen lugar en aquellos sistemas. Esta labor ha sido llevada a cabo empleando modernas y novedosas t\'ecnicas en el \'ambito de la f\'isica estad\'istica.\\

\textit{Cap\'itulo~\ref{cap0}.} Se presenta una introducci\'on que incluye una descripci\'on general de los sistemas estudiados, su lugar en la f\'isica y en la t\'ecnica, fenomenolog\'ia y problem\'atica. Asimismo se resaltan las motivaciones y el objeto principal de esta tesis.\\

\textit{Cap\'itulo~\ref{intro:DDF}.} En este cap\'itulo presentamos el modelo considerado hasta la fecha prototipo de fluido difusivo con arrastre: el Gas Reticular con Arrastre o \textit{Driven Lattice Gas} (DLG). Repasamos tanto sus propiedades m\'as relevantes como las m\'as controvertidas. Asimismo, se proporcionaron las bases te\'oricas y computacionales necesarias para los subsiguientes estudios de los Cap\'itulos~\ref{cap1} y \ref{cap2}; \'estas son modernas t\'ecnicas computacionales y teor\'ias estad\'isticas de campos.\\

\textit{Cap\'itulo~\ref{cap1}.} Describimos simulaciones tipo \textit{Monte Carlo} y desarrollos anal\'iticos de teor\'ia de campos que pretenden ilustrar c\'omo, a diferencia de lo que ocurre en equilibrio, ligeras modificaciones en la din\'amica microsc\'opica pueden influenciar, incluso cuantitativamente, tanto los estados estacionarios como la cin\'etica de relajaci\'on. Con este prop\'osito, se tom\'o como referencia el DLG introducido en el cap\'itulo anterior. Presentamos modelos reticulares y no reticulares relativos al DLG en los cuales las part\'iculas, al igual que en el DLG, interaccionan de acuerdo con ciertas reglas locales anisotr\'opicas. Estas reglas facilitan el movimiento preferente en una direcci\'on espacial, de tal modo que se establece una corriente a trav\'es del sistema si as\'i lo permiten las condiciones de contorno. En concreto, se discuti\'o las similitudes y diferencias entre el DLG y su ``an\'alogo" no-reticular, a saber, un modelo en el cual las coordenadas de las part\'iculas var\'ian de forma continua y en el que \'estas interaccionan entre s\'i con un potencial atractivo de tipo Lennard-Jones 12-6 (tambi\'en conocido como potencial Weeks-Chandler-Andersen). La detallada comparaci\'on de la morfolog\'ia observada nos llev\'o a considerar ciertas propiedades del DLG como poco realistas y excepcionales, a pesar de haber sido considerado protot\'ipico. Concluimos:
\begin{itemize}
\item
El modelo continuo no-reticular presenta, al igual que su an\'alogo el DLG, una corriente neta no nula de part\'iculas, una interfase fluido-vapor altamente anisotr\'opica (franjiforme) y un cambio de fase de segundo orden (continuo).
\item
Sin embargo, estos difieren en algunas propiedades de car\'acter esencial. Contrariamente a lo que ocurre en el DLG, su an\'alogo no-reticular exhibe una temperatura cr\'itica que decrece con la intensidad del campo externo. Es m\'as, a diferencia del DLG, no existe ning\'un cambio de fase en el l\'imite de campo externo saturante. En lo concerniente a los procesos cin\'eticos de formaci\'on de estructuras macrosc\'opicas u ordenaci\'on (nucleaci\'on), se observaron agregaciones de part\'iculas con forma triangular apuntando en el sentido del campo externo, justamente al contrario que en el modelo discreto.
\item
Estas observaciones sugieren una falta de universalidad en el modelo reticular originario. Demostramos as\'i que el comportamiento del DLG est\'a condicionado en gran medida por la geometr\'ia del ret\'iculo. \'Este es el agente que favorece la ordenaci\'on, y no el campo en s\'i mismo. Por tanto, la caracter\'istica sorprendente y contraintuitiva de que en el DLG ``el campo externo" eleva la temperatura cr\'itica respecto a la del equilibrio no es sino un efecto geom\'etrico o topol\'ogico y no din\'amico.
\item
De aqu\'i se deriva, en concreto, que el DLG no tiene un hom\'ologo no-reticular (m\'as ``realista") inmediato.
\item
Para profundizar en esta situaci\'on, se estudi\'o el DLG bajo la acci\'on de un campo externo saturante a la vez que se aument\'o el rango de las interacciones a nuevos vecinos. El modelo resultante, definido como NDLG\footnote{DLG con interacci\'on a \textit{vecinos siguientes} o \textit{next-nearest-neighbors} DLG.}, se asemeja m\'as al del caso no-reticular: la temperatura cr\'itica decrece con el campo y las anisotrop\'ias triangulares observadas en el proceso de nucleaci\'on a tiempos cortos se orientan en el sentido del campo.
\item
No obstante, mostramos que, a\'un ampliando el radio de interacci\'on entre part\'iculas, tanto el decaimiento potencial de las correlaciones espaciales como la criticalidad permanecen invariantes. Las primeras fueron monitorizadas por la funci\'on de correlaci\'on a dos cuerpos y por el factor de estructura, mientras que las propiedades cr\'iticas se determinaron empleando t\'ecnicas est\'andar de an\'alisis de tama\~no finito y corroboradas por la teor\'ia mesosc\'opica de campos conocida como \textit{anisotropic driven system} (ADS).
\item
Por consiguiente, al extender la din\'amica en el DLG se modifica de forma esencial el diagrama de fases ---el cual no puede ser determinado por crudos argumentos de simetr\'ias--- pero no las propiedades intr\'insecas relacionadas con la naturaleza de no-equilibrio del fen\'omeno ---como el decaimiento de las correlaciones. 
\end{itemize}
La conclusiones de este cap\'itulo est\'an complementadas apropiadamente por los resultados de criticalidad del modelo af\'in expuesto en el Cap\'itulo~\ref{cap2}. Una posible y recomendable continuaci\'on del trabajo desarrollado en \'este cap\'itulo es el estudio de la segregaci\'on de fases en el modelo NDLG desde el punto de vista de la teor\'ia de campos empleando el enfoque de la teor\'ia ADS. Otros problemas son el estudio detallado de la estabilidad de la interfase bajo campos intensos (v\'ease Ap\'endice~\ref{apendC}) y la extensi\'on de nuestros m\'etodos a otros modelos reticulares como, por ejemplo al \textit{Time Asymmetric Exclusion Process} (TASEP) \cite{liggett,antal}.\\

\textit{Cap\'itulo~\ref{cap2}.} Puesto que, como fue demostrado en el cap\'itulo anterior, el DLG es poco realista en ciertos aspectos esenciales, presentamos aqu\'i un modelo novedoso y m\'as realista de fluido con arrastre, ideal para el estudio de fen\'omenos anisotr\'opicos. Se trata de un modelo no-reticular lejos del equilibrio en el que las part\'iculas, que interact\'uan entre s\'i v\'ia un potencial tipo Lennard-Jones 12-6, son conducidas por la acci\'on de un campo externo, cuya energ\'ia se disipa en un ba\~no t\'ermico. Estudiamos la cin\'etica de separaci\'on de fases a tiempos cortos y sus diferentes fases estacionarias (altamente anisotr\'opicas), a saber, s\'olido (mono- y policristalino), l\'iquido y gas. En concreto, describimos los procesos de nucleaci\'on, monitorizados por el exceso de energ\'ia, el cual da cuenta de la energ\'ia interfacial; las propiedades estructurales de las diferentes fases en el estado estacionario, caracterizadas por las funciones de distribuci\'on radial y azimutal y el grado de anisotrop\'ia; y propiedades de transporte. Elaboramos tambi\'en una precisa estimaci\'on de la curva de coexistencia l\'iquido--vapor y de los par\'ametros cr\'iticos asociados. Las principales conclusiones son:
\begin{itemize}
\item
El modelo propuesto parece contener la f\'isica esencial y necesaria para resultar \'util como un modelo protot\'ipico destinado a describir fen\'omenos anisotr\'opicos observados en la naturaleza.
\item
Este modelo es m\'as conveniente para prop\'ositos simulacionales en fluidos con arrastre que otros, como, por ejemplo, simulaciones est\'andar de din\'amica molecular, que son m\'as costosas.
\item
El modelo es la extensi\'on natural a casos anisotr\'opicos lejos del equilibrio del conocido fluido de Lennard-Jones, que describe acertadamente el comportamiento de gases reales como el Ar. En efecto, nuestro modelo se reduce a \'este para campo cero. De otro modo, exhibe una corriente neta de part\'iculas y estructuras macrosc\'opicas franjeadas por debajo de un punto cr\'itico.
\item
En cuanto a las propiedades cin\'eticas, y a pesar de la anisotrop\'ia inherente del modelo en las etapas tard\'ias de la descomposici\'on espinodal, la nucleaci\'on se produce por coagulaci\'on tipo Smoluchowski y agregaci\'on de Ostwald, las cuales son conocidas por describir los procesos del nucleaci\'on de fluidos reales en equilibrio.
\item
Inesperadamente, encontramos que el comportamiento cr\'itico del modelo es consistente con la clase de universalidad de Ising (equilibrio), pero no con la del DLG. La principal raz\'on de este aparente desacuerdo y de otros detallados en este cap\'itulo, podr\'ia ser la falta de simetr\'ia part\'icula-hueco en el modelo Lennard-Jones con arrastre.
\item
De ah\'i inferimos que la discretizaci\'on espacial puede cambiar de forma significativa no s\'olo las propiedades morfol\'ogicas y cin\'eticas sino tambi\'en las propiedades cr\'iticas. Esto contrasta fuertemente con el concepto de universalidad en sistemas en equilibrio, donde el comportamiento de estos cerca de los puntos cr\'iticos es independiente de los detalles microsc\'opicos.
\item
Finalmente, creemos que este modelo motivar\'a nuevos experimentos de descomposici\'on espinodal anisotr\'opica, adem\'as de nuevos trabajos te\'oricos. De hecho, nuestras observaciones, por ejemplo, en las propiedades estructurales y en las primeras etapas de la cin\'etica de segregaci\'on de fases, son f\'acilmente accesibles por experimentos de espectroscop\'ia y microcalorim\'etricos, respectivamente.
\end{itemize}
Muchas cuestiones quedan a\'un por desarrollar. Investigaciones adicionales, las cuales est\'an siendo elaboradas actualmente, involucran estudios detallados de las propiedades interfaciales y de los efectos de tama\~no finito. Otros problemas que ser\'ia conveniente abordar en un futuro son dilucidar de forma precisa el papel de la simetr\'ia hueco-part\'icula y de la forma del potencial de interacci\'on en las propiedades cr\'iticas; analizar las etapas tard\'ias del proceso de segregaci\'on de fases, ya estudiadas en otros sistemas en y fuera del equilibrio, \cite{marro2,bray} y \cite{hurtado,levine}, respectivamente; y estudiar, desde el punto de vista de la teor\'ia de vidrios las caracter\'isticas morfol\'ogicas y estructurales de las fases s\'olida y vidri\'atica (s\'olido policristalino) que presenta nuestro modelo.\\

\textit{Cap\'itulo~\ref{intro:Granos}.} En \'este cap\'itulo proporcionamos la base adecuada para nuestro estudio de gases granulares desarrollada en los Cap\'itulos~\ref{cap3} y \ref{cap4}. Discutimos algunas controversias, problemas abiertos y avances recientes en fluidos granulares, en particular, de la aplicabilidad de teor\'ias hidrodin\'amicas. Tambi\'en presentamos los ingredientes b\'asicos de los m\'etodos computaciones implementados en los siguientes cap\'itulos.\\

\textit{Cap\'itulo~\ref{cap3}.} Se presentaron c\'alculos anal\'iticos, adem\'as de simulaciones de din\'amica molecular por eventos, para un granulado monodisperso confinado en un recipiente anular sin gravedad y forzado por una pared ``termalizada". Consideramos el l\'imite cuasiel\'astico en una descripci\'on hidrodin\'amica que emplea coeficientes de transporte tipo Chapman-Enskog \cite{jenkins}, derivables de primeros principios. Desarrollamos un estudio exhaustivo de las inestabilidades de agregaci\'on heterog\'enea de part\'iculas, ruptura de simetr\'ia y separaci\'on de fases que aparecen en el modelo, y que nos permiten determinar con exactitud su diagrama de fases. Para comprobar y complementar las predicciones te\'oricas implementamos simulaciones de din\'amica molecular por eventos para este sistema. Nuestro estudio nos lleva a desarrollar algunas consideraciones sobre la f\'isica de las inestabilidades que tienen lugar el modelo. De \'este cap\'itulo concluimos:
\begin{itemize}
\item
El enfoque hidrost\'atico qued\'o completamente determinado por tres par\'a-metros adimensionales: el de disipaci\'on de energ\'ia, la fracci\'on de \'area ocupada por los granos y la raz\'on de aspecto del sistema.
\item
Nuestras simulaciones de din\'amica molecular mostraron (m\'as claramente para un n\'umero suficientemente elevado de part\'iculas) la formaci\'on de estados agregados con simetr\'ia azimutal alejados de la pared t\'ermica que fuerza al granulado. A pesar del (relativamente) escueto n\'umero de part\'iculas empleado, estos estados anulares fueron descritos de forma precisa y para densidades moderadas por la soluci\'on num\'erica de las ecuaciones hidrost\'aticas propuestas, las cuales emplean relaciones constitutivas tipo Enskog.
\item
Mediante un an\'alisis de estabilidad marginal se obtuvieron soluciones que bifurcan del estado m\'as b\'asico (el estado anular). Dicho de otro modo, estas soluciones predicen la ruptura espont\'anea de simetr\'ia de aquellos estados con simetr\'ia azimutal (inestables respecto de peque\~nas perturbaciones), hecho que ocurre s\'olamente en un determinado rango de par\'ametros. Determinamos las curvas de estabilidad marginal y las comparamos con las fronteras de la region espinodal (de compresibilidad negativa) del sistema.
\item
Se concluye que el mecanismo f\'isico de la inestabilidad de separaci\'on de fases es la compresibilidad negativa del granulado en la direcci\'on azimutal, causada por las p\'erdidas de energ\'ia en las colisiones. Matem\'aticamente, la separaci\'on de fases se manifiesta en la existencia de soluciones adicionales en cierta regi\'on del espacio de par\'ametros a la ecuaci\'on de densidad.
\item
Tambi\'en las simulaciones mostraron la existencia de separaci\'on de fases, predichas por la descripci\'on hidrodin\'amica, pero enmascaradas por grandes fluctuaciones espacio-temporales. Midiendo la distribuci\'on de probabilidad para la amplitud del modo de fundamental del desarrollo de Fourier del espectro azimutal de la densidad de part\'iculas, identificamos con claridad el cambio de estados sim\'etricos a estados con separaci\'on de fases.
\item
Adem\'as encontramos que la regi\'on del espacio de par\'ametros asociada a la inestabilidad, predicha por la descripci\'on hidrodin\'amica, se localiza dentro de la regi\'on de separaci\'on de fases observada en nuestras simulaciones. Esto implica la presencia de una regi\'on binodal (de coexistencia de fases) donde el estado anular es metaestable.
\item
Esperamos que las inestabilidades de agregaci\'on, ruptura de simetr\'ia y separaci\'on de fases sean f\'acilmente observables en experimentos. De hecho, el modelo aqu\'i propuesto es experimentalmente accesible, de modo que podr\'ia motivar nuevos experimentos en estas inestabilidades si atrajese la atenci\'on de la comunidad pertinente.
\end{itemize}
Es claro que la comparaci\'on de las predicciones de nuestra teor\'ia y simulaciones con experimentos ser\'ia altamente interesante. Al proponer la geometr\'ia anular esperamos motivar estudios experimentales de estas inestabilidades que podr\'ian resultar ventajosos en esta geometr\'ia. Un dispositivo experimental anular no presenta paredes laterales (como en una geometr\'ia rectangular) y por tanto evitar\'ia las innecesarias, e inevitables por otro lado, p\'erdidas por colisi\'on con las part\'iculas. Un posible experimento podr\'ia emplear esferas met\'alicas deslizando sobre una superficie cuasi-plana (ligeramente c\'oncava). La pared t\'ermica podr\'ia ser implementada por una anillo ligeramente exc\'entico (y quiz\'as rugoso) rotando r\'apidamente. Igualmente interesante para estudios subsiguientes ser\'ia determinar la naturaleza de las soluciones que se bifurcan (?`sub- o supercr\'iticamente?) del estado agregado con simetr\'ia azimutal. Otros estudios podr\'ian explotar las similitudes entre la fenomenolog\'ia observada en nuestro modelo y la de los anillos planetarios, en los cuales, tambi\'en se observan inestabilidades de agregaci\'on y de ruptura de simetr\'ia, entre otras.\\

\textit{Cap\'itulo~\ref{cap4}.} En este cap\'itulo tratamos con hidrodin\'amica granular y fluctuaciones en un sencillo gas granular bidimensional donde, debido a densidades elevadas (y no a inelasticidad elevada), descripciones continuas basadas en primeros principios dejan de tener validez. Los enfoques hidrodin\'amicos fueron puestos a prueba en un caso extremo empleando simulaciones de din\'amica molecular por eventos: cuando se forman agregados granulares cuya densidad es pr\'oxima al valor de empaquetamiento m\'aximo (denominados macropart\'iculas). El sistema modelo considerado est\'a formado por discos r\'igidos inel\'asticos encerrados en un recipiente circular cuya pared est\'a termalizada forzando al granulado y llev\'andolo lejos del equilibrio. Las principales conclusiones se resumen en:
\begin{itemize}
\item
Simulaciones de din\'amica molecular mostraron, a causa del enfriamiento por colisiones, un agregado fuertemente empaquetado de part\'iculas, con una interfase cuasicircular abrupta, con fluctuaciones espaciales y aislado de la pared t\'ermica por un gas muy diluido. La formaci\'on de la macropart\'icula requiere un cierto n\'umero de part\'iculas y la transici\'on, en funci\'on de la densidad, desde un estado (cuasi) homog\'eneo a la macropart\'i-cula es abrupta pero continua.
\item
Se resolvi\'o num\'ericamente un conjunto de ecuaciones granulares hidrost\'ati-cas que emplean como relaciones constitutivas las fenomenol\'ogicas de Grossman \textit{et al.} \cite{grossman}. Encontramos que, en un amplio rango de par\'ametros, el ajuste ---incluyendo la parte de m\'aximo empaquetamiento--- con la soluci\'on con simetr\'ia azimutal de las ecuaciones hidrost\'aticas es casi perfecto. Tambi\'en demostramos que, para el mismo sistema, el empleo de relaciones constitutivas tipo Enskog \cite{jenkins}, derivables de primeros principios, no describe adecuadamente ---mediante comparaci\'on directa con los resultados de simulaci\'on--- la regi\'on de mayor densidad del granulado.
\item
Estos resultados, que est\'an de acuerdo con resultados previos en geometr\'ias planas con y sin gravedad, extienden la aplicabilidad de las relaciones constitutivas de Grossman \textit{et al.} a coeficientes de restituci\'on ligeramente menores y a geometr\'ias no planas.
\item
Un an\'alisis de estabilidad marginal demostr\'o ---independientemente de las relaciones constitutivas empleadas--- que no hay soluciones estacionarias que bifurquen linealmente de la de simetr\'ia azimutal, esto es, no hay estados que manifiesten ruptura de simetr\'ia alguna. Esto est\'a de acuerdo con la persistencia de macropart\'iculas circulares y cuasicirculares que se observan en nuestras simulaciones.
\item
Estudiamos, por primera vez, las fluctuaciones de las macropart\'iculas, midiendo la distribuci\'on de probabilidad radial del centro de masas del sistema. Las fluctuaciones resultan ser gaussianas.
\item
Las fluctuaciones gaussianas observadas del centro de masas sugieren una descripci\'on mesosc\'opica tipo Langevin en t\'erminos de una macropart\'icula confinada un potencial atractivo de naturaleza hidrodin\'amica y forzada por ruido de n\'umero finito ---queremos decir que el agregado efect\'ua un movimiento browniano dentro del sistema. Se determin\'o la correspondiente ecuaci\'on de Langevin.
\item
En contra de lo que ocurre en sistemas en equilibrio ---y en muchos lejos del equilibrio---, en los cuales la magnitud relativa de las fluctuaciones decrece al incrementar el n\'umero de part\'iculas\footnote{Para un gas ideal en equilibrio: $\sigma \sim {\cal O}(N^{-1/2})$, donde $\sigma$ es la magnitud relativa de las fluctuaciones y $N$ es el n\'umero de part\'iculas.}, encontramos que las fluctuaciones persisten al incrementar el n\'umero de part\'iculas.
\end{itemize}
A\'un quedan algunos problemas en nuestro modelo circular. Una cuesti\'on que deber\'ia ser tratada es si existen inestabilidades adicionales en este sistema modelo. Algo de intuici\'on podr\'ia obtenerse, por ejemplo, mediante medidas de orden posicional y orientacional del agregado, siguiendo la analog\'ia con la teor\'ia KTHNY \cite{melting} para fusi\'on o licuefacci\'on en sistemas en equilibrio. Concretamente, \'estas podr\'ian llevarse a cabo monitorizando las respectivas funciones de correlaci\'on. Es posible que tanto una soluci\'on general dependiente del tiempo como un estudio de estabilidad no-lineal contribuyan a la b\'usqueda. Otros posibles trabajos futuros son el estudio de inestabilidades en modelos tridimensionales relacionados (en el Ap\'endice~\ref{apendD} se incluye una primera aproximaci\'on a este tema) y de la fenomenolog\'ia mostrada por gases granulares polidispersos en este tipo de sistemas.\\

A pesar de que com\'unmente los ap\'endices se relegan a un segundo plano, creemos que nuestros Ap\'endices~\ref{apendC} y \ref{apendD} merecen sus propias conclusiones. \\

\textit{Ap\'endice~\ref{apendC}.} Recordamos que en este ap\'endice presentamos un modelo difusivo y con arrastre cuyo prop\'osito es comprender mejor el balance entre campo externo y ba\~no t\'ermico que ocurre en el DLG y en otros modelos relacionados (concretamente, los modelos reticulares estudiados en el Cap\'itulo~\ref{cap1}). Los principales resultados son:
\begin{itemize}
\item
Estudiamos la estabilidad de la interfase (orientada seg\'un la direcci\'on del campo) en el l\'imite de campos saturantes, encontrando un cambio de fase de segundo orden \textit{geom\'etrico} o \textit{topol\'ogico} en funci\'on de la conectividad de los sitios del ret\'iculo. Para bajas conectividades existe orden, dominan los efectos t\'ermicos y la interfase es estable, mientras que al incrementar la conectividad (por encima de un valor cr\'itico) el orden desaparece, domina el campo externo y la interfase se vuelve inestable.
\item
Este cambio de fase \textit{geom\'etrico} confirma que el balance campo--temperatura no es tan sencillo como la din\'amica microsc\'opica sugiere, es decir, no es una mera superposici\'on lineal de ambos. Este hecho es una manifestaci\'on clara de una din\'amica colectiva emergente.
\item
Adem\'as se determinaron los par\'ametros cr\'iticos del cambio de fase.
\end{itemize}

\textit{Ap\'endice~\ref{apendD}.} En este ap\'endice desarrollamos una serie de estudios te\'oricos para mostrar la existencia de inestabilidades de agregaci\'on y ruptura de simetr\'ia en flujos granulares tridimensionales. De hecho, se trata de una sencilla extensi\'on a tres dimensiones de la geometr\'ia circular estudiada en el Cap\'itulo~\ref{cap4}. As\'i, el modelo es el de un gas de esf\'eras r\'igidas colisionando inel\'asticamente en el interior de un cilindro cuya pared curva est\'a termalizada. Para este estudio, se hizo necesario proponer un nuevo conjunto de relaciones constitutivas.
\begin{itemize}
\item
Derivamos relaciones constitutivas para gases granulares tridimensionales empleando argumentos de \textit{volumen libre} en la vecindad de la densidad de m\'aximo empaquetamiento, y sugerimos una interpolaci\'on entre el l\'imite de m\'axima densidad y las relaciones usuales en el l\'imite dilu\'ido. Para ello nos basamos en la estrategia seguida por Grossman \textit{et al.} \cite{grossman} en dos dimensiones, que emplea t\'ecnicas de volumen exclu\'ido para determinar las relaciones constitutivas. 
\item
Mediante los m\'etodos desarrollados en los Cap\'itulos~\ref{cap3} y \ref{cap4} pronosticamos y caracterizamos las inestabilidades de agregaci\'on y ruptura de simetr\'ia en un recipiente cil\'indrico. La presencia de esta \'ultima supone una gran diferencia con lo observado en el Cap\'itulo~\ref{cap4}, donde su hom\'ologo bidimensional no presentaba ruptura de simetr\'ia alguna.
\end{itemize}

\subsubsection{Publicaciones\label{resumen:public}}
Los estudios descritos en esta tesis han dado lugar a las siguientes publicaciones:

\begin{itemize}
\item
Manuel D\'iez-Minguito, Pedro L. Garrido, and Joaqu\'in Marro, \emph{Lennard-Jones and lattice models of driven fluids}, Phys. Rev. E \textbf{72}, 026103 (2005).
\item
Joaqu\'in Marro, Pedro L. Garrido, and Manuel D{\'i}ez-Minguito, \emph{Nonequili-brium anisotropic phases, nucleation and critical behavior in a driven Lennard-Jones fluid}, Phys. Rev. B \textbf{73}, 184115 (2006).
\item
Manuel D\'iez-Minguito, Pedro L. Garrido, and Joaqu\'in Marro, \emph{Lattice versus Lennard-Jones models with a net particle flow} in \textit{Traffic and Granular Flows '05}, Edited by A. Schadschneider, T. P\"oschel, R. K\"uhne, M. Schreckenberg, D. E. Wolf, (Springer, Berlin, 2007).
\item
Baruch Meerson, Manuel D\'iez-Minguito, Thomas Schwager, and Thorsten P\"{o}schel, \emph{Close-packed granular clusters: hydrostatics and persistent Gaussian fluctuations}, Granular Matter, Granular Matter \textbf{10}(1), 21-27 (2007).
\item
Manuel D\'iez-Minguito and Baruch Meerson, \emph{Phase separation of a driven granular gas in annular geometry}, Phys. Rev. E \textbf{75}, 011304 (2007).
\item
Manuel D\'iez-Minguito, J. Marro, and P. L. Garrido, \emph{On the similarities and differences between lattice and off-lattice models of driven fluids} in \textit{Workshop on Complex Systems: New Trends and Expectations}, Euro. Phys. J.-Special Topics \textbf{143}, 269 (2007).
\item
Manuel D\'iez-Minguito, Pedro L. Garrido, Joaqu\'in Marro, and Francisco de los Santos, \emph{Driven two-dimensional Lennard-Jones fluid} in \textit{Modeling Cooperative Behavior in the Social Sciences}, P. L. Garrido, J. Marro, M. A. Mu\~noz (eds.), AIP Conference Proceedings \textbf{779}, 199 (2005).
\item
Manuel D\'iez-Minguito, Pedro L. Garrido, Joaqu\'in Marro, and Francisco de los Santos, \emph{Simple computational Lennard-Jones fluid driven out-of-equilibrium}, in \textit{XIII Congreso de F\'isica Estad\'istica' FISES'05}, Abstract Book (2005).
\item
Manuel D\'iez-Minguito and Baruch Meerson, \emph{Phase separation of a driven granular gas in annular geometry}, in \textit{XIV Congreso de F\'isica Estad\'istica FISES'06}, Abstract Book (2006).
\end{itemize}

\vspace*{1cm}
\begin{center}
Este trabajo ha sido financiado parcialmente por el Ministerio de Educaci\'on y Ciencia (MEC) de Espa\~na y fondos FEDER (proyecto FIS2005-00791).
\end{center}

\newpage
\thispagestyle{empty}
\vspace*{6cm}

\end{document}